\documentclass[a4paper,11pt]{article}
\usepackage{jheppub}
\usepackage[T1]{fontenc}
\usepackage[utf8]{inputenc}
\usepackage{lmodern,mathrsfs,amsthm,mathtools}
\usepackage{microtype,enumitem,xcolor,slashed,xurl}
\renewcommand{\afterTocRuleSpace}{\bigskip}
\usepackage{float}

\newcommand{\be}{\begin{equation}}
\newcommand{\ee}{\end{equation}}

\numberwithin{equation}{section}

\newcommand{\dd}{\mathrm d}

\DeclareMathOperator*{\Res}{Res}

\theoremstyle{remark}

\makeatletter
\newcommand{\eqalias}[2]{%
  \begingroup
  \protected@edef\@currentlabel{\p@equation\theequation}%
  \edef\eqalias@page{\getpagerefnumber{#2}}%
  \protected@write\@auxout{}{%
    \string\newlabel{#1}{{\@currentlabel}{\eqalias@page}%
      {\@currentlabelname}{\@currentHref}{}}}%
  \endgroup
}

\makeatother

\title{Type IIB instanton corrections from pure modular unfolding}
\author[a,b]{Amineh Mohseni}
\affiliation[a]{Jefferson Physical Laboratory, Harvard University,\\
Cambridge, MA 02138, USA}
\affiliation[b]{INFN, Sezione di Milano,\\
Via G. Celoria 16, 20133 Milano, Italy}
\emailAdd{Amineh.Mohseni@mi.infn.it}
\hypersetup{pdftitle={Instanton corrections from pure modular unfolding},
pdfauthor={Amineh Mohseni}}

\abstract{We study the reconstruction of instanton corrections to type IIB
higher-derivative interactions from their low Fourier modes using modular
symmetry, without invoking supersymmetry differential equations. This provides a framework for studying non-BPS interactions
through their modular coefficient functions. For the $D^6R^4$ coefficient, Rankin--Selberg unfolding
determines the continuous spectral overlap from the complete
zero mode. Assuming the conjectured simplicity of the even
cusp spectrum, we extract the cusp amplitudes from Mellin
residues of the complete first mode. We combine both
contributions to reconstruct all higher Fourier modes,
using Hecke relations for the cusp contribution. Using the Kuznetsov trace formula and bounds on the spectral
tail, we derive the $g_s^{-1/2}$ and constant terms in the weak-coupling
instanton prefactor at every fixed positive charge. The leading term
arises collectively from the infinite cusp sum, and both coefficients
agree with the known expansion. With incomplete low-mode data, we
construct modular ambiguities and give sufficient conditions for the
stability of leading higher-charge coefficients. Applied to the supplied
$D^8R^4$ data, these results identify the additional information needed
for higher-charge predictions.}

\begin{document}
\maketitle
\section{Introduction}
\label{intro:scope}

Ten-dimensional type IIB string theory has a nonperturbative
$SL(2,\mathbb{Z})$ duality. This symmetry relates the perturbative terms
in the low-energy four-graviton amplitude to its D-instanton corrections. In
Einstein frame, the coefficients of higher-derivative interactions are
modular functions of the axiodilaton. Their Fourier expansion in the
axion separates sectors by net instanton charge, as the $R^4$ coupling
already illustrates~\cite{GreenGutperle1997}. Configurations with the same net charge can have different classical
actions, so a single Fourier mode can contain several exponential
orders at weak coupling. Our aim is to use modular symmetry to
reconstruct the higher Fourier modes from the complete zero and first
modes, without invoking supersymmetry differential equations. The reconstruction does not require BPS protection, so it
also provides a framework for studying non-BPS interactions
with modular coefficient functions.

The $R^4$ and $D^4R^4$ coefficients are Eisenstein series, whose Fourier
expansions make their instanton terms explicit~\cite{GRV2008}. The
$D^6R^4$ coefficient also has a Maass cusp contribution, invisible in its
zero Fourier mode. Green, Miller and Vanhove (GMV) obtained the exact
Fourier modes from an inhomogeneous Laplace equation and appropriate
growth conditions~\cite{GMV}. Klinger-Logan, Miller and Radchenko developed
a direct Poincar\'e-series construction of the same
coefficient~\cite{KLMR}. In their spectral treatment of square-integrable
modular observables in $\mathcal N=4$ SYM, Collier and Perlmutter explained
how the complete zero and first Fourier modes determine all higher modes,
assuming a nondegenerate cusp spectrum~\cite{CP}. They also used the inhomogeneous $D^6R^4$ Laplace equation to express
the cusp overlaps in terms of Hecke $L$-values.

Building on this reconstruction principle, we explicitly calculate the
continuous and cusp spectral amplitudes of the $D^6R^4$ coefficient
from its zero and first Fourier modes without using supersymmetry
differential equations. At each fixed positive charge, we use the
resulting spectral representation and bounds on its tail to extract
the first two terms multiplying the leading instanton exponential at
weak coupling. The first scales as the inverse square root of the
string coupling, and the second is constant.

The inputs are the complete GMV functions $f_0(y)$ and $f_1(y)$ for all
$y=g_s^{-1}>0$, including neutral pairs in the zero mode and
opposite-charge pairs in the first. Their original derivation used the
supersymmetry equation. The higher Fourier modes computed by GMV
provide a check on our reconstruction.

We subtract the Eisenstein terms fixed by the prescribed growth and
use regularized Rankin--Selberg unfolding of the complete zero mode
to determine the continuous spectral overlap. Removing the first
Fourier modes of the Eisenstein subtraction and the continuous
contribution from $f_1$ leaves its cusp part. The residues of its
Mellin transform determine the total amplitude in each even cusp
eigenspace. We evaluate these residues in terms of completed Hecke
$L$-values by applying Motohashi's additive-divisor
formula~\cite{Motohashi1994} to the shifted divisor sum arising from
the opposite-charge pairs. Assuming that, as conjectured, the even Maass cusp
spectrum of the full modular group is simple, each residue determines
an individual cusp amplitude. Hecke relations then fix its
contribution to every higher Fourier mode.

At fixed positive charge $k$, the weak-coupling limit requires control
of the full Maass cusp sum. After factoring out $e^{-2\pi ky}$, each
individual term remains bounded as $y\to\infty$. The Bessel kernel,
however, probes a growing range of spectral parameters, with cusp
forms of $t_j$ up to order $\sqrt y$ contributing collectively to
the $\sqrt y$ term in the prefactor. A uniform Gaussian approximation
to the kernel, together with bounds on the spectral tail, reduces the
calculation to a spectral moment involving both the cusp sum and the
Eisenstein integral. We obtain its meromorphic continuation using the
first-moment identity of Hoffstein, Lee and Nastasescu, which follows
from the Kuznetsov trace formula~\cite{Kuznetsov1981,HLN}. The residue of the spectral moment fixes the square-root coefficient.
Its finite value at the origin and the restored Eisenstein terms give
the constant term.

For each fixed positive charge $k$, we find the following weak-coupling
expansion for the reconstructed $D^6R^4$ Fourier mode in the GMV
normalization:
\be
 f_k(y)=e^{-2\pi ky}\left[
 \frac{8\zeta(3)\sigma_{-2}(k)}{\sqrt{k}}\,\sqrt{y}
 +B_k+o(1)\right],
 \qquad y\to\infty,
 \label{paper:intro_result}
\ee
where $\sigma_a(k)=\sum_{d\mid k}d^a$ and $B_k$ is the explicit finite
sum over positive charge pairs in~\eqref{d6a:constant_general}. Here
$o(1)$ denotes a remainder in the instanton prefactor that tends to
zero as $y\to\infty$ at fixed $k$. We give a sharper estimate for this remainder in~\eqref{d6a:opening_result}. Both coefficients agree with those in the known Fourier
expansion~\cite{GMV,KLMR}. The $\sqrt{y}$ term arises entirely from
the infinite sum over Maass cusp forms. The constant $B_k$ combines
the contribution of the spectral moment with that of the restored
Eisenstein terms. 

Complete low-mode functions contain more information than perturbative
expansions or finitely many instanton coefficients. We compare modular
functions that agree on the available low-mode data to quantify the
remaining freedom in their continuous spectral overlaps and cusp
amplitudes. Sufficient bounds on the differences in spectral weights
guarantee that the leading coefficient of a growing
instanton prefactor is preserved. We also show that by
adding an infinite sum of Maass cusp forms, we can change
the instanton prefactor at any prescribed higher charge.
This leaves both the complete zero mode and specified
asymptotic data at charge one unchanged. A second modular
construction changes the prefactor at charge two while
preserving the full asymptotic expansion about a single
instanton in the first mode. In the zero mode, it preserves
both the perturbative expansion and every order in the
prefactor of the leading instanton--anti-instanton
contribution. Additional constraints from string theory may restrict
these modular ambiguities.

For $D^8R^4$, perturbative calculations give several terms in the
weak-coupling expansion of the zero Fourier
mode~\cite{DHokerGreen,CD2412,DGP1806,Pioline1810}, while the disk
amplitude fixes the leading one-instanton coefficient~\cite{Agmon}.
Among modular functions that reproduce these data, the Fourier mode
at any prescribed charge $k\ge2$ can vary at order $y e^{-2\pi ky}$.
Our stability bounds identify what additional control of the spectral
weights would suffice to predict the coefficient at that order.

The paper is organized as follows. In Section~\ref{rev:framework} we describe how to reconstruct an even
modular function from its low Fourier modes. We apply this framework
to the $D^6R^4$ coefficient in Section~\ref{rev:cusp_reconstruction}
and derive its weak-coupling instanton asymptotics at fixed charge
in Section~\ref{d6a:section}. Section~\ref{rev:incomplete} examines
the ambiguities left by incomplete low-mode data and gives sufficient
conditions for the stability of higher-charge coefficients. In
Section~\ref{d8e:input} we apply these results to $D^8R^4$.
We conclude in Section~\ref{paper:conclusion}. The appendices give the analytic continuation
arguments, arithmetic evaluations and amplitude conventions used in
the analysis.

\section{Modular spectral reconstruction}\label{rev:framework}

Before turning to a particular interaction, we fix the physical
conventions and explain how the low Fourier modes determine the
spectral weights shared by all instanton charges. After subtracting the prescribed weak-coupling growth, we use
regularized Rankin--Selberg unfolding of the complete zero mode to
determine the continuous Eisenstein contribution at every nonzero
charge. The first mode
supplies the complementary information about Maass cusp forms, whose
zero modes vanish. Under the conjectured simplicity of the even cusp
spectrum, Mellin residues of the cusp part of the first mode determine
the individual amplitudes, and Hecke relations fix their contributions
at higher charge.

\subsection{Physical and spectral conventions}\label{rev:fourier_conventions}
We work in ten-dimensional type IIB string theory with axiodilaton
$\tau=x+iy=C_0+i/g_s$. In Einstein frame, the coefficient functions
in the low-energy four-graviton amplitude are invariant under the
action of $\Gamma=\mathrm{PSL}_2(\mathbb Z)$ on $\tau$. We express
the Mandelstam invariants $S,T,U$ and the threshold scales in Planck
units.

Following~\cite{Agmon}, we normalize the four-graviton amplitude by
its tree-level supergravity value. The resulting ratio
$M=\mathcal A/\mathcal A_{\rm SUGRA}$ has the low-energy expansion
\begin{align}
 M={}&1+STU\bigl[\mathcal F_0
 +\mathcal F_4(S^2+T^2+U^2)
 +\mathcal F_6(S^3+T^3+U^3)\notag\\
 &\hspace{16mm}+\mathcal F_8^{(\mu)}(S^4+T^4+U^4)\bigr]
 +\cdots .
 \label{intro:amplitude_expansion}
\end{align}
Here $\mathcal F_n$ denotes the coefficient of the $D^nR^4$
interaction for $n=0,4,6,8$. Since $S+T+U=0$, there is no
independent local $D^2R^4$ term. The ellipsis includes
local interactions beyond $D^8R^4$ as well as nonanalytic contributions
from massless thresholds. At order $D^8R^4$, the local coefficient
$\mathcal F_8^{(\mu)}$ depends on the scale $\mu$ used to separate
these threshold contributions from the local term, as described in
Section~\ref{d8e:conventions}. For the spectral analysis we introduce the rescaled couplings
\be
 F_{R^4}=64\,\mathcal F_0,\qquad
 F_{D^4R^4}=1024\,\mathcal F_4,\qquad
 f=4096\,\mathcal F_6,
 \label{intro:amplitude_functions}
\ee
where $f$ is the $D^6R^4$ coefficient in the GMV normalization.

For a modular coefficient $F$, we have the following Fourier expansion
in the axion $x=C_0$:
\be
 F(x+iy)=\sum_{k\in\mathbb Z}F_k(y)e^{2\pi ikx},\qquad
 F_k(y)=\int_0^1F(x+iy)e^{-2\pi ikx}\,\dd x.
 \label{intro:fourier}
\ee
For a real coefficient function even in $x$, the Fourier coefficients
satisfy $F_{-k}(y)=F_k(y)\in\mathbb R$, so the coefficient of
$\cos(2\pi kx)$ is $2F_k(y)$ for $k>0$. The Fourier index $k$ labels
the net D-instanton charge. A sector with $n_+$ units of instanton
charge and $n_-$ units of anti-instanton charge carries the factor
$e^{2\pi i(n_+-n_-)x}e^{-2\pi(n_++n_-)y}$. The zero mode can
therefore contain neutral instanton--anti-instanton pairs in addition
to the perturbative terms, while the sectors $(1,0),(2,1),\ldots$
all belong to the first mode. In Einstein frame, a genus-$h$ contribution to the perturbative zero
mode of a $D^{2r}R^4$ coupling scales as $y^{(r+3)/2-2h}$, with
logarithms when required by massless thresholds. We call a low mode complete when its Fourier coefficient
$F_k(y)$ is known for all $y>0$.

We compute the spectral overlaps on the standard fundamental domain
of $\Gamma$ in the upper half-plane,
\be
 \mathcal F=\left\{\tau=x+iy\in\mathbb H:
 |x|\le\tfrac12,\ |\tau|\ge1\right\},
 \qquad \operatorname{vol}_{\rm hyp}(\mathcal F)=\frac{\pi}{3}.
 \label{intro:fundamental_domain}
\ee
The invariant hyperbolic measure, the weight-zero Petersson inner
product and the nonnegative hyperbolic Laplacian are
\be
 \dd\mu=\frac{\dd x\,\dd y}{y^2},\qquad
 \langle g,h\rangle=\int_{\mathcal F}g\,\overline h\,\dd\mu,\qquad
 \Delta=-y^2(\partial_x^2+\partial_y^2).
 \label{intro:inner_product}
\ee
\eqalias{intro:laplacian}{intro:inner_product}
A modular function $g$ is square integrable if
$\|g\|_2^2=\langle g,g\rangle<\infty$. For a smooth coefficient,
convergence need only be checked at the weak-coupling cusp
$y\to\infty$, where a power $y^a$ is square integrable precisely
when $\Re a<\tfrac12$. We subtract a modular function $F_{\rm sub}$
that matches the non-square-integrable growth of $F$ and apply the
spectral decomposition to the remainder $g=F-F_{\rm sub}$.

\subsection{Eisenstein series and cusp forms}
\label{rev:eisenstein_normalizations}\label{rev:cusp_background}

Eisenstein series enter both the subtraction of the prescribed
weak-coupling growth and the spectral expansion of the remainder.
On the line $\Re s=\tfrac12$, they furnish the continuous spectrum.
The nonzero Fourier modes can also receive a discrete contribution
from Maass cusp forms. 

We use the normalization $E_s$ for spectral overlaps and denote the
lattice-normalized Eisenstein series by $\mathbb E_s$. As we recall
below, the protected $R^4$ and $D^4R^4$ coefficient functions are
given by $\mathbb E_{3/2}$ and $\tfrac12\mathbb E_{5/2}$,
respectively~\cite{GRV2008}. For $\Re s>1$,
both are given by absolutely convergent sums,
\begin{align}
 E_s(\tau)&=\sum_{\gamma\in\Gamma_\infty\backslash\Gamma}
 \bigl(\Im(\gamma\tau)\bigr)^s,
 \label{intro:Eorbit}\\
 \mathbb E_s(\tau)&=
 \sum_{(m,n)\ne(0,0)}\frac{y^s}{|m\tau+n|^{2s}}
 =2\zeta(2s)E_s(\tau).
 \label{intro:lattice}
\end{align}
Here $\Gamma_\infty$ is the subgroup generated by
$\tau\mapsto\tau+1$. The series $E_s$ satisfies
$\Delta E_s=s(1-s)E_s$ and extends meromorphically in $s$.
Writing
\be
 \Lambda(s)=\pi^{-s}\Gamma(s)\zeta(2s),\qquad
 \varphi(s)=\frac{\Lambda(1-s)}{\Lambda(s)},
 \label{intro:Ezero}
\ee
\eqalias{d6c:Lambda_phi}{intro:Ezero}
In this normalization, the Fourier coefficients of $E_s$ are~\cite{CP,GMV}
\begin{align}
 (E_s)_0(y)&=y^s+\varphi(s)y^{1-s},
 \label{d6c:E_zero}\\
 (E_s)_k(y)&=\frac{2\sqrt y}{\Lambda(s)}|k|^{s-1/2}
 \sigma_{1-2s}(|k|)K_{s-1/2}(2\pi|k|y),\qquad k\ne0,
 \label{intro:Ezero0}
\end{align}
\eqalias{d6c:eq:Ek}{intro:Ezero0}
where $\sigma_a(n)=\sum_{d\mid n}d^a$. With
$E_s^*=\Lambda(s)E_s$, the functional equations take the form
\be
 \Lambda(s)=\Lambda(\tfrac12-s),\qquad E_s^*=E_{1-s}^*.
 \label{intro:Ecompleted}
\ee

The protected couplings provide two examples in which a single
Eisenstein series determines the contributions at every charge.
We have, $F_{R^4}=\mathbb E_{3/2}$ and
$F_{D^4R^4}=\tfrac12\mathbb E_{5/2}$~\cite{GRV2008}. For $R^4$,
the Fourier modes are
\begin{align}
 (F_{R^4})_0(y)&=2\zeta(3)y^{3/2}+4\zeta(2)y^{-1/2},\notag\\
 (F_{R^4})_k(y)&=8\pi|k|\sigma_{-2}(|k|)\sqrt y
 K_1(2\pi|k|y),\qquad k\ne0,
 \label{intro:R4_modes}
\end{align}
while for $D^4R^4$ they are
\begin{align}
 (F_{D^4R^4})_0(y)&=\zeta(5)y^{5/2}
 +\frac43\zeta(4)y^{-3/2},\notag\\
 (F_{D^4R^4})_k(y)&=\frac{8\pi^2}{3}|k|^2\sigma_{-4}(|k|)
 \sqrt yK_2(2\pi|k|y),\qquad k\ne0.
 \label{intro:protected}
\end{align}
At fixed charge $k>0$ and fixed Bessel order $\nu$, the weak-coupling
behavior follows from
\be
 \sqrt yK_\nu(2\pi ky)=\frac{e^{-2\pi ky}}{2\sqrt k}
 \left[1+\frac{4\nu^2-1}{16\pi ky}+O_{\nu,k}(y^{-2})\right],
 \qquad y\to\infty.
 \label{intro:fixedK}
\ee
\eqalias{d6c:ambiguity_fixed_order}{intro:fixedK}

For $R^4$ and $D^4R^4$, the charged modes follow directly from the
Fourier expansion of the corresponding Eisenstein series. Modular
invariance and the complete zero mode alone leave room for an even
Maass cusp form, whose addition could change the Fourier mode at a given charge $k\ne0$. To describe such contributions, we choose an
orthonormal basis of Maass cusp forms $u_j$ satisfying
\be
 \Delta u_j=(\tfrac14+t_j^2)u_j,\qquad (u_j)_0=0,\qquad
 \langle u_i,u_j\rangle=\delta_{ij}.
 \label{intro:cusp_definition}
\ee
For the full modular group, all spectral parameters $t_j$
are real and positive~\cite{IwaniecSpectral}. The
Fourier expansion of each basis function is
\be
 u_j(x+iy)=\sum_{k\ne0}\rho_j(k)\sqrt y
 K_{it_j}(2\pi|k|y)e^{2\pi ikx}.
 \label{intro:cusps}
\ee
Here the Bessel order varies with the Laplace eigenvalue. Applying
the fixed-order expansion~\eqref{intro:fixedK} to an infinite cusp
sum therefore requires control of the spectral tail.

\paragraph{Roelcke--Selberg spectral theorem.}
Every $g\in L^2(\Gamma\backslash\mathbb H)$ has a unique orthogonal
decomposition into a constant, a continuous Eisenstein component and
a discrete cuspidal component~\cite{IwaniecSpectral},
\be
 g=C_g+\frac1{4\pi i}\int_{\Re s=1/2}
 \langle g,E_s\rangle E_s\,\dd s
 +\sum_j\langle g,u_j\rangle u_j,
 \qquad C_g=\frac3\pi\int_{\mathcal F}g\,\dd\mu.
 \label{intro:spectral}
\ee
\eqalias{intro:constant_projection}{intro:spectral}
The contour runs upward along $s=\tfrac12+it$, with measure
$\dd t/(4\pi)$ over $t\in\mathbb R$, or equivalently $\dd t/(2\pi)$
after folding the integral onto $t>0$. The overlaps
$\langle g,E_s\rangle$ and $\langle g,u_j\rangle$ give the Eisenstein
and cusp coefficients, respectively. The constant $C_g$ is the average
of $g$ over the fundamental domain.

\subsection{Rankin--Selberg unfolding}\label{rev:growth_framework}
Rankin--Selberg unfolding relates the Eisenstein overlap to a Mellin
transform of the zero Fourier mode. For a modular function $g$ that
decreases rapidly at the cusp, it gives
\be
 R_g(s):=\int_{\mathcal F}gE_s\,\dd\mu
 =\int_0^\infty g_0(y)y^{s-2}\,\dd y,
 \qquad \Re s>1.
 \label{intro:RS_unfold}
\ee
Using the modular invariance of $g$, the Eisenstein sum unfolds the
integral over $\mathcal F$ to the strip $-\tfrac12\le x<\tfrac12$,
$y>0$. Integration over one period in $x$ then selects the zero
Fourier mode $g_0(y)$. With the convention
$\mathcal M_{\psi}(z)=\int_0^\infty \psi(y)y^{z-1}\,\dd y$,
this gives $R_g(s)=\mathcal M_{g_0}(s-1)$. Meromorphic continuation to the
spectral line then yields
\be
 \langle g,E_s\rangle=R_g(1-s),\qquad \Re s=\tfrac12,
 \label{intro:RS}
\ee
where $\bar s=1-s$ accounts for complex conjugation in the Hermitian
inner product. In this rapidly decreasing case,
$C_g=\Res_{s=1}R_g(s)$.

\paragraph{Perturbative growth and regularization.}
We now allow perturbative growth at the cusp, described by a finite
sum of powers and logarithms,
\be
 F(x+iy)=P_F(y)+O_N(y^{-N}),\qquad
 P_F(y)=\sum_{a,m}c_{a,m}y^a(\log y)^m,
 \label{intro:growth_expansion}
\ee
for every $N>0$, thus $P_F$ contains all power and logarithmic terms
in the large-$y$ expansion. Write $F_0=P_F+N_F$.\footnote{As $y\to0$, we assume that $N_F$
grows at most as an inverse power of $y$. This ensures convergence
at the lower endpoint of the Mellin integral for sufficiently
large $\Re s$.} To obtain a square-integrable remainder,
we subtract a modular function $F_{\rm sub}$ built from Eisenstein
series and their derivatives at the exponents with $\Re a>1/2$.
At $a=1$, we use $\widehat E_1$, obtained by subtracting the pole
$3/[\pi(s-1)]$ from $E_s$ before taking $s\to1$. We require the subtraction to leave a remainder satisfying
\be
 g=F-F_{\rm sub}=O(y^\beta),\qquad \beta<\tfrac12.
 \label{intro:growth_remainder}
\ee

The modular subtraction makes $g$ square integrable. Regularized
unfolding also requires subtracting \emph{all} powers and logarithms
in the cusp expansion of the zero mode inside the Mellin integral,
including those compatible with square integrability. Applied to
$F_0$, this gives
\be
 R_F^{\rm reg}(s)=\int_0^\infty[F_0(y)-P_F(y)]y^{s-2}\,\dd y
 =\int_0^\infty N_F(y)y^{s-2}\,\dd y.
 \label{intro:RS_regularized}
\ee
Thus $R_F^{\rm reg}(s)=\mathcal M_{N_F}(s-1)$, initially for
sufficiently large $\Re s$. The zero mode of $F_{\rm sub}$ consists entirely of powers and
logarithms. Its subtraction therefore changes only the power and
logarithmic part of the zero mode. Removing that entire part from
either $F_0$ or $g_0$ leaves the same remainder $N_F$, so their
regularized transforms agree,
$R_g^{\rm reg}(s)=R_F^{\rm reg}(s)$.

Zagier's regularized unfolding~\cite{Zagier} shows that, after
meromorphic continuation, this transform equals
$\int_{\mathcal F}gE_s\,\dd\mu$ in the strip
$\beta<\Re s<1-\beta$, away from Eisenstein poles.
On the spectral line, this gives the Hermitian overlap
\be
 \langle g,E_s\rangle=R_F^{\rm reg}(1-s),\qquad \Re s=\tfrac12.
 \label{intro:RS_reg_overlap}
\ee
The constant projection $C_g$ is determined from the residue at
$s=1$, with the correction described below. Adding $F_{\rm sub}$
to the spectral expansion of $g$ then recovers $F$.

To determine the constant projection, write the power and logarithmic
part of $g_0$ as
\be
 g_0(y)\sim\sum_{a,m}b_{a,m}y^a(\log y)^m,
 \qquad y\to\infty.
\ee
Introduce the truncated fundamental domain $\mathcal F_T$ and collect
the integrated asymptotic terms in $H_g(T;s)$,
\be
 \mathcal F_T=\{\tau\in\mathcal F:\Im\tau\le T\},\qquad
 H_g(T;s)=\sum_{a,m}b_{a,m}
 \partial_a^m\frac{T^{s+a-1}}{s+a-1}.
\ee
Regularized unfolding then gives
\be
 R_g^{\rm reg}(s)=\lim_{T\to\infty}
 \left[\int_{\mathcal F_T}gE_s\,\dd\mu
 -H_g(T;s)-\varphi(s)H_g(T;1-s)\right].
 \label{intro:RS_truncated}
\ee
At $s=1$, the Eisenstein series has residue $3/\pi$. The constant
term $b=b_{0,0}$ in the cusp expansion of $g_0$ contributes residue
$b$ to $H_g(T;s)$. Pure logarithms give higher-order poles with zero
residue, while the residue of the reflected subtraction vanishes
as $T\to\infty$. Hence
\be
 C_g=\Res_{s=1}R_F^{\rm reg}(s)+b.
 \label{intro:RS_reg_average}
\ee
This correction affects only charge zero. At every nonzero charge,
the continuous contribution is determined by the zero mode through
\be
 (g_{\rm cont})_k(y)=\frac1{4\pi i}\int_{\Re s=1/2}
 R_F^{\rm reg}(1-s)(E_s)_k(y)\,\dd s,
 \qquad k\ne0.
 \label{intro:RS_higher}
\ee

\subsection{First-mode residues and Hecke reconstruction}
\label{rev:first_residual_framework}
Cusp forms have vanishing zero mode, so their amplitudes remain
undetermined by the unfolding above. We extract them from Mellin
residues of the cusp part of the first Fourier mode.
The gravitational couplings considered here satisfy
$F(-x+iy)=F(x+iy)$, so only even Maass cusp forms
contribute.\footnote{This symmetry follows from $(-1)^{F_L}$, which
reverses the Ramond--Ramond potentials while preserving the metric
and dilaton~\cite{TachikawaYonekura2019}.}

The zero-mode analysis fixes the first Fourier modes of the modular
subtraction $F_{\rm sub}$ and the continuous contribution
$g_{\rm cont}$. Subtracting these from $F_1(y)$ isolates the first
mode of the cusp contribution, which we denote by $h_1(y)$,
\begin{align}
 h_1(y)&:=F_1(y)-(F_{\rm sub})_1(y)-(g_{\rm cont})_1(y)
 =\sqrt y\sum_{j\ {\rm even}}d_jK_{it_j}(2\pi y),\\
 d_j&=\langle g,u_j\rangle\rho_j(1).
 \label{intro:first_residual}
\end{align}
To extract the cusp amplitudes from $h_1$, it is useful to introduce
the Mellin transform of the Bessel kernel,
\be
 \mathscr K(s,t):=\int_0^\infty y^{s-1}K_{it}(2\pi y)\,\dd y
 =\frac{\pi^{-s}}4
 \Gamma\left(\frac{s+it}{2}\right)
 \Gamma\left(\frac{s-it}{2}\right),\qquad \Re s>0.
 \label{intro:Bessel_Mellin}
\ee
\eqalias{d6c:ordinary_kernel}{intro:Bessel_Mellin}
For $t>0$ the kernel has poles at $s=\pm it-2n$, $n=0,1,\ldots$.
Write $\mathcal M_h$ for the Mellin transform of $h_1/\sqrt y$,
\be
 \mathcal M_h(s)=\int_0^\infty y^{s-3/2}h_1(y)\,\dd y.
\ee
For each even spectral parameter $T>0$, its residue gives the total
amplitude in that eigenspace,
\be
 \mathcal D_F(T):=\sum_{j:t_j=T}d_j
 =\frac{2\pi^{iT}}{\Gamma(iT)}\Res_{s=iT}\mathcal M_h(s).
 \label{intro:cusp_residue}
\ee
Throughout the $D^6R^4$ reconstruction and its asymptotic analysis,
we assume the conjectured simplicity of the even Maass cusp
spectrum~\cite{Terras}. Each eigenspace then contains
a single basis form, so $\mathcal D_F(t_j)=d_j$. 

\paragraph{Damped Mellin transform.}
To justify taking residues term by term, we introduce the
exponentially damped Mellin transform
\begin{align}
 \mathcal M_{h,d}(s)
 &:=\int_0^\infty y^{s-3/2}e^{-2\pi y}h_1(y)\,\dd y
 =\sum_jd_j\mathscr K_d(s,t_j),
 \qquad \Re s>\tfrac12,
 \notag\\
 \mathscr K_d(s,t)
 &=\frac{\sqrt\pi}{(4\pi)^s}
 \frac{\Gamma(s+it)\Gamma(s-it)}{\Gamma(s+1/2)}.
 \label{intro:damped_Mellin}
\end{align}
\eqalias{d6c:damped_Mellin}{intro:damped_Mellin}
For $\Re s>1/2$, this transform is an $L^2$ inner product
with a Poincar\'e series. The Cauchy--Schwarz inequality
and bounds on $\mathscr K_d$ give locally uniform convergence
of the spectral sum away from its poles. This allows residues
to be taken term by term without a prior formula for $d_j$.
If $h_1$ decays faster than every inverse power at infinity,
the absolute unfolding bound near zero then shows that
$\mathcal M_h-\mathcal M_{h,d}$ is holomorphic for
$\Re s>-1/2$. For $T>0$, the ordinary and damped kernels
$\mathscr K(s,T)$ and $\mathscr K_d(s,T)$ have the same
residue $\Gamma(iT)/(2\pi^{iT})$ at $s=iT$.
Thus damping preserves the residue formula.
Appendix~\ref{d6c:Mellin_justification} proves these statements
and verifies the required decay for $D^6R^4$.

For any fixed $Y>0$, we may split the transform as
\be
 \mathcal M_{h,d}(s)=
 \int_0^Y y^{s-3/2}e^{-2\pi y}h_1(y)\,\dd y
 +\int_Y^\infty y^{s-3/2}e^{-2\pi y}h_1(y)\,\dd y.
 \label{intro:Mellin_split}
\ee
The second integral is entire in $s$. The residues therefore come
from continuing the first integral, which a finite weak-coupling
expansion of the first mode leaves undetermined.

\paragraph{Reconstruction.}To reconstruct the higher modes, we choose a cusp basis that
simultaneously diagonalizes the commuting self-adjoint Hecke operators
\be
 (T_kF)(\tau)=\frac1{\sqrt k}
 \sum_{\substack{ad=k\\a,d>0}}\sum_{b=0}^{d-1}
 F\left(\frac{a\tau+b}{d}\right).
 \label{intro:Hecke_operator}
\ee
The first Fourier mode of $T_ku_j$ is
$(T_ku_j)_1=\rho_j(k)\sqrt yK_{it_j}(2\pi y)$.
Comparing this with the first mode of the eigenvalue equation
gives~\cite{Terras}
\be
 T_ku_j=\lambda_j(k)u_j,\qquad
 \rho_j(k)=\rho_j(1)\lambda_j(k),\qquad \lambda_j(1)=1.
 \label{intro:Hecke_coefficients}
\ee
In particular, $\rho_j(1)\ne0$, since otherwise all Fourier
coefficients of $u_j$ would vanish. The Hecke relations then determine the cusp contribution at every
higher charge from the amplitudes $d_j$. Once the individual $d_j$ are known, we have
\begin{align}
 F_k(y)={}&(F_{\rm sub})_k(y)
 +\frac1{4\pi i}\int_{\Re s=1/2}
 R_F^{\rm reg}(1-s)(E_s)_k(y)\,\dd s\notag\\
 &+\sqrt y\sum_{j\ {\rm even}}d_j\lambda_j(k)K_{it_j}(2\pi ky),
 \qquad k\ge1.
 \label{intro:reconstructed_modes}
\end{align}

The complete first mode also yields a uniqueness result. Two smooth,
even modular functions with $L^2$ difference and the same complete
first mode have identical continuous contributions, and their zero
modes differ at most by a constant, regardless of cusp degeneracies.
Under simplicity, the functions themselves differ only by that
constant, fixed by matching the constant terms in their zero-mode
asymptotics. Appendix~\ref{d6c:complete_first_uniqueness} gives the
proof. In the explicit reconstruction, we determine the continuous
overlap directly by unfolding the complete zero mode.

\section{Reconstructing the \texorpdfstring{$D^6R^4$}{D6R4} coefficient}
\label{d6c:chapter}\label{rev:cusp_reconstruction}
We now reconstruct the $D^6R^4$ coefficient $f=4096\mathcal F_6$
from its complete zero and first Fourier modes, known for all
$y>0$ from GMV~\cite{GMV}, without invoking supersymmetry
differential equations. The coefficient is a smooth, even modular
function with uniform cusp asymptotics
\be
 f(x+iy)=Ay^3+By+O(y^{-1}),\qquad
 A=\frac23\zeta(3)^2,\qquad B=\frac43\zeta(2)\zeta(3).
 \label{d6c:cusp_growth}
\ee
The growing terms specify the Eisenstein subtraction used below.
Rankin--Selberg unfolding of the complete zero mode determines
the continuous overlap, while Mellin residues of the residual
first mode give the cusp amplitudes. Hecke relations determine
their contribution at higher charge. Combining this with the
continuous contribution and the restored Eisenstein terms gives
the full higher Fourier modes.

\subsection{The zero mode and the continuous spectrum}
\label{d6c:input}\label{d6c:zero}\label{rev:d6_growth}\label{d6c:continuous_reconstruction}
We begin with the complete zero mode, which we write as
\begin{align}
 f_0(y)&=P(y)+N(y),
 \label{d6c:zero_split}
\end{align}
where the perturbative part $P(y)$ and the sum of neutral
nonperturbative contributions $N(y)$ are given by
\begin{align}
P(y)&=Ay^3+By+4\zeta(4)y^{-1}+\frac4{27}\zeta(6)y^{-3},
 \label{d6c:zero_polynomial}\\
 N(y)&=\frac{32\pi}{315}\sum_{n\ne0}
 \frac{\sigma_2(|n|)^2}{|n|^3}
 \sum_{i,j=0}^1r^{ij}(\pi|n|y)
 K_i(2\pi|n|y)K_j(2\pi|n|y).
 \label{d6c:neutral_input}
\end{align}
The functions $r^{ij}$ are given in Appendix~\ref{paper:GMV_data}.
The sum includes both orderings of each neutral charge pair $(n,-n)$.
At weak coupling, its leading exponential factor is $e^{-4\pi y}$,
arising from $n=\pm1$. Thus $N(y)$ lies beyond all orders of the
perturbative genus expansion.

Only the two growing powers $Ay^3$ and $By$ obstruct square
integrability. We remove them with the following modular subtraction
\be
 F_{\rm sub}=AE_3+B\widehat E_1,\qquad
 g=f-F_{\rm sub},\qquad
 \widehat E_1:=\lim_{s\to1}\left(E_s-\frac3{\pi(s-1)}\right).
 \label{d6c:eq:g}
\ee
\eqalias{d6c:Ehat_definition}{d6c:eq:g}
The Fourier coefficients of the regularized Eisenstein series
$\widehat E_1$ are
\begin{align}
 (\widehat E_1)_0(y)&=y-\frac3\pi\log y+\omega,
 \label{d6c:Ehat_zero}\\
 \omega&=\lim_{s\to1}\left[\varphi(s)-\frac{3}{\pi(s-1)}\right]=\frac6\pi\bigl[1-12\zeta'(-1)-\log(4\pi)\bigr],
 \label{d6c:Ehat_constant}\\
 (\widehat E_1)_k(y)&=\frac6\pi\sigma_{-1}(|k|)e^{-2\pi|k|y},
 \qquad k\ne0.
 \label{d6c:Ehat_nonzero}
\end{align}
Expanding $\varphi(s)y^{1-s}$ about $s=1$ and subtracting the pole
leaves the logarithmic term~\cite{Terras}.
The zero mode of $g$ is therefore
\be
 g_0(y)=\frac{3B}{\pi}\log y-B\omega
 +4\zeta(4)y^{-1}-A\varphi(3)y^{-2}
 +\frac4{27}\zeta(6)y^{-3}+N(y).
 \label{intro:d6_remainder_zero}
\ee
The logarithmic growth of $g$ is mild enough for square integrability
and for the regularized unfolding of
Section~\ref{rev:growth_framework} to apply. The continued transform
gives its ordinary Eisenstein overlap on the spectral line.

Subtracting the logarithm, constant and inverse powers from $g_0$
inside the Mellin integral leaves only $N(y)$,
\be
 R_f(s):=R_f^{\rm reg}(s)=R_g^{\rm reg}(s)
 =\int_0^\infty N(y)y^{s-2}\,\dd y,
 \qquad \Re s>4.
 \label{d6c:zero_RS_definition}
\ee
Substituting $N(y)$ into the Mellin transform gives integrals of
Bessel products summed over the neutral charges. We evaluate
these integrals and the charge sum in Appendix~\ref{d6c:gamma},
obtaining
\be
 R_f(s)=\frac{16\pi^2}{12+s(1-s)}
 \frac{\Lambda(s/2-1)\Lambda(s/2)^2\Lambda(s/2+1)}{\Lambda(s)}.
 \label{d6c:eq:Rf}
\ee
In particular, the denominator $12+s(1-s)$ arises from the Mellin
integrals of the supplied zero-mode kernel, without imposing a
differential equation on $f$.
The completed Rankin-Selberg transform obeys
\be
 \mathcal B(s):=\Lambda(s)R_f(s),\qquad
 \mathcal B(s)=\mathcal B(1-s),
 \label{d6c:RS_reflection}
\ee
as required by the Eisenstein functional equation.
Regularized unfolding gives
\be
 \langle g,E_s\rangle=R_f(1-s),\qquad \Re s=\tfrac12.
 \label{d6c:continuous_overlap}
\ee
The constant projection contributes only to charge zero.
Equation~\eqref{intro:RS_reg_average} gives
\be C_g=\Res_{s=1}R_f(s)-B\omega.\label{rev:d6_constant_projection}\ee
Moreover,
unfolding against a cusp form gives
$\langle E_s,u_j\rangle=0$, and continuation extends this to
$\widehat E_1$. The subtraction therefore preserves cusp overlaps.

We now use the calculated overlap to determine the continuous
contribution at charge $k>0$. On the spectral line $s=1/2+it$,
the Eisenstein Fourier coefficients contain the factor
\be
 \chi_k(t):=k^{it}\sigma_{-2it}(k)
 =\sum_{d\mid k}e^{it\log(k/d^2)},\qquad
 |\chi_k(t)|\le d(k):=\sigma_0(k).
 \label{d6a:divisor_factor}
\ee
 The bound  will be used to estimate the continuous contribution.
Substituting~\eqref{d6c:eq:Rf} and the Eisenstein
coefficients~\eqref{intro:Ezero0} into~\eqref{intro:RS_higher}
and combining the equal contributions at $t$ and $-t$ gives
\be
 C_k(y):=(g_{\rm cont})_k(y)
 =\frac{\sqrt y}{\pi}\int_0^\infty
 W(t)\chi_k(t)K_{it}(2\pi ky)\,\dd t,
 \label{d6a:continuous}
\ee
where
\begin{align}
 W(t)&:=\frac{R_f(1/2-it)}{\Lambda(1/2+it)}
 =\frac{\mathcal B(1/2+it)}{|\Lambda(1/2+it)|^2},
 \label{d6c:a_weight}\\
 &=\frac{16\pi^2}{t^2+49/4}
 \left|\frac{\Lambda(5/4+it/2)\Lambda(1/4+it/2)}
 {\Lambda(1/2+it)}\right|^2.
 \label{d6c:positiveweight}
\end{align}
This weight is real, even and nonnegative. As $t\to0$, $\Lambda(1/2+it)=1/(2it)+O(1)$, while the
numerator in~\eqref{d6c:positiveweight} remains finite. Thus
\be
 W(t)=O(t^2),\qquad W(0)=0.
 \label{d6c:weight_origin}
\ee
\eqalias{d6a:weight_at_zero}{d6c:weight_origin}
The spectral integral is therefore regular at its lower endpoint. At charge one, $\chi_1(t)=1$, so
\be
 C_1(y)=\frac{\sqrt y}{\pi}\int_0^\infty
 W(t)K_{it}(2\pi y)\,\dd t.
 \label{d6c:C1_halfline}
\ee
This is the continuous contribution to be subtracted from $f_1$
when isolating its cusp part.

\subsection{The first mode and its Mellin poles}
\label{d6c:subtractone}\label{d6c:weighted}
The complete first mode known from GMV~\cite{GMV} can be written as
\be
 f_1(y)=2f^P_{0,1}(y)+2\sum_{m\ge1}f^P_{-m,m+1}(y),
 \qquad y>0.
 \label{d6c:app_input_particular_sum}
\ee
The indices label the charges in each contribution. Besides the
$(0,1)$ term, the sum includes opposite charges $-m$ and $m+1$,
which also have net charge one. The factors of two account for
both orderings. The terms proportional to $\sqrt yK_{7/2}(2\pi y)$ in
GMV's representation cancel in the complete sum by a divisor
identity~\cite{FKR}.

For the opposite-charge terms, set $p=m+1$ and $q=-m$, with
$m\ge1$. The functions appearing in the sum are
\begin{align}
 f^P_{q,p}(y)=f^P_{p,q}(y)
 ={}&\frac{32\pi\sigma_2(p)\sigma_2(|q|)}{3|pq|}
 \sum_{i,j=0}^1 Q_{ij}(\pi y;p,q)\notag\\
 &\hspace{12mm}\times K_i(2\pi py)K_j(2\pi|q|y).
 \label{d6c:first_mode_pair}
\end{align}
Appendix~\ref{paper:GMV_data} gives the coefficient functions
$Q_{ij}$ and the expression for $f^P_{0,1}$, and proves the
cancellation and convergence. Although the opposite-charge sum
is $O(e^{-6\pi y})$ at weak coupling, its large-$m$ tail is essential
for the Mellin residues that determine the cusp amplitudes.

Subtracting $S_1$ and $C_1$, both fixed by the zero mode,
from $f_1$ isolates its cusp contribution,
\be
 h_1(y):=f_1(y)-S_1(y)-C_1(y),\qquad
 S_1=(F_{\rm sub})_1.
 \label{d6c:residual_definition}
\ee
The first Fourier mode of the Eisenstein subtraction is
$S_1=A(E_3)_1+B(\widehat E_1)_1$. Explicitly,
\be
 S_1(y)=e^{-2\pi y}\left[
 \frac{315\zeta(3)^2}{\pi^3}
 \left(1+\frac3{2\pi y}+\frac3{4\pi^2y^2}\right)
 +\frac{4\pi\zeta(3)}3\right].
 \label{d6c:S1_explicit}
\ee
Following Section~\ref{rev:first_residual_framework}, we extract
the cusp amplitudes from the residues of the Mellin transform
of the residual first mode,
\be
 \mathcal M_h(s):=\int_0^\infty y^{s-3/2}h_1(y)\,\dd y,
 \qquad \Re s>\tfrac12.
 \label{d6c:eq:ordinaryM}
\ee
The cusp residues at $s=iT$, $T>0$, come entirely from the
opposite-charge sum $2\sum_{m\ge1}f^P_{-m,m+1}(y)$: the Mellin
transforms of $f^P_{0,1}$, $S_1$ and $C_1$ have no poles at
these points. The infinite sum is essential here. Any finite
truncation also has no such Mellin poles; the cusp poles arise
from the analytic continuation of the full sum. 
We consequently examine the large-$m$ behavior of
$f^P_{-m,m+1}$ below. The absence of poles at $s=iT$, $T>0$, in these individual
Mellin transforms is established in
Appendices~\ref{d6c:subtraction_proof} and~\ref{d6c:uniform}.

For the large-$m$ expansion, we introduce
\be
 u=m+\tfrac12,\qquad X=2\pi uy,\qquad
 a_m:=\sigma_{-2}(m)\sigma_{-2}(m+1).
 \label{am}
\ee
The Bessel arguments are then $X(1\pm1/(2u))$. At fixed $X$,
the large-$m$ limit probes the region $y=O(1/m)$ of the Mellin
integral. Combining the four Bessel terms before expanding in
$u^{-1}$ gives
\begin{align}
 \frac{f^P_{-m,m+1}(X/(2\pi u))}{a_mu}
 &=F_0(X)+O_X(u^{-2}),\notag\\
 F_0(X)&=\frac{32\pi}{315}\sum_{i,j=0}^1r^{ij}(X/2)K_i(X)K_j(X).
 \label{d6c:pair_leading_profile}
\end{align}
The leading profile involves the same Bessel kernel as the neutral
contribution to the zero mode in~\eqref{d6c:neutral_input}.
Its Mellin transform is
\be
 \widehat F_0(s):=\int_0^\infty X^{s-3/2}F_0(X)\,\dd X
 =\frac{32\pi I_{11}(s+1/2)}{49/4-s^2},\qquad \Re s>\tfrac72,
 \label{d6c:leading_Mellin_coefficient}
\ee
where the common gamma factor is
\be
 I_{11}(a)=\frac{2^{a-3}}{\Gamma(a)}
 \Gamma\left(\frac a2+1\right)
 \Gamma\left(\frac a2\right)^2
 \Gamma\left(\frac a2-1\right).
 \label{d6c:I11_explicit}
\ee
In Appendix~\ref{d6c:uniform}, we combine the Mellin transforms
of the four Bessel terms to obtain the rational prefactor
multiplying $I_{11}$.

The sum over $m$ contributes the factor $D(s-3/2)$ to the
leading Mellin term, where $D$ is the Dirichlet
series
\be
 D(w):=\sum_{m\ge1}a_m(m+\tfrac12)^{-w},\qquad \Re w>1.
 \label{d6c:eq:D}
\ee
With the change of variables $X=2\pi uy$, the leading large-$m$
contribution to $\mathcal M_h(s)$ takes the form
\be
 2\sum_{m\ge1}\int_0^\infty y^{s-3/2}a_muF_0(2\pi uy)\,\dd y
 =2(2\pi)^{1/2-s}\widehat F_0(s)D(s-3/2).
 \label{d6c:pair_Mellin_scaling}
\ee
 The subleading terms do not contribute to the cusp residues at
$s=iT$, $T>0$. The normalized expansion
in~\eqref{d6c:pair_leading_profile} contains only even powers of
$u^{-1}$, so the first correction is of order $u^{-2}$.
After Mellin transformation and summation over $m$, this
correction contains the factor $D(s+1/2)$ and is regular near
these points (Appendix~\ref{d6c:shift_residues}).

Once this correction is separated, the remaining sum of Mellin
transforms converges locally uniformly and defines a holomorphic
function near each $s=iT$, as proved in
Appendix~\ref{d6c:uniform}. We combine these regular contributions
with the continued Mellin transforms of $2f^P_{0,1}-S_1-C_1$
in a remainder $\mathcal R(s)$, giving
\begin{align}
 \mathcal M_h(s)
 &=2(2\pi)^{1/2-s}\widehat F_0(s)D(s-3/2)+\mathcal R(s),
 \label{d6c:exact_Mellin_separation}\\
 &=\frac{64\pi(2\pi)^{1/2-s}I_{11}(s+1/2)}{49/4-s^2}
 D(s-3/2)+\mathcal R(s).
 \label{d6c:Mellin_pole_part}
\end{align}
Initially valid for $\Re s>7/2$, this identity extends by analytic
continuation to a neighborhood of each $s=iT$, $T>0$, where
$\mathcal R(s)$ is holomorphic. The prefactor of $D(s-3/2)$ is
finite and nonzero there. Consequently, $\mathcal M_h(s)$ has
a pole at $s=iT$ precisely when $D(w)$ has a pole at
$w=-3/2+iT$.

\subsection{Cusp amplitudes and higher Fourier modes}
\label{d6c:complete_cusp_reconstruction}\label{rev:cusp_conventions}
\label{rev:higher_modes}\label{d6c:secondcontinuous}\label{d6c:hecke}

We now evaluate the residues of $D(w)$ in terms of the Hecke
$L$-functions of the Maass cusp forms. Let $\epsilon_j=0$ for
even forms and $\epsilon_j=1$ for odd forms. The Hecke $L$-function associated with $u_j$ and its completed
form are defined by
\begin{align}
 L_j(w)&=\sum_{n\ge1}\frac{\lambda_j(n)}{n^w},
 \label{d6c:L_series}\\
 \widehat L_j(w)&=\pi^{-w}
 \Gamma\left(\frac{w+\epsilon_j+it_j}{2}\right)
 \Gamma\left(\frac{w+\epsilon_j-it_j}{2}\right)L_j(w).
 \label{d6c:eq:Lhat}
\end{align}
\eqalias{d6a:L_completion}{d6c:eq:Lhat}
The Dirichlet series converges initially for sufficiently large
$\Re w$. The completed function is entire and satisfies
$\widehat L_j(w)=(-1)^{\epsilon_j}\widehat L_j(1-w)$.
Thus $\widehat L_j(1/2)=0$ for odd forms.

Using Motohashi's additive-divisor formula~\cite{Motohashi1994},
we evaluate the residue of $D(w)$ entering
\eqref{d6c:Mellin_pole_part}. For $T>0$,
\be
 \Res_{w=-3/2+iT}D(w)
 =\frac{2^{iT-9/2}\sqrt\pi\,\Gamma(iT)}{I_{11}(iT+1/2)}
 \mathcal A(T),
 \label{d6c:shift_residue_Bessel}
\ee
where
\be
 \mathcal A(T):=\sum_{\substack{j:t_j=T\\j\ {\rm even}}}
 |\rho_j(1)|^2\widehat L_j(\tfrac12)\widehat L_j(\tfrac52).
 \label{d6c:eq:AT}
\ee
The divisor product $a_m=\sigma_{-2}(m)\sigma_{-2}(m+1)$
fixes the factor $\widehat L_j(1/2)\widehat L_j(5/2)$ in
Motohashi's Maass contribution. Only terms with $t_j=T$
contribute to the residue at $w=-3/2+iT$; the remaining
terms are regular there. Appendix~\ref{d6c:shift_residues}
derives this specialization and evaluates the residue.

Substituting~\eqref{d6c:shift_residue_Bessel}
into~\eqref{d6c:Mellin_pole_part} and cancelling the factor
$I_{11}$ gives
\be
 \Res_{s=iT}\mathcal M_h(s)
 =\frac{4\pi^2\pi^{-iT}\Gamma(iT)}{T^2+49/4}\mathcal A(T).
 \label{d6c:target_direct_residue}
\ee
\eqalias{d6c:period_target_residue}{d6c:target_direct_residue}
Using the first-mode residue relation, we obtain the sum of cusp
amplitudes with spectral parameter $t_j=T$:
\begin{align}
 \mathcal D_f(T)&:=\sum_{j:t_j=T}d_j
 =\frac{2\pi^{iT}}{\Gamma(iT)}\Res_{s=iT}\mathcal M_h(s),
 \label{d6c:eq:aggregate}\\
 &=\frac{8\pi^2}{T^2+49/4}\mathcal A(T).
 \label{d6c:eq:answer}
\end{align}
\eqalias{d6c:cusp_amplitudes_preview}{d6c:eq:answer}
Assuming the conjectured simplicity of the even cusp spectrum,
each sum contains a single term, giving the individual amplitudes
\be
 d_j=\frac{8\pi^2|\rho_j(1)|^2
 \widehat L_j(1/2)\widehat L_j(5/2)}{t_j^2+49/4},
 \qquad j\ \hbox{even}.
 \label{d6c:canonical}
\ee

Using these amplitudes, the Hecke relations determine the cusp
contribution at each positive charge:
\be
 U_k(y)=\sqrt y\sum_{j\ {\rm even}}
 d_j\lambda_j(k)K_{it_j}(2\pi ky).
 \label{d6a:cusp}
\ee
Combining this with the continuous contribution $C_k$ and the
restored Eisenstein terms $S_k$ gives the full Fourier mode,
\be
 f_k(y)=S_k(y)+C_k(y)+U_k(y),
 \qquad k\ge1,\qquad y>0.
 \label{d6a:split}
\ee
With $X=2\pi ky$, the restored Eisenstein contribution is
\be
 S_k(y)=e^{-X}\left[
 \frac{315\zeta(3)^2}{\pi^3}k^2\sigma_{-5}(k)
 \left(1+\frac3X+\frac3{X^2}\right)
 +\frac{4\pi\zeta(3)}3\sigma_{-1}(k)\right].
 \label{d6a:restored}
\ee
The continuous contribution $C_k$ is given
in~\eqref{d6a:continuous}, and evenness gives $f_{-k}=f_k$.
Appendix~\ref{rev:selected_numerics} compares the full Fourier
modes at $k=5,10$ numerically with GMV at finite $y$ and
verifies analytic agreement of the square-root and constant
coefficients derived below at every positive charge.

\section{Instanton asymptotics from the spectral sum}
\label{d6a:section}

We now derive the weak-coupling expansion of the reconstructed
$D^6R^4$ Fourier modes at fixed positive charge, through constant
order in the instanton prefactor. The main issue is taking this limit through the
infinite cusp sum in~\eqref{d6a:split}. For a fixed finite set
$\mathcal I$ of cusp forms, the large-argument Bessel expansion gives
\begin{align}
 U_{k,\mathcal I}(y)
 &:=\sqrt y\sum_{j\in\mathcal I}
 d_j\lambda_j(k)K_{it_j}(2\pi ky)\notag\\
 &=\frac{e^{-2\pi ky}}{2\sqrt k}
 \sum_{j\in\mathcal I}d_j\lambda_j(k)
 +O_{\mathcal I,k}(y^{-1}e^{-2\pi ky}).
 \label{d6c:finite_cusp_asymptotic}
\end{align}
The prefactor of any such finite sum tends to a constant. However, the corrections to the fixed-order Bessel expansion are
no longer small when $t_j$ is comparable to $\sqrt y$. We therefore
retain the dependence on $t_j/\sqrt y$ when evaluating the infinite
cusp sum.

We first combine the cusp and continuous contributions and identify
the Gaussian approximation to their common Bessel kernel.
Using the Kuznetsov trace formula, with bounds on the approximation
and the infinite spectral tail, we derive the $\sqrt y$ and constant
coefficients of the full instanton prefactor, including the restored
Eisenstein terms.

\subsection{Spectral weights and the Bessel kernel}
\label{d6a:weights}

The cusp sum and the continuous spectrum have the same Bessel
kernel. We combine their coefficients in a signed spectral measure,
defined by
\begin{align}
 \mu_k([0,T])
 &:=\sum_{\substack{j\ {\rm even}\\t_j\leq T}}
 d_j\lambda_j(k)
 +\frac1\pi\int_0^T W(t)\chi_k(t)\,\dd t,
 \label{d6a:cumulative_spectral_weight}
\end{align}
so that for a bounded test function $\Phi$ on $[0,T]$,
\begin{align}
 \int_{[0,T]}\Phi(t)\,\dd\mu_k(t)
 &=\sum_{\substack{j\ {\rm even}\\t_j\leq T}}
 d_j\lambda_j(k)\Phi(t_j)
 +\frac1\pi\int_0^T W(t)\chi_k(t)\Phi(t)\,\dd t.
 \label{d6a:jointmeasure}
\end{align}
\eqalias{d6a:delta_sum_identity}{d6a:jointmeasure}
The discrete part retains the individual cusp parameters and their
weights. With $X=2\pi ky$, the exact mode becomes
\be
 f_k(y)=S_k(y)+\sqrt y\int_0^\infty K_{it}(X)\,\dd\mu_k(t).
 \label{d8e:jointmeasure}
\ee

We separate the gamma factors from the arithmetic dependence of the
cusp and continuous weights. The cusp amplitudes
in~\eqref{d6c:canonical} contain the normalization factor
$|\rho_j(1)|^2$. For an orthonormal even cusp form,
\be
 |\rho_j(1)|^2
 =\frac{2\cosh(\pi t_j)}{L_j(1,\operatorname{sym}^2)},
 \label{d6a:cusp_norm_for_weight}
\ee
where $L_j(s,\operatorname{sym}^2)$ is the symmetric-square
$L$-function associated with $u_j$.
Appendix~\ref{rev:analytic_norm} derives this identity.
Using this identity and writing out the gamma factors in the
cusp amplitudes and the continuous weight~\eqref{d6c:positiveweight},
we obtain
\be
 d_j\lambda_j(k)=r(t_j)b_j(k),
 \qquad
 W(t)=r(t)P(t),
 \label{d6a:commonweight}
\ee
with
\begin{align}
 r(t)&=\frac{4}{\pi}\cosh(\pi t)
 \frac{t^2+1/4}{t^2+49/4}
 \bigl[\Gamma(1/4+it/2)\Gamma(1/4-it/2)\bigr]^2,
 \label{d6a:r}\\
 b_j(k)&=\frac{\lambda_j(k)L_j(1/2)L_j(5/2)}
 {L_j(1,\operatorname{sym}^2)},
 \label{d6a:arithmetic_cusp}\\
 P(t)&=\frac{|\zeta(1/2+it)\zeta(5/2+it)|^2}
 {|\zeta(1+2it)|^2}.
 \label{d6a:rP}
\end{align}
Here $P(0)$ is defined by continuity. The factorization uses
the gamma identities
\begin{align}
 \Gamma(5/4+it/2)\Gamma(5/4-it/2)
 &=\frac{t^2+1/4}{4}
 \Gamma(1/4+it/2)\Gamma(1/4-it/2),\notag\\
 \Gamma(1/2+it)\Gamma(1/2-it)
 &=\frac{\pi}{\cosh(\pi t)}.
 \label{d6a:gamma_reduction}
\end{align}
The common factor $r(t)$ is positive and even. Stirling's formula
gives its large-$t$ behavior,
\be
 r(t)=\frac{16\pi}{t}+O(t^{-3}),\qquad t\to+\infty.
 \label{d6a:stirling}
\ee

To estimate the continuous contribution at weak coupling, we bound
its spectral weight. Standard zeta estimates give
$P(t)\ll t^{1/2}\log^2(2+t)$ for $t\geq2$
\cite{DLMF,Yang}.
Combining $W=rP$ with~\eqref{d6a:stirling} and the behavior
$W(t)=O(t^2)$ near the origin yields
\begin{align}
 W(t)&\ll_\epsilon(1+t)^{-1/2+\epsilon},
 \qquad t\geq0,\quad\epsilon>0,
 \label{d6a:Wbound}\\
 \int_0^T W(t)|\chi_k(t)|\,\dd t
 &\ll_{k,\epsilon}(1+T)^{1/2+\epsilon},
 \label{d6a:continuous_mass_bound}
\end{align}
where the second estimate uses $|\chi_k(t)|\leq d(k)$.
Together with the decay of the Bessel kernel, these bounds will
show that the continuous contribution is subleading, allowing us
to identify the infinite cusp spectrum as the source of the leading
growth of the instanton prefactor
(Section~\ref{rev:joint_limit}).

To isolate the instanton prefactor, we introduce the normalized
Bessel kernel and its spectral integral,
\begin{align}
 H_X(t)&:=\sqrt{\frac{2X}{\pi}}e^XK_{it}(X),\qquad
 \mathscr H_k(X):=\int_0^\infty H_X(t)\,\dd\mu_k(t).
 \label{d6a:normalized_Bessel}
\end{align}
The reconstructed Fourier mode then satisfies
\be
 e^Xf_k(y)=e^XS_k(y)+\frac{\mathscr H_k(X)}{2\sqrt k}.
 \label{d6a:joint_Bessel}
\ee
To identify the large-$X$ behavior of the kernel while retaining
its dependence on $t/\sqrt X$, we use the integral representation
\be
 K_{it}(X)=\frac12\int_{\mathbb R}e^{-X\cosh u+itu}\,\dd u.
 \label{d6a:besselintegral}
\ee
For large $X$, the exponential weight is concentrated near
$u=0$ on a scale of order $X^{-1/2}$. Rescaling
$u=v/\sqrt X$ therefore gives
\be
 H_X(t)=\frac1{\sqrt{2\pi}}\int_{\mathbb R}
 \exp\!\left(-X[\cosh(v/\sqrt X)-1]\right)e^{itv/\sqrt X}\,\dd v.
 \label{d6a:rescaled_Bessel_integral}
\ee
Keeping the quadratic term in the exponent gives a Gaussian
in $t/\sqrt X$,
\be
 H_X(t)\approx\frac1{\sqrt{2\pi}}\int_{\mathbb R}
 e^{-v^2/2+itv/\sqrt X}\,\dd v=e^{-t^2/(2X)}.
 \label{d6a:Gaussian_transform}
\ee
The Gaussian suppresses spectral parameters much larger than
$\sqrt X$, so the range sampled by the kernel increases with $X$.
We first derive the asymptotics of the spectral integral with this
Gaussian kernel. The calculation also yields bounds on the absolute
spectral weights. Together with estimates of the large-$t$ tail,
these bounds will justify the approximation and establish the
asymptotics of the exact Bessel integral in
Section~\ref{d6a:leading}.

\subsection{The spectral moment from the Kuznetsov trace formula}
\label{d6a:leading}

 To evaluate the Gaussian integral by Mellin inversion, we introduce
the spectral moment, a weighted integral over the cusp and
continuous spectra,
\be
 Z_k(z):=\int_0^\infty(4+t^2)^{-z/2}\,\dd\mu_k(t),
 \qquad \Re z>c_*.
 \label{d6a:shifted_regulator}
\ee
Here $c_*>1$ is chosen so that the spectral integral converges
absolutely for $\Re z>c_*$. The existence of such a $c_*$ follows from the polynomial
bounds on the absolute spectral weights established
independently in Appendix~\ref{d6a:initial_spectral_convergence}.
The shift by four keeps the weight regular at $t=0$.
Since the unweighted spectral integral need not converge, $Z_k(0)$
will denote the value obtained by analytic continuation.
 Mellin inversion gives
\begin{align}
 \mathscr H_k(X)
 &\approx \int_0^\infty e^{-t^2/(2X)}\,\dd\mu_k(t)\notag\\
 &=\frac{e^{2/X}}{2\pi i}\int_{(c)}
 \frac{\Gamma(z/2)}2(2X)^{z/2}Z_k(z)\,\dd z,
 \qquad c>c_*.
 \label{d6a:heat_mellin}
\end{align}
Here $(c)$ denotes the upward vertical contour $\Re z=c$. The approximation replaces the Bessel kernel by its Gaussian form,
while the Mellin representation is exact.

\paragraph{Applying the Kuznetsov trace formula.}
We use the first-moment identity of Hoffstein--Lee--Nastasescu
(HLN)~\cite{HLN}, derived from the Kuznetsov trace formula by
summing one Fourier index against a divisor weight.
The Hecke identity
\be
 \zeta(a+b)\sum_{m\geq1}
 \frac{\lambda_j(m)\sigma_{a-b}(m)}{m^a}
 =L_j(a)L_j(b),\qquad \Re a,\Re b>\frac32,
 \label{d6a:main_divisor_matching}
\ee
produces the two cusp $L$-values, while the corresponding
Eisenstein sum produces four zeta factors.
After continuation to $(a,b)=(1/2,5/2)$, described in
Appendix~\ref{d6a:kuznetsov}, the HLN identity takes the form
\begin{align}
 &\sum_{j\ {\rm even}}h(t_j)b_j(k)
 +\frac1\pi\int_0^\infty h(t)P(t)\chi_k(t)\,\dd t\notag\\
 &\qquad=
 \frac{\zeta(3)\sigma_{-2}(k)}{\pi^2\sqrt k}
 \int_0^\infty h(t)t\tanh(\pi t)\,\dd t
 +\mathcal E_k[h].
 \label{d6a:main_HLN_intro}
\end{align}
Here $h(t)$ is an even test function weighting the cusp sum and
the continuous integral. Odd cusp forms do not contribute because
$L_j(1/2)=0$ for odd parity. The first term on the right is the
diagonal contribution of the Kuznetsov trace formula.
The remainder $\mathcal E_k[h]$ collects the continued
Kloosterman contribution and the corrections from contour shifts.
The normalization and analytic continuation are worked out in
Appendix~\ref{rev:constant} for the test functions used below.

To apply the trace formula to $Z_k(z)$, we start with the test
function $h(t)=r(t)(4+t^2)^{-z/2}$. Choosing $\Re z$ sufficiently
large ensures that it decays rapidly enough at large $|t|$.
The trace formula also requires $h(\pm i/2)=0$. We enforce these
zeros by multiplying the weight by a factor that tends to one
at large real $t$,
\be
 q_2(t):=\frac{t^2+1/4}{t^2+4},\qquad
 h(t)=r(t)q_2(t)(4+t^2)^{-z/2},
 \label{d6a:q2_definition}
\ee
and define the corresponding modified moment,
\be
 G_k(z):=\int_0^\infty q_2(t)(4+t^2)^{-z/2}\,\dd\mu_k(t).
 \label{d6a:corrected_joint_moment}
\ee
Since $1-q_2(t)=15/[4(4+t^2)]$, the difference $Z_k(z)-G_k(z)$
has an extra factor $(4+t^2)^{-1}$, shifting $z$ to $z+2$ and giving 
\be
 Z_k(z)=G_k(z)+\frac{15}{4}Z_k(z+2).
 \label{d6a:addback}
\ee
Applying the HLN identity~\eqref{d6a:main_HLN_intro} with $h(t)$
from~\eqref{d6a:q2_definition} expresses the modified moment
$G_k(z)$ as
\begin{align}
 G_k(z)
 &=\frac{\zeta(3)\sigma_{-2}(k)}{\pi^2\sqrt k}
 \int_0^\infty r(t)q_2(t)t\tanh(\pi t)
 (4+t^2)^{-z/2}\,\dd t\notag\\
 &\quad+\mathcal E_{k,q}(z).
 \label{d6a:main_diagonal_split}
\end{align}
This identity holds initially for $\Re z>c_*$.
Appendix~\ref{d6a:existence} shows that $\mathcal E_{k,q}(z)$ is
holomorphic for $\Re z>-1/2$. The diagonal density satisfies
$r(t)q_2(t)t\tanh(\pi t)=16\pi+O(t^{-2})$ as $t\to\infty$.
Separating off this constant contribution continues $G_k(z)$
meromorphically to $\Re z>-1/2$, with a single simple pole at
$z=1$, whose residue is
\be
 a_k:=\Res_{z=1}G_k(z)
 =\frac{16\zeta(3)\sigma_{-2}(k)}{\pi\sqrt k}.
 \label{d6a:moment_properties}
\ee
The difference $G_k(z)-a_k/(z-1)$ is holomorphic for
$\Re z>-1/2$. The finite value $G_k(0)$ is the sum of the
continued diagonal contribution and $\mathcal E_{k,q}(0)$.

\paragraph{Asymptotics of the Gaussian spectral integral.}
We use the same Mellin inversion as above, with the modified
moment $G_k(z)$ replacing $Z_k(z)$:
\begin{align}
 \mathscr H_{k,q}(X)
 &:=\int_0^\infty q_2(t)e^{-t^2/(2X)}\,\dd\mu_k(t)\notag\\
 &=\frac{e^{2/X}}{2\pi i}\int_{(c)}
 \frac{\Gamma(z/2)}2(2X)^{z/2}G_k(z)\,\dd z,
 \qquad c>c_*.
 \label{d6a:corrected_heat_mellin}
\end{align}
\eqalias{d6a:corrected_heat_integral}{d6a:corrected_heat_mellin}
Here $\mathscr H_{k,q}(X)$ is the Gaussian spectral integral
with the auxiliary factor $q_2$ included. The bound on $G_k(z)$
along vertical lines in Appendix~\ref{d6a:existence}, together
with the exponential decay of $\Gamma(z/2)$, justifies shifting
the contour to $\Re z=-\delta$, with $0<\delta<1/2$.
The pole of $G_k(z)$ at $z=1$ produces the $\sqrt X$ term,
while the pole of $\Gamma(z/2)/2$ at $z=0$ gives the constant
$G_k(0)$. Thus
\be
 \mathscr H_{k,q}(X)
 =a_k\sqrt{\frac{\pi X}{2}}+G_k(0)
 +O_{k,\delta}(X^{-\delta/2}).
 \label{d6a:corrected_heat_asymptotic}
\ee

We now remove the auxiliary factor $q_2$. Since
$1-q_2(t)=15/[4(4+t^2)]$, the correction carries the weight
$(4+t^2)^{-1}$. To control its integral, and later the comparison
with the exact Bessel kernel, we first bound the total absolute
spectral weight.

At charge one, positivity allows the Gaussian
estimate~\eqref{d6a:corrected_heat_asymptotic}, with $X=T^2$
and $T\ge1$, to bound the total weight below $T$ by $O(1+T)$.
Comparison with the charge-one weights then gives
$|\mu_k|([0,T])\ll_k1+T$ for each fixed $k$
(Appendix~\ref{d6a:accumulated_weight_proof}).
In particular, the moment
\be
 Z_k(2)=\int_0^\infty\frac{\dd\mu_k(t)}{4+t^2}
 \label{d6a:convergent_addback}
\ee
is absolutely convergent. Evaluating the continued identity~\eqref{d6a:addback} at $z=0$
gives
\be
 Z_k(0)=G_k(0)+\frac{15}{4}Z_k(2).
 \label{d6a:existence_value_at_zero}
\ee

Restoring the contribution proportional to $1-q_2$ in the
Gaussian integral gives
\begin{align}
 \int_0^\infty e^{-t^2/(2X)}\,\dd\mu_k(t)
 &=\mathscr H_{k,q}(X)
 +\frac{15}{4}\int_0^\infty
 \frac{e^{-t^2/(2X)}}{4+t^2}\,\dd\mu_k(t)\notag\\
 &=\mathscr H_{k,q}(X)+\frac{15}{4}Z_k(2)+O_k(X^{-1/2}).
 \label{d6a:heat_addback}
\end{align}
The last equality uses $|\mu_k|([0,T])\ll_k1+T$ to bound
the effect of replacing the Gaussian by one in the correction
integral. Removing $q_2$ therefore changes the constant from
$G_k(0)$ to $Z_k(0)$, while leaving the leading coefficient
unchanged.

Appendix~\ref{rev:bessel_estimates} bounds the difference between
the exact Bessel and Gaussian integrals by
$O_k(\log X/\sqrt X)$. Combining this estimate with
\eqref{d6a:corrected_heat_asymptotic} and the contribution
proportional to $1-q_2$ evaluated in~\eqref{d6a:heat_addback} gives
\begin{align}
 \mathscr H_k(X)
 &=\int_0^\infty e^{-t^2/(2X)}\,\dd\mu_k(t)
 +O_k(\log X/\sqrt X)\notag\\
 &=a_k\sqrt{\frac{\pi X}{2}}+Z_k(0)
 +O_{k,\delta}(X^{-\delta/2}),\qquad 0<\delta<\frac12.
 \label{d6a:jointheat}
\end{align}
\eqalias{d6a:proved_heat_asymptotic}{d6a:jointheat}
The corrections from removing $q_2$ and passing to the exact
Bessel integral are both absorbed into the stated remainder.

\subsection{The weak-coupling expansion}
\label{rev:joint_limit}

We now include the restored Eisenstein terms to obtain the
weak-coupling expansion of the full Fourier mode. Their
contribution to the instanton prefactor begins at constant order:
\be
 e^{2\pi ky}S_k(y)
 =\frac{315\zeta(3)^2}{\pi^3}k^2\sigma_{-5}(k)
 +\frac{4\pi\zeta(3)}3\sigma_{-1}(k)+O_k(y^{-1}).
 \label{d6a:restored_constant}
\ee
Substituting~\eqref{d6a:jointheat} and~\eqref{d6a:restored_constant}
into~\eqref{d6a:joint_Bessel}, with $X=2\pi ky$, gives the
square-root and constant coefficients of the full instanton
prefactor:
\begin{align}
 A_k&=\frac\pi2a_k
 =\frac{8\zeta(3)\sigma_{-2}(k)}{\sqrt k},
 \label{d6a:existenceresult}\\
 B_k&=\frac{315\zeta(3)^2}{\pi^3}k^2\sigma_{-5}(k)
 +\frac{4\pi\zeta(3)}3\sigma_{-1}(k)
 +\frac{Z_k(0)}{2\sqrt k}.
 \label{d6a:Bzeta}
\end{align}
\eqalias{d6a:constant_from_corrected_test}{d6a:Bzeta}

To evaluate $B_k$, we compute $Z_k(0)$ from the continued HLN
identity and combine its contribution with the restored
Eisenstein terms in~\eqref{d6a:Bzeta}. As detailed in
Appendix~\ref{rev:constant}, the arithmetic evaluation uses
the divisor-convolution identity~\cite{FKR} to reduce the result
to a finite sum over positive charge pairs:
\be
 B_k=\frac{64\pi^2}{3k^2}
 \sum_{\substack{a,b\geq1\\a+b=k}}
 (ab)^{3/2}\sigma_{-2}(a)\sigma_{-2}(b)
 \left[1-2\left(\frac{b-a}{k}\right)^2\right].
 \label{d6a:constant_pair_kernel}
\ee
Using $a+b=k$ and $\sigma_2(n)=n^2\sigma_{-2}(n)$, this becomes
\be
 B_k=\frac{64\pi^2}{3k^4}
 \sum_{\substack{a,b\geq1\\a+b=k}}
 \frac{\sigma_2(a)\sigma_2(b)}{\sqrt{ab}}
 (6ab-a^2-b^2).
 \label{d6a:constant_general}
\ee
This gives the constant term in the instanton prefactor of the
full reconstructed Fourier mode.
The reconstructed Fourier modes with $k\ge1$ therefore have
the weak-coupling expansion
\be
 f_k(y)=e^{-2\pi ky}\left[
 \frac{8\zeta(3)\sigma_{-2}(k)}{\sqrt k}\sqrt y+B_k
 +O_{k,\epsilon}(y^{-1/4+\epsilon})\right],
 \qquad 0<\epsilon<\frac14,
 \label{d6a:opening_result}
\ee
\eqalias{d6a:final}{d6a:opening_result}
where $B_k$ is given by~\eqref{d6a:constant_general} and we have
set $\delta=1/2-2\epsilon$ in~\eqref{d6a:jointheat}.\footnote{The
$\epsilon$-dependent power in~\eqref{d6a:opening_result} gives
an asymptotic bound on the remainder and does not specify
the next term in the expansion.}
Both the square-root and constant coefficients are in exact
agreement with the GMV Fourier expansion at every positive
charge~\cite{GMV,KLMR}. This agreement is established analytically
in Appendix~\ref{check:GMV_analytic}.

The square-root term comes entirely from the infinite cusp sum.
The continuous contribution is
$o(\sqrt y\,e^{-2\pi ky})$
(Appendix~\ref{rev:bessel_estimates}), while the restored
Eisenstein terms enter at constant order.
The Gaussian kernel captures the growing range of cusp
parameters $t_j=O(\sqrt y)$ responsible for this growth.

\section{Reconstruction from incomplete low-mode data}
\label{rev:incomplete}\label{d6c:alternatives}

The $D^6R^4$ reconstruction relied on the complete zero and first
Fourier modes. We now ask what can be determined
from partial knowledge of the weak-coupling expansions of these
two modes. Neutral sectors omitted from the zero-mode data
affect the continuous contribution and, through the first-mode
subtraction, the extraction of cusp amplitudes as well.
We also examine how modular functions can share the same partial weak-coupling data for the zero
and first Fourier modes while differing in their higher Fourier modes.
We derive sufficient bounds on the missing spectral weights
to ensure that a leading instanton coefficient at higher charge
remains unchanged.

\subsection{Cusp ambiguities from finite first-mode data}
\label{d6c:finite_first_mode_ambiguity}

We keep the complete zero mode fixed and construct a cusp
addition that preserves the first $N$ coefficients of the
first-mode prefactor, from the constant term through
$y^{-(N-1)}$. We then use the Hecke relations to determine
whether the same coefficients remain unchanged in higher
Fourier modes.

For $N\ge1$, choose $N+1$ even Hecke--Maass forms $u_j$ with
distinct $t_j^2$ and normalize them as follows:
\be
 v_j:=\frac{u_j}{\rho_j(1)}, \qquad (v_j)_0=0, \qquad
 (v_j)_k(y)=\lambda_j(k)\sqrt y K_{it_j}(2\pi ky).
 \label{d6c:ambiguity_normalized_form}
\ee
Consider the finite cusp addition
\be
 \delta f:=\sum_{j=1}^{N+1}\beta_jv_j,
 \qquad \beta_j\in\mathbb R.
 \label{d6c:ambiguity_cusp_addition}
\ee
Adding $\delta f$ preserves reality, evenness and modular
invariance, and leaves the complete zero mode unchanged.
For the first Fourier mode, the large-$y$ Bessel
expansion~\cite{DLMF} gives
\be
 e^{2\pi y}(\delta f)_1(y)
 \sim\sum_{r\ge0}\left[\sum_{j=1}^{N+1}
       \beta_jc_r(t_j^2)\right]y^{-r}.
\ee
The coefficients in the Bessel expansion are polynomials
in the squared spectral parameter:
\be
 c_r(x)=\frac{(-1)^r}{2r!(16\pi)^r}
       \prod_{\ell=1}^r\bigl[4x+(2\ell-1)^2\bigr].
 \label{d6c:ambiguity_coefficient_polynomials}
\ee
Each $c_r$ is a polynomial of degree $r$.
We choose the weights $\beta_j$ so that the first $N$ terms vanish:
\be
 \sum_{j=1}^{N+1}\beta_jc_r(t_j^2)=0,
 \qquad r=0,\ldots,N-1.
 \label{d6c:ambiguity_finite_constraints}
\ee
Since $c_r(x)$ has degree $r$ and the values $t_j^2$ are
distinct, fixing $\beta_1\ne0$ uniquely determines the
remaining weights. Every $\beta_j$ is then nonzero, and the
first surviving term occurs at order $y^{-N}$:
\be
 e^{2\pi y}(\delta f)_1(y)=C_Ny^{-N}
 +O(y^{-N-1}),
 \qquad C_N\ne0.
 \label{d6c:ambiguity_finite_remainder}
\ee

In the $k$th Fourier mode, with $k\ge2$, the Hecke relations
multiply each weight $\beta_j$ by $\lambda_j(k)$.
Since the cancellation conditions fix the relative weights
and every $\beta_j$ is nonzero, the first $N$ terms still
vanish exactly when all $\lambda_j(k)$ agree:
\be
 e^{2\pi ky}(\delta f)_k(y)=O(y^{-N})
 \quad\Longleftrightarrow\quad
 \lambda_1(k)=\cdots=\lambda_{N+1}(k).
 \label{d6c:ambiguity_retained_orders}
\ee
Distinct normalized Hecke--Maass forms must differ in at
least one Hecke eigenvalue~\cite{JacquetShalika1981}. At least one of these cancellations therefore fails for some
$k\ge2$, giving
\be
 e^{2\pi ky}(\delta f)_k(y)
 =a\,y^{-r_*}+O(y^{-r_*-1}),
 \qquad a\ne0,\qquad 0\le r_*\le N-1.
 \label{d6c:ambiguity_higher_finite_difference}
\ee
This addition preserves the complete zero mode and the first
$N$ coefficients of the first-mode prefactor, while changing
at least one of these orders in a higher Fourier mode.

\subsection{Neutral sectors in the spectral reconstruction}
\label{d6c:missing_neutral_data}

We now compare modular functions with the same full perturbative expansion, allowing their neutral contributions to differ.
Their difference decays rapidly at the cusp. Let $\delta f_0$
denote the difference of their complete zero modes. Unfolding
determines the change in the Rankin--Selberg transform and hence
in the continuous Fourier modes:
\begin{align}
 \delta R(s)&=\int_0^\infty\delta f_0(y)y^{s-2}\,\dd y,
       \qquad \Re s>1,
 \label{d6c:neutral_response_Mellin}\\
 \delta C_k(y)&=\frac1{4\pi i}\int_{\Re s=1/2}
       \delta R(1-s)(E_s)_k(y)\,\dd s.
 \label{d6c:neutral_response_continuous}
\end{align}
The second equation uses the analytic continuation of
$\delta R(s)$ to the spectral line $\Re s=1/2$.

To illustrate the contribution of a single neutral term, we
consider the profile $e^{-4\pi My}$. By linearity, its continuous
response can be evaluated separately using
\eqref{d6c:neutral_response_Mellin}
and~\eqref{d6c:neutral_response_continuous}, giving
\be
 \delta C_k(y)=e^{-2\pi ky}
       \left[b_{k,M}+O_{k,M}(y^{-1})\right],
 \qquad k,M\ge1.
 \label{d6c:exponential_continuous_asymptotic}
\ee
A neutral term in the zero mode can therefore contribute at
the leading charged exponential in the continuous spectrum.
In this example, its contribution to the instanton prefactor
approaches the constant $b_{k,M}$, evaluated as an absolutely
convergent Bessel series in
Appendix~\ref{d6c:exponential_profile_details}.\footnote{Modularity
requires the Mellin transform of the complete zero-mode
variation to satisfy the reflection
relation~\eqref{d6c:neutral_response_scattering}; the isolated
exponential fails this condition.
Appendix~\ref{d6c:neutral_truncation_details} shows that finite
truncations of the GMV neutral profiles also fail this condition
and explains why an infinite neutral sum requires separate
estimates.}

Recall that the continuous contribution also enters the
first-mode subtraction used to determine the cusp amplitudes.
For a compatible modular variation, the changes in the positive
Fourier modes and the first-mode residual are
\be
 \delta f_k=\delta C_k+\delta U_k,
 \qquad \delta h_1=\delta f_1-\delta C_1.
 \label{d6c:neutral_response_charged_input}
\ee
Preserving the specified first-mode coefficients requires
the corresponding terms in $\delta C_1+\delta U_1$ to vanish.
The effect of a change in the neutral zero-mode data on higher
Fourier modes therefore also depends on the accompanying
change in the cusp amplitudes.\footnote{For even modular
candidates with an $L^2$ difference and the same perturbative
zero mode, equality of the complete first modes implies
equality of the complete zero modes; see
Appendix~\ref{d6c:complete_first_uniqueness}.}

\subsection{Stability of leading instanton coefficients}
\label{d8e:d6inputaudit}

Let $\delta f=\widetilde f-f$ be the difference between two
compatible modular candidates with the same Eisenstein
subtraction. For a given $\alpha>0$, we seek sufficient bounds
on the differences in their spectral weights to ensure
$e^{2\pi ky}\delta f_k(y)=o(y^\alpha)$, so that the coefficient
of the leading term $y^\alpha e^{-2\pi ky}$ remains unchanged. Let $\delta d_j$ and $\delta W$ denote the differences in the
cusp and continuous weights, respectively. With $X=2\pi ky$
and the normalized Bessel kernel
$H_X(t)=\sqrt{2X/\pi}\,e^XK_{it}(X)$, the difference in the
$k$th Fourier mode takes the form
\begin{align}
 e^{2\pi ky}\delta f_k(y)
 &=\frac1{2\sqrt k}\bigg[
 \sum_{j\ {\rm even}}\delta d_j\lambda_j(k)H_X(t_j)\notag\\
 &\qquad+\frac1\pi\int_0^\infty
       \delta W(t)\chi_k(t)H_X(t)\,\dd t\bigg].
 \label{d6c:missing_exact_spectrum}
\end{align}
To track the cusp and continuous weights up to spectral
parameter $T$, we define
\begin{align}
 R_k(T)&:=\sum_{\substack{j\ {\rm even}\\t_j\le T}}
       \delta d_j\lambda_j(k)
       +\frac1\pi\int_0^T\delta W(t)\chi_k(t)\,\dd t,\notag\\
 V_k(T)&:=\sum_{\substack{j\ {\rm even}\\t_j\le T}}
       |\delta d_j\lambda_j(k)|
       +\frac1\pi\int_0^T|\delta W(t)\chi_k(t)|\,\dd t.
 \label{d6c:missing_cumulative_weights}
\end{align}
We assume that $V_k(T)$ is finite for every finite $T$.
Contributions of opposite sign can cancel in $R_k(T)$,
whereas $V_k(T)$ adds their absolute values.

We replace $H_X(t)$ in~\eqref{d6c:missing_exact_spectrum}
by its Gaussian approximation $e^{-t^2/(2X)}$.
Under the bound $V_k(T)=O(1+T^2)$, the exact and Gaussian
spectral expressions differ by
$O((\log X)^2)=o(X^\alpha)$ for every $\alpha>0$.
The replacement therefore preserves the coefficient of
$y^\alpha e^{-2\pi ky}$ being tested.
Writing the Gaussian contribution in terms of $R_k(t)$ gives
\be
 e^{2\pi ky}\delta f_k(y)
 =\frac1{2\sqrt k\,X}\int_0^\infty
 t e^{-t^2/(2X)}R_k(t)\,\dd t+O((\log X)^2).
 \label{d6c:missing_exact_bound}
\ee
After rescaling $t=\sqrt X\,v$ and dividing by $X^\alpha$,
the integral depends on $X$ only through
$R_k(\sqrt X\,v)/X^\alpha$. If $R_k(T)=o(T^{2\alpha})$,
this ratio tends to zero for each fixed $v>0$.
The same growth condition, together with local finiteness,
bounds the absolute integrand by
$Cve^{-v^2/2}(1+v^{2\alpha})$ for $X\ge1$.
This bound allows us to take the large-$X$ limit inside
the integral, which therefore tends to zero. The remaining term $(\log X)^2/X^\alpha$ also
vanishes for $\alpha>0$. We obtain
the sufficient criterion
\begin{align}
 &R_k(T)=o(T^{2\alpha}),\qquad
 V_k(T)=O(1+T^2),\qquad \alpha>0,\notag\\
 &\qquad\Longrightarrow\quad
 e^{2\pi ky}\delta f_k(y)=o(y^\alpha).
 \label{d6c:missing_leading_coefficient}
\end{align}
The bound on $V_k$ justifies the Gaussian replacement in
the cusp sum and continuous integral, including the
contributions from large spectral parameters. The bound
on $R_k$ controls the resulting integral and hence the
change in the instanton prefactor.

Under the same bound $V_k(T)=O(1+T^2)$, we have
\be
 |R_k(T)|=O(1+T^\beta)
 \quad\Longrightarrow\quad
 e^{2\pi ky}\delta f_k(y)=O(y^{\beta/2}),\qquad\beta>0.
 \label{d6c:missing_power_bound}
\ee
Thus $\beta<1$ leaves the $\sqrt y$ coefficient unchanged,
while $\beta<2$ leaves the $y$ coefficient unchanged.
These estimates are proved in
Appendix~\ref{d6c:stability_proof}.

\paragraph{Spectral truncation in $D^6R^4$.}
We truncate the $D^6R^4$ reconstruction at a fixed cutoff
$T_0$: the cusp sum retains terms with $t_j\le T_0$,
and the continuous integral runs from $0$ to $T_0$.
We keep $S_k$ unchanged and denote the omitted
contribution by $\delta f$.

The omitted weights satisfy $V_k(T)\ll_k1+T$ and
$R_k(T)=a_kT+o(T)$, with $a_k$ defined
in~\eqref{d6a:moment_properties}.
Appendix~\ref{d6c:stability_proof} establishes these
relations using the spectral moment of
Section~\ref{d6a:leading}.
Equation~\eqref{d6c:missing_exact_bound} therefore gives
\be
 e^{2\pi ky}\delta f_k(y)
 =\frac{\pi a_k}{2}\sqrt y+o(\sqrt y).
\ee
The coefficient $\pi a_k/2$ equals the square-root
coefficient of the full instanton prefactor.
Since the omitted continuous contribution is subleading
at this order, the omitted cusp sum supplies the entire
leading square-root term. This truncation changes the
complete zero and first Fourier modes.

\subsection{Modular ambiguities in higher Fourier modes}
\label{d6c:infinite_cusp_ambiguity}

We now construct explicit modular additions that preserve
specified partial data for the zero and first Fourier modes
while changing leading instanton coefficients in higher modes.
The two constructions keep different sets of low-mode
data fixed.

\paragraph{Preserving the complete zero mode.}
For a chosen positive Fourier mode $m$, we define
the following cusp addition:
\be
 \Psi_m(\tau)=\frac{\pi}{\sqrt m}
 \sum_{j\ {\rm even}}q_2(t_j)
 \frac{\overline{\rho_j(m)}}{\cosh(\pi t_j)}u_j(\tau),
 \qquad q_2(t)=\frac{t^2+1/4}{t^2+4}.
 \label{d6c:infinite_cusp_construction}
\ee
The series converges to a smooth, real, even modular
function in the cuspidal $L^2$ subspace.
Applying the even Kuznetsov formula to the Gaussian
spectral sum gives a diagonal term only for $k=m$.
Both convergence and the following Fourier asymptotics
are established in Appendix~\ref{d6c:infinite_cusp_proof}. We have
\be
 e^{2\pi ky}(\Psi_m)_k(y)
 =\delta_{mk}y+O_{m,k}\!\left(\sqrt y\log^2(2+y)\right),
 \qquad k\ge1.
 \label{d6c:infinite_cusp_asymptotic}
\ee
\eqalias{d6c:infinite_cusp_gaussian_main}{d6c:infinite_cusp_asymptotic}
The remainder is $o(y)$. Therefore we have
\be
 (\Psi_m)_0(y)=0,\qquad
 \lim_{y\to\infty}\frac{e^{2\pi ky}}{y}(\Psi_m)_k(y)
 =\delta_{km},\qquad k\ge1.
 \label{d6c:infinite_cusp_limits}
\ee
For any real $c$, let $\widetilde f=f+c\Psi_m$.
The complete zero mode is unchanged. For all positive modes,
\be
 \lim_{y\to\infty}\frac{e^{2\pi ky}}{y}
 [\widetilde f_k(y)-f_k(y)]=c\delta_{km}.
 \label{d6c:infinite_cusp_shift}
\ee
Whenever the corresponding leading limits exist for $f$,
finite linear combinations of the $\Psi_m$ allow
independent shifts in any finite set of higher modes.
Subleading terms in the first mode may change.
For $c\ne0$, this addition changes the leading coefficient
and, as expected, does not satisfy the stability condition
$R_m(T)=o(T^2)$ of Section~\ref{d8e:d6inputaudit}.

If $m\ge2$, the coefficient of $y e^{-2\pi y}$ in the
first mode is also unchanged. This choice therefore preserves the first-mode input from
the $D^8R^4$ disk amplitude~\cite{Agmon} while allowing
a leading higher-mode coefficient to vary, as discussed
further in Section~\ref{d8e:input}.

\paragraph{Preserving the full one-instanton expansion.}
The next construction preserves the full asymptotic prefactors
of $e^{-2\pi y}$ in the first mode and $e^{-4\pi y}$ in the
zero mode.
Let $\Delta_{\rm hol}$ denote the holomorphic discriminant
of weight $12$ and $E_{24}^{\rm hol}$ the holomorphic
Eisenstein series of weight $24$, normalized to have
constant term one. Define
\be
 H_{\rm amb}(\tau)=\frac{y^{24}}{E_{23}(\tau)}
 \left[\Delta_{\rm hol}(\tau)^2\overline{E_{24}^{\rm hol}(\tau)}
 +\overline{\Delta_{\rm hol}(\tau)}^{\,2}E_{24}^{\rm hol}(\tau)\right].
 \label{d8e:ambiguity_full_sector_construction}
\ee
The products in brackets have modular weights $(24,24)$,
canceled by $y^{24}$. The nonholomorphic Eisenstein series
$E_{23}$ is invariant and strictly positive, so the
denominator introduces no singularities. The resulting
function is smooth, real, even and modular.

For $\delta f=cH_{\rm amb}$ with $c\in\mathbb R$,
Appendix~\ref{d8e:full_sector_ambiguity} shows
\begin{align}
 \delta f_0(y)&=O(ye^{-8\pi y}), \notag\\
 \delta f_1(y)&=O(ye^{-6\pi y}), \notag\\
 \lim_{y\to\infty}\frac{e^{4\pi y}}{y}\delta f_2(y)&=c.
 \label{d8e:full_sector_ambiguity_limits}
\end{align}
Thus the coefficient of $y e^{-4\pi y}$ in the second
mode changes by $c$, while the specified zero- and
first-mode asymptotic sectors remain unchanged.

The two constructions preserve different low-mode data;
additional string-theory constraints may restrict this
modular freedom.

\section{Implications for \texorpdfstring{$D^8R^4$}{D8R4}}
\label{d8e:input}

The input for $D^8R^4$ consists of the known perturbative
zero-mode terms and the leading one-instanton coefficient
from the disk amplitude. We first express these results in a common
normalization. We then apply Section~\ref{rev:incomplete} to exhibit
ambiguities in higher Fourier modes and give sufficient
spectral bounds for determining the coefficient of
$y e^{-2\pi ky}$.

\subsection{Perturbative terms and the disk amplitude}
\label{d8e:conventions}

We follow the axiodilaton and complex Fourier conventions of
Section~\ref{rev:fourier_conventions}. Normalizing the amplitude
by its tree-level supergravity value, we write the local
$D^8R^4$ contribution as
\be
 M\big|_{D^8R^4,\,\mathrm{local}}
 =STU\,\mathcal F_8^{(\mu)}(\tau)(S^4+T^4+U^4),
 \qquad S+T+U=0.
 \label{d8e:ampdef}
\ee
 Massless loops generate logarithmic momentum dependence in
addition to the local $D^8R^4$ term. We write these threshold
logarithms as $\log[(-S-i0)/\mu]$, with analogous terms in
$T$ and $U$, at a fixed dimensionless scale $\mu>0$.
The resulting local coupling $\mathcal F_8^{(\mu)}$ depends
on the subtraction scale $\mu$.
A change of $\mu$ is compensated by a shift of the local
coupling, leaving the full amplitude unchanged. We choose $\mu=1$ and write
$\mathcal F_8=\mathcal F_8^{(1)}$.

In this convention, the zero mode takes the form
\begin{align}
 \mathcal F_{8,0}(y)
 ={}&\frac{\zeta(7)}{16384}y^{7/2}
 +\frac{\pi^2\zeta(3)}{92160}y^{3/2}
       \left[C+\log4-\frac12\log y\right]\notag\\
 &+y^{-1/2}\left[a_2-\frac{\pi^4}{552960}\log y\right]
 +\mathcal R_{8,0}(y),
 \label{d8e:zeroinput}\\
 C={}&\frac{\zeta'(4)}{\zeta(4)}-\frac{\zeta'(3)}{\zeta(3)}
 -\gamma_E+\frac{63}{20}-\log(2\pi).
 \label{d8e:g1DG}
\end{align}
The tree-level term follows from the expansion of the
Virasoro--Shapiro amplitude
\cite{GreenVanhove9910056}.
The genus-one calculation fixes both the finite term and
the logarithmic coefficient~\cite{DHokerGreen,CD2412}.
The genus-two results determine the displayed logarithmic
term~\cite{DGP1806,Pioline1810}.
Here $a_2$ is the nonlogarithmic genus-two coefficient. It is not fixed by the currently
known perturbative results.
The remainder $\mathcal R_{8,0}$ contains higher-genus
and neutral nonperturbative contributions. 
Appendix~\ref{d8e:amplitude_details} derives the normalization
and the conversion to our threshold convention.

A D$(-1)$-instanton is a Euclidean D-brane localized in
all ten spacetime directions. Its classical action
$2\pi/g_s-2\pi iC_0$ gives the instanton weight
$e^{2\pi i\tau}=e^{-2\pi y}e^{2\pi ix}$.
For the $D^8R^4$ interaction, the disk-amplitude calculation
of Agmon et al.~\cite{Agmon} fixes the
coefficient of the leading term $y e^{-2\pi y}$ in the
first Fourier mode:
\be
 \lim_{y\to\infty}\frac{e^{2\pi y}}{y}\mathcal F_{8,1}(y)
 =\frac{\zeta(4)}{2048}.
 \label{d8e:firstasymp}
\ee
Changing the subtraction scale $\mu$ shifts the first mode
by $O(e^{-2\pi y})$ and therefore leaves the coefficient
of the leading $y e^{-2\pi y}$ term unchanged;
see~\eqref{d8e:firstscaleshift}.

We use this first-mode coefficient together with the
perturbative zero-mode data to ask whether they determine
the corresponding coefficients of $y e^{-2\pi ky}$
in the higher Fourier modes.

\subsection{Ambiguities in higher Fourier modes}

For higher Fourier modes, we define the analogue of the
first-mode coefficient in~\eqref{d8e:firstasymp} by
\be
 a_{8,k}:=\lim_{y\to\infty}
 \frac{e^{2\pi ky}}{y}\mathcal F_{8,k}(y),
 \qquad k\ge2,
 \label{d8e:leading_coefficient}
\ee
whenever the limit exists. We ask whether modular invariance and the available zero-
and first-mode data determine $a_{8,k}$ for $k\ge2$.
These data constrain the weak-coupling behavior of the
zero mode without determining its full dependence on $y$.
The first-mode result alone does not determine the leading
power of $y$ in a higher-mode prefactor.

To exhibit the ambiguity in $a_{8,m}$ for $m\ge2$,
we use the cusp construction of
Section~\ref{d6c:infinite_cusp_ambiguity} to obtain
the alternative modular candidate
\be
 \widetilde{\mathcal F}_8=\mathcal F_8+c\Psi_m,
 \qquad c\in\mathbb R,
 \label{d8e:leading_ambiguity_addition}
\ee
whose Fourier modes satisfy
\be
 \lim_{y\to\infty}\frac{e^{2\pi ky}}{y}
   \bigl(\widetilde{\mathcal F}_{8,k}-\mathcal F_{8,k}\bigr)
 =c\delta_{km}.
 \label{d8e:leading_ambiguity_shift}
\ee
The addition preserves the complete zero mode and the
leading first-mode coefficient, while shifting $a_{8,m}$
by $c$. 

Adding a multiple of $H_{\rm amb}$, defined
in~\eqref{d8e:ambiguity_full_sector_construction},
gives another ambiguity.
It preserves the perturbative zero-mode terms and the
full asymptotic expansions multiplying $e^{-2\pi y}$
in the first mode and $e^{-4\pi y}$ in the zero mode. It shifts $a_{8,2}$ by an arbitrary constant,
while changing only more suppressed exponentials in the
zero and first modes. Further string-theory constraints on the $D^8R^4$
coupling may restrict the modular freedom described above.

\subsection{Conditions for determining a leading coefficient}
\label{d8e:endpoint}
\label{d8e:prediction_stability}

The Eisenstein subtraction removes the known growing
terms in~\eqref{d8e:zeroinput}. Its Fourier modes satisfy
$e^{2\pi ky}S_k^{(8)}(y)=O(1)$ and hence do not contribute
to $a_{8,k}$. Regularized unfolding expresses the continuous contribution
in terms of the zero-mode profile over all $y>0$.

The cusp amplitudes are extracted from the residues of
the Mellin transform of the subtracted first Fourier mode. To examine what can be calculated from the disk result,
we split the Mellin integral at $Y>1$. This separates
the large-$y$ region, where only the leading asymptotic
is known, from $0<y<Y$, where the first-mode profile
is unspecified. We can evaluate the contribution of
the known leading term to the upper part of the damped
transform in~\eqref{intro:damped_Mellin}:
\begin{align}
 \mathcal M_8^{\mathrm{disk}}(s;Y)
 &:=\int_Y^\infty y^{s-3/2}e^{-2\pi y}
       \left[a_{8,1}y e^{-2\pi y}\right]\dd y
       \notag\\
 &=\frac{\zeta(4)}{2048}(4\pi)^{-s-1/2}
      \Gamma\!\left(s+\tfrac12,4\pi Y\right).
 \label{d8e:leading_profile_transform}
\end{align}
Here $a_{8,1}=\zeta(4)/2048$ is the leading first-mode
coefficient in~\eqref{d8e:firstasymp}, and the additional
factor $e^{-2\pi y}$ provides the Mellin damping.
For fixed $Y$, this expression is entire in $s$.
More generally, under our growth assumptions, exponential
damping makes the contribution from $y\ge Y$ to the Mellin
transform entire in $s$. The cusp residues therefore require
analytic continuation of the integral over $0<y<Y$, after
subtracting the Eisenstein and continuous contributions
from the first mode.
The disk asymptotic alone determines neither these residues
nor an explicit bound on the remainder at finite $Y$.

Let $\widetilde{\mathcal F}_8$ be a modular approximation
to the physical coupling $\mathcal F_8$, chosen to match
the supplied data. We seek sufficient conditions for this
approximation to reproduce the physical coefficient $a_{8,k}$.
For the difference
$\delta\mathcal F_8=\widetilde{\mathcal F}_8-\mathcal F_8$,
define $R_k$ and $V_k$ as
in~\eqref{d6c:missing_cumulative_weights}.
Taking $\alpha=1$ in the stability criterion of
Section~\ref{d8e:d6inputaudit} gives sufficient conditions
for the difference to be subleading to $y e^{-2\pi ky}$:
\begin{align}
 &R_k(T)=o(T^2),\qquad V_k(T)=O(1+T^2) \notag\\
 &\qquad\Longrightarrow\quad
 \lim_{y\to\infty}\frac{e^{2\pi ky}}{y}
       \delta\mathcal F_{8,k}(y)=0.
 \label{d8e:D8stability}
\end{align}
Under these bounds, the coefficient of $y e^{-2\pi ky}$
calculated from $\widetilde{\mathcal F}_8$ equals the
physical coefficient $a_{8,k}$. The ambiguity constructions show that matching the supplied
low-mode asymptotics alone does not fix $a_{8,k}$.
To establish agreement through the stability criterion,
the bounds on $R_k$ and $V_k$ must be verified separately
for the difference between the approximation and the
physical coupling.

Calculating the local $D^8R^4$ contribution to the
four-graviton amplitude with two instantons of the same
sign would provide an independent second-mode coefficient
for testing a modular approximation.  Combining further amplitude data with the
spectral stability bounds offers a possible route to controlled
predictions for higher Fourier modes.

\section{Conclusion and outlook}
\label{paper:conclusion}

We have reconstructed the coefficient of the $D^6R^4$ interaction
in type IIB string theory from its complete zero and first Fourier
modes, using modular spectral theory without invoking supersymmetry
differential equations. Regularized Rankin--Selberg unfolding
determines the continuous contribution from the zero mode. Subtracting
the continuous and Eisenstein contributions from the first mode
isolates its cusp part. We evaluate the residues of its Mellin
transform using Motohashi's additive-divisor formula. Under the stated
simplicity assumption for the even cusp spectrum, these residues
determine the individual cusp amplitudes. The Hecke relations then
fix the cusp contribution to every higher Fourier mode.

The interaction-specific information enters through the two low-mode
functions, while the spectral basis and Hecke eigenvalues depend only
on the modular group. The $D^6R^4$ calculation thus provides a
benchmark for reconstructing the modular coefficients of non-BPS
interactions from their low Fourier modes.

From the exact spectral representation, we extract the
square-root and constant terms in the weak-coupling instanton
prefactor, recovering the GMV coefficients, and establish
a bound on the remainder for each positive Fourier mode.
The square-root term comes entirely from the infinite cusp sum,
through the growing range $t_j=O(\sqrt y)$; every finite cusp
sum has a bounded prefactor. Using the Kuznetsov trace formula,
we evaluate the residue of the combined cusp and continuous
spectral moment that determines the square-root coefficient.
The bounds on the continuous contribution show that it is
subleading at this order. The value of the analytically continued moment at the origin,
together with the restored Eisenstein terms, gives the
constant term. The exact reconstruction also encodes further inverse
powers of $y$ and more suppressed exponential sectors, beyond
the two orders evaluated here.

Our reconstruction uses the complete zero and first Fourier modes. With only partial weak-coupling
data, explicit modular additions demonstrate freedom in both
bounded and growing terms of higher-mode instanton prefactors.
Our spectral stability bounds give sufficient conditions for
omitted contributions to leave the leading coefficient of
a growing instanton prefactor unchanged. These conditions require control of the spectral
weights beyond agreement with the supplied low-mode asymptotics.

For $D^8R^4$, the known perturbative terms and the disk
amplitude supply partial zero-mode data and the leading
one-instanton coefficient. These results constrain the weak-coupling behavior of the
two low modes without determining their full dependence on $y$.
Our modular additions show that the supplied data leave
the higher-mode coefficients of $y e^{-2\pi ky}$ undetermined,
even when their defining limits exist, with a second-mode
ambiguity persisting even when the full one-instanton
expansion is fixed and the zero mode changes only by
$O(ye^{-8\pi y})$. Our stability bounds give sufficient conditions for a modular
approximation to reproduce the physical leading coefficients
in higher Fourier modes. Establishing these bounds for a
specific $D^8R^4$ approximation remains to be done.

Two complementary directions could extend the present analysis.
Further D-instanton amplitude calculations could refine the
low-mode input~\cite{SenSingle,Agmon} and provide independent
checks of higher Fourier modes. In particular, calculating
the local $D^8R^4$ contribution to the four-graviton amplitude
with two instantons of the same sign would provide a
second-mode coefficient for testing a modular approximation.

Resummation methods could help construct more complete
low-mode functions, while additional nonperturbative data
may be needed to resolve remaining ambiguities. Relevant
precedents include the resurgent reconstruction of
exponentially suppressed zero-mode terms in modular graph
functions~\cite{DK,DKT} and the modular-invariant Borel
resummation of large-$N$ expansions of integrated correlators
in $\mathcal N=4$ SYM~\cite{DorigoniTreilis2024}.
The Mellin reflection relation provides a further constraint
that may help reconstruct the missing zero-mode terms,
although it does not determine them uniquely.

Together with the spectral stability bounds, these approaches
could support controlled predictions for $D^8R^4$. The
framework could also be applied to other higher-derivative
interactions and modular observables whose low-mode data
and cusp growth can be determined independently.

\acknowledgments
This work began in the summer of 2025, during the author's
postdoctoral appointment at Harvard University.
The author thanks Xi Yin for valuable discussions.
This work is supported in part by the Istituto Nazionale
di Fisica Nucleare (INFN) through the ``Gauge Theories,
Strings, Supergravity'' (GSS) research project
(postdoctoral fellowship ref.~28201/2025).
AI tools were used for checks, numerical code development,
and editorial assistance.

\appendix
\section{Low Fourier modes and cusp reconstruction}
\label{d6c:Mellin_details}
In this appendix, we give the explicit GMV formulas for the zero
and first Fourier modes and evaluate the zero-mode transform
that determines the continuous contribution. We then compute
the first-mode Mellin residues from the GMV formulas and use
the spectral residue relation to determine the sum of cusp
amplitudes at each even cusp eigenvalue. We justify the
subtractions, convergence and normalization used in these
calculations and describe the remaining freedom when several
cusp forms share an eigenvalue.

\subsection{Explicit formulas for the zero and first Fourier modes}
\label{paper:GMV_data}\label{d6c:app:input}
\label{d6c:input_coefficients}\label{d6c:input_tail}
The neutral kernel in~\eqref{d6c:neutral_input} is~\cite{GMV}
\begin{align}
 r^{00}(z)&=-512z^5+48z^3-15z, \notag\\
 r^{01}(z)&=r^{10}(z)=-128z^4-12z^2-15, \notag\\
 r^{11}(z)&=512z^5+16z^3+33z-15z^{-1}.
 \label{d6c:eq:rmatrix}
\end{align}
The first mode is the sum over ordered pairs
$f_1=2f_{0,1}+2\sum_{m\ge1}f_{-m,m+1}$, with
\be
 f_{a,b}(y)=f^P_{a,b}(y)
       +\alpha_{a,b}\sqrt yK_{7/2}(2\pi y).
 \label{d6c:app_input_pair_split}
\ee
We use these pair functions, with $a+b=1$, as supplied input from
\cite{GMV}. 
For the pair with one zero charge,
\be
 f^P_{0,1}(y)=\frac8{9\pi}
 \left[q^0(z)K_0(2z)+q^1(z)K_1(2z)\right],
 \qquad z=\pi y,
 \label{d6c:app_input_zero_particular}
\ee
where
\begin{align}
 q^0(z)&=\frac{90\zeta(3)-\pi^4+9z^2\zeta(3)}{z}, \notag\\
 q^1(z)&=\frac{90\zeta(3)-\pi^4+54z^2\zeta(3)}{z^2}.
 \label{d6c:app_input_zero_coefficients}
\end{align}
For $p=m+1$, $q=-m$, $m\ge1$, the opposite-charge pair is
\begin{align}
 f^P_{q,p}(y)=f^P_{p,q}(y)
 &=\frac{32\pi\sigma_2(p)\sigma_2(|q|)}{3|pq|}
 \sum_{i,j=0}^1Q_{ij}(z;p,q) \notag\\
 &\hspace{14mm}\times K_i(2\pi py)K_j(2\pi|q|y),
 \qquad z=\pi y,
 \label{d6c:app_input_mixed_particular}
\end{align}
with
\begin{align}
 Q_{00}(z;p,q)&=4zpq(p^2+q^2-6pq)+\frac{30}{z}pq(p-q)^2,
 \label{d6c:app_input_Q00}\\
 Q_{01}(z;p,q)&=-p(13p^2q-65pq^2+p^3+19q^3) \notag\\
 &\qquad+\frac{30}{z^2}pq^2(p-q),
 \label{d6c:app_input_Q01}\\
 Q_{10}(z;p,q)&=q(13q^2p-65qp^2+q^3+19p^3) \notag\\
 &\qquad-\frac{30}{z^2}qp^2(q-p),
 \label{d6c:app_input_Q10}\\
 Q_{11}(z;p,q)&=-4zpq(p^2+q^2-6pq) \notag\\
 &\qquad-\frac{14p^3q-94p^2q^2+14pq^3+p^4+q^4}{z}.
 \label{d6c:app_input_Q11}
\end{align}
The coefficients are evaluated at $q=-m$, whereas the divisor factors
and Bessel arguments use $|q|=m$; the first Bessel index belongs to $p$.
The coefficients of the common $K_{7/2}$ term are
\be
 \alpha_{0,1}=\alpha_{1,0}
 =\frac{64(\pi^4-90\zeta(3))}{135\pi}.
 \label{d6c:app_input_zero_homogeneous}
\ee
For the opposite-charge pairs,
\begin{align}
 \alpha_{p,q}=\alpha_{q,p}
 &=\frac{128\pi\sigma_2(p)\sigma_2(|q|)}{45p^2q^2}
 \Bigl[p^5+q^5+15p^4q+15pq^4 \notag\\
 &\qquad-80p^3q^2-80p^2q^3 \notag\\
 &\qquad+60p^2q^2(p-q)\log(p/|q|)\Bigr].
 \label{d6c:app_input_mixed_homogeneous}
\end{align}

To evaluate the total coefficient of
$\sqrt yK_{7/2}(2\pi y)$, set
\begin{align}
 \Phi_m={}&-\frac14\left(\frac{p^2}{m^2}+\frac{m^2}{p^2}\right)
 +\frac72\left(\frac p m+\frac m p\right)+\frac{47}{2} \notag\\
 &\quad-15(2m+1)\log\frac p m.
 \label{d6c:homogeneous_phi}
\end{align}
Direct substitution gives
$\alpha_{-m,m+1}=-(512\pi/45)\sigma_2(m)\sigma_2(m+1)\Phi_m$.
The expansion $\Phi_m=-1/(28m^6)+O(m^{-7})$, together with
$\sigma_2(n)\le\zeta(2)n^2$, makes this series absolutely convergent.
The divisor-convolution identity of Fedosova--Klinger-Logan--Radchenko~\cite{FKR},
at total charge one with $r_1=r_2=2$, $d=1$, gives
\be
 2\sum_{m\ge1}\Phi_m\,\sigma_2(m)\sigma_2(m+1)
 =\frac{\zeta(2)}2+30\zeta'(-2).
 \label{d6c:homogeneous_divisor_identity}
\ee
The factor two counts the pair orderings. The possible holomorphic
cusp contribution to the Fedosova--Klinger-Logan--Radchenko identity has weight eight and vanishes
at level one. Using
$\zeta'(-2)=-\zeta(3)/(4\pi^2)$~\cite{DLMF}, we obtain
\be
 2\alpha_{0,1}+2\sum_{m\ge1}\alpha_{-m,m+1}=0.
 \label{d6c:homogeneous_cancellation}
\ee
Thus the full first mode is the sum of explicit Bessel terms
\eqref{d6c:app_input_particular_sum}.

For convergence, the Bessel integral representation gives
$0<K_0(x)\le K_1(x)\le\sqrt{\pi/(2x)}e^{-x+1/(2x)}$ for $x>0$
by $\cosh v\ge1+v^2/2$~\cite{DLMF}.
The coefficients above satisfy
$\sum_{i,j}|Q_{ij}(\pi y;p,-m)|\ll p^4(y+1+y^{-1}+y^{-2})$.
Since $(m,p)=1$, the divisor factors satisfy
$\sigma_2(m)\sigma_2(p)=m^2p^2\sigma_{-2}(mp)\le\zeta(2)m^2p^2$.
The two Bessel bounds contribute a factor $1/(y\sqrt{mp})$, giving
\begin{align}
 |f^P_{-m,m+1}(y)|&\ll(m+1)^5e^{-2\pi(2m+1)y},
 \qquad m\ge1,\quad y\ge1, \notag\\
 \left|2\sum_{m\ge1}f^P_{-m,m+1}(y)\right|
 &\ll e^{-6\pi y}\sum_{m\ge1}(m+1)^5e^{-4\pi(m-1)}
 =O(e^{-6\pi y}).
 \label{d6c:f1tail}
\end{align}
For any $y_0>0$ the same argument gives absolute convergence
uniformly on $y\ge y_0$, with a bound depending on $y_0$. The Mellin
integral also samples $y=O(1/m)$, where the two Bessel arguments can
stay finite as the pair charges grow.

\subsection{Mellin transforms and the large-charge expansion}
\label{d6c:gamma}\label{d6c:shifted_divisors}\label{d6c:uniform}
The neutral kernel enters both the zero-mode transform and the leading
large-charge term of the first mode. We evaluate its Mellin integral
once for both purposes. Writing $r^{ij}(z)=\sum_m c_{ijm}z^m$ and
rescaling by $X=2\pi|n|y$ separates the zero-mode transform into
\begin{align}
 R_f(s)
 &=\frac{32\pi}{315(2\pi)^{s-1}}
 \sum_{n\ne0}\frac{\sigma_2(|n|)^2}{|n|^{s+2}} \notag\\
 &\qquad\times
 \sum_{i,j,m}\frac{c_{ijm}}{2^m}
 \int_0^\infty X^{m+s-2}K_i(X)K_j(X)\,\dd X.
 \label{d6c:RS_separated}
\end{align}
Its Bessel and arithmetic factors are
\begin{align}
 I_{ij}(a)&:=\int_0^\infty X^{a-1}K_i(X)K_j(X)\,\dd X \notag\\
 &=\frac{2^{a-3}}{\Gamma(a)}
 \prod_{\epsilon,\eta=\pm1}
 \Gamma\left(\frac{a+\epsilon i+\eta j}{2}\right),
 \qquad \Re a>i+j.
 \label{d6c:zero_Bessel_Mellin}
\end{align}
\be
 \sum_{n\ge1}\frac{\sigma_2(n)^2}{n^{s+2}}
 =\frac{\zeta(s+2)\zeta(s)^2\zeta(s-2)}{\zeta(2s)},
 \qquad \Re s>3.
 \label{d6c:divisor_square_series}
\ee
The first identity is given in~\cite{DLMF}; the second is
Ramanujan's identity~\cite{GMV}. Here $a=m+s-1$. The term $m=-1$, $i=j=1$
requires $\Re s>4$ at $X=0$, while the charge sum converges for
$\Re s>3$. Exponential decay controls the large-$y$ endpoint. Together
with the small-$y$ bounds, this justifies interchanging the
sum and the Mellin integral for $\Re s>4$. Including both signs of $n$ gives
\be
 R_f(s)=\frac{8\pi^{2-s}}{315}
 \frac{\zeta(s+2)\zeta(s)^2\zeta(s-2)}{\zeta(2s)}
 \,\mathcal G(s),
 \label{d6c:RS_gamma_sum}
\ee
where
\be
 \mathcal G(s)=\sum_{i,j,m}\frac{c_{ijm}}{\Gamma(m+s-1)}
 \prod_{\epsilon,\eta=\pm1}
 \Gamma\left(\frac{m+s-1+\epsilon i+\eta j}{2}\right).
 \label{d6c:gamma_sum_definition}
\ee
The thirteen nonzero terms reduce by $\Gamma(z+1)=z\Gamma(z)$ to
\begin{align}
 Q(s)&:=\frac{\mathcal G(s)\Gamma(s-2)}{\Gamma(s/2)^4},
 \label{d6c:Q_definition}\\
 Q(s)&=-\frac{630s}{(s-4)(s-2)^2(s-1)(s+3)}.
 \label{d6c:gammarecurrence}
\end{align}
Conversion to completed zeta functions leaves the rational factor
$(s-1)(s-2)^2Q(s)/(630s)=[12+s(1-s)]^{-1}$ and gives~\eqref{d6c:eq:Rf}.
For the first mode, the leading pair kernel has Mellin integral
$\int_0^\infty X^{a-2}R_0(X)\,\dd X=2^{a-4}\mathcal G(a)$, where
$R_0(X)=\sum_{i,j}r^{ij}(X/2)K_i(X)K_j(X)$.
Dividing by $I_{11}(a)$ in~\eqref{d6c:I11_explicit} yields
$315/[(4-a)(a+3)]$ and hence~\eqref{d6c:leading_Mellin_coefficient}.

To control the remainder after summing over the pairs, we expand about the
midpoint of the charge magnitudes $m$ and $m+1$. Put $u=m+1/2$,
$\delta=(2u)^{-1}$ and $X=2\pi uy$, so $p=u(1+\delta)$, $q=-u(1-\delta)$
and $\pi y=\delta X$. With $a_m$ as in~\eqref{am}, the explicit
Bessel contribution retained for this pair is
\be
 f^P_{-m,m+1}(y)
 =\frac{\sigma_2(m)\sigma_2(m+1)}{u^3}F_P(X,\delta)
 =a_m u(1-\delta^2)^2F_P(X,\delta),
 \label{d6c:pair_normalization}
\ee
where
\be
 F_P(X,\delta)=\frac{16\pi}{3\delta(1-\delta^2)}
 \sum_{i,j=0}^1Q_{ij}(\delta X;p,q)
 K_i(X(1+\delta))K_j(X(1-\delta)).
 \label{d6c:normalized_pair_kernel}
\ee
Substituting $p=(1+\delta)/(2\delta)$ and $q=-(1-\delta)/(2\delta)$
extends this expression to $|\delta|<1$. Taylor expansion of
$K_i(X(1\pm\delta))$, using $K_0'=-K_1$ and
$K_1'=-K_0-K_1/X$, cancels its negative powers of $\delta$.
Moreover $\mathcal Q_{ij}(-\delta)=-\mathcal Q_{ji}(\delta)$ for
$\mathcal Q_{ij}(\delta)=Q_{ij}(\delta X;p,q)$; the prefactor is odd
and the Bessel arguments exchange. Thus $F_P$ is even and analytic
at $\delta=0$, with constant term $F_0$ in~\eqref{d6c:pair_leading_profile}.

The small-$X$ Bessel series~\cite{DLMF} give, uniformly
on $|\delta|\le r<1$,
\be
 F_P(X,\delta)
 =A_\delta X^{-3}+B_\delta X^{-1}
 +O_r\!\left(X(1+|\log X|^2)\right),
 \qquad X\to0^+.
 \label{d6c:eq:smallX}
\ee
The logarithms cancel at the two displayed orders, and the coefficients are
\be
 A_\delta=-\frac{16\pi}{3}
 \frac{G_\delta}{\delta^2(1-\delta^2)^2},
 \qquad
 B_\delta=
 \frac{16\pi/5+(32\pi/15)G_\delta}{(1-\delta^2)^2},
 \label{d6c:eq:ABdelta}
\ee
where the following expression has a removable singularity at the origin:
\begin{align}
 G_\delta&=\frac{15(1-\delta^2)^2}{4\delta^5}
 \log\frac{1+\delta}{1-\delta}
 -\frac{15}{2\delta^4}+\frac{25}{2\delta^2}-4 \notag\\
 &=\sum_{k\ge1}\frac{60\delta^{2k}}{(2k+1)(2k+3)(2k+5)},
 \qquad |\delta|<1.
 \label{d6c:G_series}
\end{align}
The logarithm is on its principal branch. In particular
$A_0=-64\pi/21$ and $B_0=16\pi/5$.
Subtracting these two powers at the lower endpoint isolates the real
Mellin poles of each pair. Set
$J(X,\delta)=F_P(X,\delta)-\boldsymbol1_{(0,1)}(X)
(A_\delta X^{-3}+B_\delta X^{-1})$.
Here $\boldsymbol1_{(0,1)}(X)$ equals one for $0<X<1$
and zero otherwise, so the singular terms are subtracted
only on $0<X<1$.
The pair integral $\widehat F_P(s,\delta)=\int_0^\infty X^{s-3/2}F_P(X,\delta)\dd X$,
initially defined for $\Re s>7/2$, continues as
\be
 \widehat F_P(s,\delta)=
 \int_0^\infty X^{s-3/2}J(X,\delta)\,\dd X
 +\frac{A_\delta}{s-7/2}
 +\frac{B_\delta}{s-3/2}.
 \label{d6c:eq:addbacks}
\ee
Its subtracted integrand is
$O_r(X^{\Re s-1/2}(1+|\log X|^2))$ at zero, so the integral is
holomorphic for $\Re s>-1/2$. Only the displayed real poles remain.

For $m\ge1$ all physical values of $\delta$ lie in $(0,1/3]$.
Choose $1/3<r<R<1$ to bound their Taylor remainders uniformly.
The integral representation implies
$|K_i(X(1\pm\delta))|\le K_i(X(1-R))$ on $|\delta|\le R$.
On $|\delta|=R$ the matrix grows at most linearly in $X$, while the
Bessel product is $O_R(X^{-1}e^{-2(1-R)X})$.
The maximum modulus principle and Cauchy's estimates therefore give
$|\partial_\delta^nF_P(X,\delta)|\le C_{r,R,n}e^{-2(1-R)X}$ for
$X\ge1$, $|\delta|\le r$. At zero the same argument applies to $J$,
using~\eqref{d6c:eq:smallX} on the outer circle. Differentiation under
the subtracted Mellin integral is consequently valid, and
\be
 (1-\delta^2)^2\widehat F_P(s,\delta)
 =\widehat F_0(s)
 +\delta^2\bigl[\widehat F_2(s)-2\widehat F_0(s)\bigr]
 +\delta^4\widetilde E(s,\delta),
 \label{d6c:full_pair_Taylor}
\ee
Here $\widehat F_2(s)=\tfrac12\partial_\delta^2\widehat F_P(s,0)$.
The remainder $\widetilde E(s,\delta)$ is holomorphic in $s$
for $\Re s>-1/2$, except possibly at $s=3/2,7/2$.
Let $\Omega$ be any compact subset of this half-plane that
excludes these two points. Then $\widetilde E(s,\delta)$ is
uniformly bounded for $s\in\Omega$ and $|\delta|\le r$.
The arithmetic bound $a_m=\sigma_{-2}(m(m+1))\le\zeta(2)$ gives
\begin{align}
 \sup_{s\in\Omega}\sum_{m\ge1}
 \left|a_m u^{3/2-s}\delta^4\widetilde E(s,\delta)\right|
 &\le\frac{\zeta(2)C_\Omega}{16}
 \sum_{m\ge1}\left(m+\frac12\right)^{-5/2-\sigma_\Omega} \notag\\
 &<\infty,
 \label{d6c:charge_sum_remainder}
\end{align}
with $\sigma_\Omega=\inf_\Omega\Re s$. Keeping only $F_0$ would instead
leave $O(a_mu^{-1/2-\Re s})$, which is not summable on the imaginary axis.

With both pair orderings, the rescaling $y=X/(2\pi u)$ now gives
\begin{align}
 \mathcal M_{f_1}(s)
 ={}&J_f(s)+2(2\pi)^{1/2-s}\widehat F_0(s)D(s-3/2) \notag\\
 &+F_2^*(s)D(s+1/2)+H_f(s),
 \label{d6c:eq:targetMexp}
\end{align}
initially for $\Re s>7/2$, where
$J_f(s)=2\int_0^\infty y^{s-3/2}f^P_{0,1}(y)\,\dd y$ and
\begin{align}
 F_2^*(s)&:=\frac{(2\pi)^{1/2-s}}2
 \left[\widehat F_2(s)-2\widehat F_0(s)\right], \notag\\
 H_f(s)&:=\frac{(2\pi)^{1/2-s}}8
 \sum_{m\ge1}a_m u^{-s-5/2}\widetilde E(s,\delta).
 \label{d6c:target_nonleading_terms}
\end{align}
The finite Bessel sum $J_f$ has only real Mellin poles, while
$F_2^*$ and $H_f$ are holomorphic near every $s=iT$, $T>0$.
The two shifted-divisor factors now play different roles. The mean of
the divisor coefficients suffices to continue $D(s+1/2)$ across the
imaginary $s$-axis. To extract the residue from $D(s-3/2)$ we need
Motohashi's formula.

\subsection{Analytic continuation and residues of the shifted divisor series}
\label{d6c:shift_residues}\label{rev:divisor_continuation}
The bound $1\le a_m\le\zeta(2)$ gives absolute convergence of $D(w)$
for $\Re w>1$. Expanding the two divisor sums and counting
$m\equiv0\pmod d$, $m\equiv-1\pmod e$ gives
\begin{align}
 \sum_{m\le X}a_m
 &=\sum_{\substack{d,e\ge1\\(d,e)=1}}
   \frac1{d^2e^2}\left(\frac{X}{de}+O(1)\right)
 =\mathcal C X+O(1), \notag\\
 \mathcal C&=\sum_{(d,e)=1}d^{-3}e^{-3}
 =\frac{\zeta(3)^2}{\zeta(6)}.
 \label{d6c:shift_elementary_mean}
\end{align}
For each coprime pair $(d,e)$, the count differs from
$X/(de)$ by at most one. After multiplication by
$d^{-2}e^{-2}$ and summation, the total discrepancy is
bounded by $\sum_{d,e\ge1}d^{-2}e^{-2}=\zeta(2)^2$,
independently of $X$.
Partial summation continues $\sum a_mm^{-w}$ to $\Re w>0$ with only
a pole at one. The centering difference
$\sum a_m[(m+1/2)^{-w}-m^{-w}]$ converges locally uniformly there.
Thus $D(s+1/2)$ is regular near $s=iT$, and the quadratic term in
\eqref{d6c:eq:targetMexp} contributes no residue.

For the leading term we continue $D(w)$ to $w=-3/2+iT$. We first apply Motohashi's formula with a smooth weight
$V\in C_c^\infty(0,\infty)$, which vanishes outside a
finite interval contained in $(0,\infty)$.
At divisor exponents $\alpha=\beta=-2$, the discrete Maass part of
Motohashi's shift-one formula is
\be
 \left[\sum_{m\ge1}a_mV(m)\right]_{\mathrm{Maass}}
 =\frac1{4\pi^3}\sum_j
 \frac{|\rho_j(1)|^2}{\cosh(\pi t_j)}
 L_j(\tfrac52)L_j(\tfrac12)
 \Psi_{\epsilon_j}[V](t_j),
 \label{d6c:shift_discrete_formula}
\ee
where $\epsilon_j=0,1$ denotes parity and
\begin{align}
 \Phi_t[V]
 &:=\frac{\Gamma(\tfrac52+it)\Gamma(\tfrac12+it)}
 {\Gamma(1+2it)}
 \int_0^\infty V(x)x^{-5/2-it} \notag\\
 &\qquad\times{}_2F_1\!\left(\tfrac52+it,\tfrac12+it;1+2it;-\frac1x\right)
 \,\dd x, \notag\\
 \Psi_{\epsilon}[V](t)
 &:=\frac\pi4\left[
 \frac{\Phi_t[V]-\Phi_{-t}[V]}{i\sinh(\pi t)}
 -(-1)^\epsilon\bigl(\Phi_t[V]+\Phi_{-t}[V]\bigr)
 \right].
 \label{d6c:shift_hypergeometric_kernel}
\end{align}
Here ${}_2F_1$ denotes the Gauss hypergeometric function.
The value at $t=0$ is defined by continuity. The prefactor $1/(4\pi^3)$ is
$2(2\pi)^{\beta-1}$ in~\cite{Motohashi1994}, in the same
complex Fourier normalization. The kernels follow from its
Eqs.~(3.24), (3.26), (3.27), (3.31), and (3.32) by gamma reflection.

We start at $(\alpha,\beta)=(-2,-5)$, which lies in
Motohashi's initial convergence domain~\cite{Motohashi1994},
and continue $\beta$ to $-2$ using the contour prescription
in Section~4 of that reference. The arguments of the two
$L$-functions run from $(4,2)$ to $(5/2,1/2)$; the continuous zeta poles
cross when the second argument passes through one at $\beta=-3$. Their finitely many
residues occur at real $\xi=it$ and hence give only real Mellin poles.
At coincident parameter poles the arithmetic main terms and
contour residues must be combined before taking the endpoint limit;
the resulting logarithms still give only real Mellin poles.
The cusp kernel stays regular along this continuation.

Choose smooth $\eta$ equal to zero for $x\le1/2$ and one for $x\ge1$,
and put $V_w(x)=\eta(x)(x+1/2)^{-w}$. Its arithmetic sum is exactly
$D(w)$. For large $\Re w$, an upper cutoff can be removed from the
compactly supported formula by absolute arithmetic convergence and
the spectral estimates below. For fixed $t>0$, the large-$x$ tail of
$\Phi_{-t}[V_w]$ is its gamma prefactor times
$x^{-w-5/2+it}(1+O(x^{-1}))$. Since its leading integral is
$(w+3/2-it)^{-1}$ and $\Phi_t$ is regular at $w=-3/2+it$,
\be
 \Res_{w=-3/2+it}\Psi_0[V_w](t)
 =-\frac\pi4\left(1-\frac{i}{\sinh(\pi t)}\right)
 \frac{\Gamma(\tfrac52-it)\Gamma(\tfrac12-it)}
 {\Gamma(1-2it)}.
 \label{d6c:shift_kernel_residue}
\ee

For the spectral summation, let $q_t$ denote the factor
$x^{-5/2-it}{}_2F_1$ in~\eqref{d6c:shift_hypergeometric_kernel}.
The Gauss differential identity for this trace-formula kernel gives
\be
 x(x+1)q_t''+(6x+3)q_t'
 +(\tfrac{25}{4}+t^2)q_t=0.
 \label{d6c:shift_kernel_equation}
\ee
The adjoint operator is $\mathcal L^*V=x(x+1)V''-(2x+1)V'+9V/4$.
Using $x(x+1)=(x+1/2)^2-1/4$ gives
$\mathcal L^*(x+1/2)^{-w}=(w+3/2)^2(x+1/2)^{-w}
-w(w+1)(x+1/2)^{-w-2}/4$.
Thus two integrations by parts yield
\be
 \bigl[t^2+(w+\tfrac32)^2\bigr]\Psi_\epsilon[V_w](t)
 =\frac{w(w+1)}4\Psi_\epsilon[V_{w+2}](t)
 -\Psi_\epsilon[B_w](t),
 \label{d6c:shift_kernel_continuation}
\ee
where $B_w=\mathcal L^*(\eta(x)(x+1/2)^{-w})
-\eta(x)\mathcal L^*(x+1/2)^{-w}$ is supported in $[1/2,1]$.
The original integral is holomorphic for $\Re w>-3/2$; the right side
extends to $\Re w>-7/2$, giving continuation across $-3/2\pm it$.
Euler's integral bounds the kernel, including its gamma prefactor,
by a polynomial in $t$ times $x^{-5/2}$ for $x\ge1/2$.
Repeated integration by parts with $\mathcal L^*$ gives arbitrary
inverse powers of $t$ in $\Re w>-3/2$. The recurrence preserves this
decay locally after continuation, outside the finitely many poles
in a bounded spectral interval. Polynomial bounds on the spectral
coefficients then give local uniform convergence of the remaining sum.

The other terms in Motohashi's formula are regular at the same nonreal
point. The continuous contour is deformed locally past the moving
poles $w=-3/2\pm it$ within neighborhoods where its arithmetic weight
is analytic. The holomorphic cusp terms also have real $\xi=it$;
in weight $2k$, Euler's integral contains
$[v(1-v)/(x+v)]^{k-1}$, bounded by $(2-\sqrt3)^{k-1}$ for
$0<v<1$, $x\ge1/2$. Retaining a fixed integrable power gives local
uniform exponential decay of the large-$k$ tail. Arithmetic main terms
and the parameter-continuation residues have only real Mellin poles.
Only $t_j=T$ therefore contributes at $w=-3/2+iT$.

Odd forms have $L_j(1/2)=0$. Inserting~\eqref{d6c:shift_kernel_residue}
into~\eqref{d6c:shift_discrete_formula} and applying gamma reflection gives
\begin{align}
 \Res_{w=-3/2+iT}D(w)
 ={}&-\frac{(1+i\sinh\pi T)\Gamma(2iT)
 \Gamma(\tfrac52-iT)\Gamma(\tfrac12-iT)}{8\pi^3} \notag\\
 &\times\sum_{\substack{j:t_j=T\\j\ \mathrm{even}}}
 |\rho_j(1)|^2L_j(\tfrac52)L_j(\tfrac12).
 \label{d6c:shift_residue_uncompleted}
\end{align}
Gamma duplication and the completion~\eqref{d6c:eq:Lhat} convert this
to~\eqref{d6c:shift_residue_Bessel}, with the eigenspace sum
$\mathcal A(T)$ in~\eqref{d6c:eq:AT}.

\subsection{Hecke identities and norm conversion}
\label{d6c:spectral_normalizations}\label{d6c:Hecke_L_identities}
\label{rev:analytic_norm}
The residue above contains the first Fourier coefficient of a unit-norm
cusp form, whereas the Hecke $L$-series has first coefficient one. To
relate the two normalizations, recall that the unit-norm basis obeys
$u_j(-x+iy)=(-1)^{\epsilon_j}u_j(x+iy)$ and
$\rho_j(-n)=(-1)^{\epsilon_j}\rho_j(n)$.
For even forms choose $\rho_j(1)>0$ and set
$v_j=u_j/\rho_j(1)$, $N_j=\langle v_j,v_j\rangle=|\rho_j(1)|^{-2}$.
The completion~\eqref{d6c:eq:Lhat} follows by Mellin transforming
$u_j(iy)$ for even parity and
$w_j(y)=\partial_xu_j(x+iy)|_{x=0}$ for odd parity:
\begin{align}
 \int_0^\infty u_j(iy)y^{w-3/2}\,\dd y
 &=\frac{\rho_j(1)}2\widehat L_j(w),\qquad \epsilon_j=0,
 \notag\\
 \int_0^\infty w_j(y)y^{w-1/2}\,\dd y
 &=i\rho_j(1)\widehat L_j(w),\qquad \epsilon_j=1.
 \label{d6c:Hecke_Mellin_pair}
\end{align}
Cusp decay and inversion,
$u_j(iy)=u_j(i/y)$, $w_j(y)=-y^{-2}w_j(1/y)$,
make these integrals entire and give
$\widehat L_j(w)=(-1)^{\epsilon_j}\widehat L_j(1-w)$~\cite{LewisZagier2001}.

The Hecke product relation and its prime-power generating function are
\begin{align}
 \lambda_j(m)\lambda_j(n)
 &=\sum_{d\mid(m,n)}\lambda_j(mn/d^2), \notag\\
 F_{j,p}(z):=\sum_{\ell\ge0}\lambda_j(p^\ell)z^\ell
 &=\bigl[1-\lambda_j(p)z+z^2\bigr]^{-1}.
 \label{d6c:period_Hecke_product}
\end{align}
The normalized Hecke average
$\mathcal A_n=\sqrt n\,T_n/\sigma_1(n)$ preserves the invariant
integral and has $\mathcal A_n1=1$. Cauchy--Schwarz gives
$\|\mathcal A_n\psi\|_2^2\le\int_{\mathcal F}\mathcal A_n(|\psi|^2)\dd\mu
=\|\psi\|_2^2$, hence
\be
 |\lambda_j(n)|\leq\frac{\sigma_1(n)}{\sqrt n}.
 \label{d6a:elementary_Hecke_bound}
\ee
At $5/2$, each Euler denominator satisfies
$1-\lambda_j(p)p^{-5/2}+p^{-5}\ge(1-p^{-2})(1-p^{-3})>0$.
Absolute convergence therefore gives
\be
 0<L_j(5/2)
 \leq\sum_{n\geq1}\frac{|\lambda_j(n)|}{n^{5/2}}
 \leq\sum_{n\geq1}\frac{\sigma_1(n)}{n^3}
 =\zeta(2)\zeta(3).
 \label{d6a:Lfive_bounds}
\ee
For the divisor-weighted series, the local factor is
$[F_{j,p}(z)-qF_{j,p}(qz)]/(1-q)$ with $q=p^\beta$, interpreted
by continuity at $q=1$.
Combining its two rational terms produces numerator $1-qz^2$,
so multiplication over primes yields
\be
 \sum_{n\ge1}\frac{\lambda_j(n)\sigma_\beta(n)}{n^w}
 =\frac{L_j(w)L_j(w-\beta)}{\zeta(2w-\beta)},
 \label{d6c:period_divisor_identity}
\ee
for $\Re w>3/2$ and $\Re(w-\beta)>3/2$. Indeed,
$|\sigma_\beta(n)|\le d(n)n^{\max(0,\Re\beta)}$ and
\eqref{d6a:elementary_Hecke_bound} gives
$|\lambda_j(n)|\le\sqrt n\,d(n)$, so the divisor-weighted series
converges absolutely in this domain.

Unfolding the square of the Hecke-normalized form $v_j$ relates its squared norm
$N_j$ to the symmetric-square $L$-function. Define
\be
 L_j(s,\operatorname{sym}^2):=
 \prod_p\frac1{(1-p^{-s})
 \left[1-(\lambda_j(p)^2-2)p^{-s}+p^{-2s}\right]},
 \qquad \Re s>2.
 \label{d6c:numerical_sym2_definition}
\ee
The same prime-power recurrence gives
\begin{align}
 \mathcal D_j(s):=\sum_{n\ge1}\lambda_j(n)^2n^{-s}
 &=\frac{\zeta(s)L_j(s,\operatorname{sym}^2)}{\zeta(2s)},
 \notag\\
 L_j(s,\operatorname{sym}^2)
 &=\zeta(2s)\sum_{n\ge1}\lambda_j(n^2)n^{-s},
 \qquad \Re s>2.
 \label{d6a:symmetric_square}
\end{align}
Unfolding $\mathcal Z_j(s)=\int_{\mathcal F}|v_j|^2E_s\,\dd\mu$
for $\Re s>2$ gives
$2(2\pi)^{-s}\mathcal D_j(s)\int_0^\infty X^{s-1}K_{it_j}(X)^2\dd X$;
the factor two counts the two Fourier signs.
The Bessel-product transform then yields
\be
 \mathcal Z_j(s)
 =\frac{\Gamma(s/2)^2\Gamma(s/2+it_j)\Gamma(s/2-it_j)}
 {4\pi^s\Gamma(s)}
 \frac{\zeta(s)L_j(s,\operatorname{sym}^2)}{\zeta(2s)}.
 \label{d6c:numerical_norm_unfolding}
\ee
At $s=1$, use analytic continuation~\cite{BHS},
$\Res E_s=3/\pi$, $\Res\zeta(s)=1$ and
$|\Gamma(1/2+it)|^2=\pi/\cosh(\pi t)$.
Comparing the residues gives~\cite{BHS}
\be
 \boxed{N_j=\frac{L_j(1,\operatorname{sym}^2)}
 {2\cosh(\pi t_j)}.}
 \label{d6a:norm_identity}
\ee
\eqalias{d6c:numerical_sym2_norm}{d6a:norm_identity}

\subsection{Cusp amplitudes from first-mode Mellin residues}
\label{d6c:subtraction_proof}\label{d6c:Mellin_justification}
To relate the arithmetic residue to the sum of cusp amplitudes in an
eigenspace, we first show that the two terms subtracted from $f_1$
are regular at $s=iT$ for
$T>0$ after Mellin transformation. The explicit $S_1$
in~\eqref{d6c:S1_explicit} is a finite sum of
powers times $e^{-2\pi y}$, hence has only real Mellin poles.
For the continuous term, write its weight meromorphically as
\be
 W(t)=\frac{16\pi^2}{t^2+49/4}
 \frac{\Lambda(-3/4+it/2)\Lambda(1/4+it/2)^2
 \Lambda(5/4+it/2)}
 {\Lambda(1/2+it)\Lambda(1/2-it)}.
 \label{d6c:continuous_weight_meromorphic}
\ee
It is analytic near each real $\pm T\ne0$, because
$\zeta(1\pm2iT)\ne0$~\cite{DLMF}, and is $O(t^2)$
at zero. The gamma and zeta estimates give the real-axis
bound~\eqref{d6a:Wbound}. Shifting the Fourier integral for $K_{it}$
to $\mathbb R+i\theta$ gives
\be
 |K_{it}(X)|\le e^{-\theta t}K_0(X\cos\theta),
 \qquad X>0,\quad t\ge0,\quad 0<\theta<\frac\pi2.
 \label{d6c:Mellin_Bessel_decay}
\ee
The integrand is entire and decays at both ends. Taking $\theta=\pi/4$
in~\eqref{d6c:C1_halfline} gives
$|C_1(y)|\le\sqrt yK_0(\sqrt2\pi y)\pi^{-1}
\int_0^\infty W(t)e^{-\pi t/4}\dd t$.
The finite integral implies $C_1(y)=O(e^{-\sqrt2\pi y})$ at infinity
and $O(\sqrt y(1+|\log y|))$ at zero.

The damped kernel $\mathscr K_d$ of~\eqref{intro:damped_Mellin}
follows by inserting the Fourier integral for $K_{it}$, performing
the $y$ integral and setting $r=e^v$: the remaining beta integral is
$\Gamma(s)\Gamma(s+it)\Gamma(s-it)/(2\pi^s\Gamma(2s))$.
Gamma duplication gives the stated kernel. Thus
\be
 \mathcal C_d(s):=\int_0^\infty
 y^{s-3/2}e^{-2\pi y}C_1(y)\,\dd y
 =\frac1{2\pi}\int_{\mathbb R}
 W(t)\mathscr K_d(s,t)\,\dd t,
 \qquad \Re s>0.
 \label{d6c:continuous_damped_Mellin}
\ee
The Bessel bound justifies exchanging the two integrals, and Stirling's
formula gives locally
uniform exponential decay of the spectral tails.
Near $s=iT$, the moving poles $t=-is$ and $t=is$ lie respectively
below $T$ and above $-T$ when $\Re s>0$. Deform the contour upward
near $T$ and downward near $-T$, within disks where $W$ is analytic,
leaving the tails fixed. This crosses no pole in the original
half-plane and gives a holomorphic continuation near $iT$.
The ordinary transform minus $\mathcal C_d(s)$ is
$\int_0^\infty y^{s-3/2}(1-e^{-2\pi y})C_1(y)\dd y$,
holomorphic for $\Re s>-1$ by the endpoint bounds.
Thus neither known subtraction changes the residue at $iT$.

It remains to justify taking the residue inside the cusp sum.
The exponential damping realizes the Mellin transform as an $L^2$
overlap, so we can use orthogonality before the amplitudes are known.
Let $g_{\rm cusp}$ have first coefficient $h_1$, set
$c_j=\langle g,u_j\rangle$ and $d_j=c_j\rho_j(1)$, and define for
$\sigma=\Re w>1/2$
\begin{align}
 P_w(\tau)&=\sum_{\gamma\in\Gamma_\infty\backslash\Gamma}
 [\Im(\gamma\tau)]^{w+1/2}e^{2\pi i\gamma\tau}, \notag\\
 \mathcal A_\sigma(\tau)&=
 \sum_{\gamma\in\Gamma_\infty\backslash\Gamma}
 [\Im(\gamma\tau)]^{\sigma+1/2}
 e^{-2\pi\Im(\gamma\tau)},\qquad \sigma=\Re w.
 \label{d6c:Mellin_Poincare}
\end{align}
Both series converge absolutely, with
$|P_w(\tau)|\le\mathcal A_\sigma(\tau)$ for
$\sigma=\Re w$.
Bounding the nonidentity terms by $E_{\sigma+1/2}$ gives
$\mathcal A_\sigma(x+iy)=y^{\sigma+1/2}e^{-2\pi y}
+O_\sigma(y^{1/2-\sigma})$. The integrand defining its squared norm is
$O_\sigma(y^{-1-2\sigma})$ as $y\to\infty$,
so both series are in $L^2$. Absolute unfolding and Cauchy--Schwarz give
\be
 \int_0^\infty |h_1(y)|y^{\sigma-3/2}e^{-2\pi y}\,\dd y
 \le\langle |g_{\rm cusp}|,\mathcal A_\sigma\rangle
 \le\|g_{\rm cusp}\|_2\|\mathcal A_\sigma\|_2<\infty,
 \qquad \sigma>\tfrac12.
 \label{d6c:Mellin_absolute_bound}
\ee
Unfolding $P_{\bar s}$ then gives
\be
 \mathcal M_{h,d}(s)
 =\langle g_{\rm cusp},P_{\bar s}\rangle
 =\sum_jd_j\mathscr K_d(s,t_j),\qquad \Re s>\frac12.
 \label{d6c:damped_spectral_expansion}
\ee
For real $\sigma_0>1/2$, Bessel's inequality gives
\be
 \sum_j|c_j|^2\le\|g_{\rm cusp}\|_2^2,
 \qquad
 \sum_j|\rho_j(1)\mathscr K_d(\sigma_0,t_j)|^2
 \le\|P_{\sigma_0}\|_2^2.
 \label{d6c:Mellin_Bessel_inequalities}
\ee
For compact $\Omega$ away from kernel poles, choose
$\sigma_0>\max(1/2,\sup_\Omega\Re s)$.
Stirling's formula bounds $\sup_\Omega|\mathscr K_d(s,t)/\mathscr K_d(\sigma_0,t)|$
by $O_{\Omega,\sigma_0}((1+t)^{2(\sup_\Omega\Re s-\sigma_0)})$.
Cauchy--Schwarz therefore bounds the spectral tail uniformly by
\begin{align}
 \sup_{s\in\Omega}\sum_{t_j>R}|d_j\mathscr K_d(s,t_j)|
 &\ll_{\Omega,\sigma_0}
 \left(\sum_{t_j>R}|c_j|^2\right)^{1/2} \notag\\
 &\qquad\times
 \left(\sum_{t_j>R}|\rho_j(1)\mathscr K_d(\sigma_0,t_j)|^2\right)^{1/2}
 \longrightarrow0.
 \label{d6c:Mellin_uniform_tail}
\end{align}
Only finitely many kernels have poles in a bounded region; residues
can thus be taken termwise without estimating the unknown $d_j$ separately.

For $D^6R^4$, $h_1=f_1-S_1-C_1$ has exponential decay at infinity by
\eqref{d6c:f1tail} and the preceding bounds. Its ordinary and damped
transforms differ by
\begin{align}
 H(s)&:=\mathcal M_h(s)-\mathcal M_{h,d}(s) \notag\\
 &=\int_0^\infty y^{s-1}\frac{h_1(y)}{\sqrt y}
 (1-e^{-2\pi y})\,\dd y.
 \label{d6c:eq:weightedtransfer}
\end{align}
At zero, $|1-e^{-2\pi y}|\le2\pi y$ and the absolute unfolding bound
with $\sigma=\Re s+1$ make this integral holomorphic for
$\Re s>-1/2$; on compact subsets a margin in $\Re s$ controls logarithmic
derivatives. Exponential decay controls the integral at infinity for every $s$.
Gamma duplication gives both kernel residues as
$\Res_{s=iT}\mathscr K(s,T)=\Res_{s=iT}\mathscr K_d(s,T)
=\Gamma(iT)/(2\pi^{iT})$.
Together with local uniform convergence, this establishes
the residue formula~\eqref{d6c:eq:aggregate} for the ordinary
Mellin transform.

\subsection{Uniqueness and degenerate eigenspaces}
\label{d6c:complete_first_uniqueness}\label{d6c:degeneracy}

Let $h=\widetilde F-F$ be the difference of two smooth,
even modular-invariant functions, and suppose that
$h\in L^2(\Gamma\backslash\mathbb H)$. Denote its complete
first Fourier mode by $h_1$ and its constant spectral
component by $C_h$. We show that $h_1(y)=0$ for all $y>0$
implies
\begin{align}
 &h_{\rm cont}=0,\qquad h_0(y)=C_h, \notag\\
 &\sum_{j:t_j=T}
 \langle h,u_j\rangle\rho_j(1)=0
 \quad\text{for every }T.
 \label{d6c:complete_first_conclusion}
\end{align}
The proof shows that the vanishing first mode forces its
spectral measure to vanish. The continuous density and
the total weight at each cusp eigenvalue must then vanish
separately, without any assumption of spectral simplicity.

Write $a_j=\langle h,u_j\rangle$ and
$a(t)=\langle h,E_{1/2+it}\rangle$. The Plancherel formula
expresses $\|h\|_2^2$ as the sum of the constant contribution,
the squared cusp coefficients, and the integral of $|a(t)|^2$
with the continuous spectral measure. Since $h\in L^2$,
it gives $\sum_j|a_j|^2<\infty$ and $a\in L^2(0,\infty)$.
Using the first Eisenstein Fourier coefficient and folding
the continuous integral onto $t>0$, we obtain the locally
finite measure
\be
 \dd\omega_h(t)=
 \sum_{j\ {\rm even}}a_j\rho_j(1)\delta_{t_j}(\dd t)
 +\frac{a(t)}{\pi\Lambda(1/2+it)}\,\dd t.
 \label{d6c:complete_first_measure}
\ee
Here $\delta_{t_j}$ denotes the unit point mass at $t_j$.

The absolute-convergence argument based on the Poincar\'e
series also applies to $h$. Since $h_1=0$, unfolding $P_s$
and using the spectral decomposition give, for real $s>1$,
\be
 0=\int_0^\infty h_1(y)y^{s-3/2}e^{-2\pi y}\,\dd y
 =\langle h,P_s\rangle
 =\int_0^\infty\mathscr K_d(s,t)\,\dd\omega_h(t).
 \label{d6c:complete_first_damped_identity}
\ee
The constant component has zero overlap with $P_s$.
At half-integer values of $s$, these identities give
vanishing polynomial moments. To conclude that the measure
itself vanishes, we first establish an exponential moment
bound.

Restricting the integral defining $\|u_j\|_2^2=1$ to
$-1/2\le x\le1/2$, $1\le y\le2$, and retaining its first
Fourier mode gives
\be
 1\ge|\rho_j(1)|^2\int_1^2K_{it_j}(2\pi y)^2\,\frac{\dd y}{y}.
 \label{d6c:complete_first_strip_bound}
\ee
The power series for $I_{\pm it}$ and their relation to
$K_{it}$ give, uniformly for $1\le y\le2$ as
$t\to\infty$~\cite{DLMF},
\be
 K_{it}(2\pi y)=
 -\sqrt{\frac{\pi}{t\sinh(\pi t)}}
 \left\{\sin\!\left[t\log(\pi y)-\arg\Gamma(1+it)\right]
 +O(t^{-1})\right\}.
 \label{d6c:complete_first_order_bound}
\ee
Integrating the squared sine with measure $\dd y/y$ gives
$\log2/2+O(t^{-1})$, and hence
$|\rho_j(1)|\ll(1+t_j)^{1/2}e^{\pi t_j/2}$.
For the continuous contribution,
$|\Lambda(1/2+it)|^{-1}
=\sqrt{\cosh(\pi t)}/|\zeta(1+2it)|
\ll e^{\pi t/2}\log(2+t)$ for
$t\ge1$~\cite{Yang}, while this reciprocal is
$O(t)$ at zero. These bounds give an exponential moment
estimate for the weighted measure:
\begin{align}
 \dd\nu_h(t)&:=\frac{\pi(t^2+1/4)}{\cosh(\pi t)}\,\dd\omega_h(t),
 \notag\\
 \int_0^\infty e^{\eta t}|\dd\nu_h(t)|&<\infty,
 \qquad 0<\eta<\frac\pi2.
 \label{d6c:complete_first_exponential_moment}
\end{align}
For the cusp contribution, Cauchy--Schwarz and
$\sum_j|a_j|^2<\infty$ reduce this estimate to the
summability of
$(1+t_j)^5e^{-(\pi-2\eta)t_j}$.
The Weyl bound states that the number of cusp forms with
$t_j\le T$, counted with multiplicity, is
$O((1+T)^2)$~\cite{IwaniecSpectral}. This polynomial growth
is dominated by the exponential decay in the summand,
so the series converges.
For the continuous contribution, the same argument uses
$a\in L^2(0,\infty)$ and the integrability of the
corresponding exponentially decaying weight.

Now set $s=n+3/2$, $n=0,1,\ldots$, in
\eqref{d6c:complete_first_damped_identity}.
Gamma recurrence and reflection give
\begin{align}
 \Gamma(n+3/2+it)\Gamma(n+3/2-it)
 &=\frac{\pi(t^2+1/4)}{\cosh(\pi t)}p_n(t^2), \notag\\
 p_n(x)&:=\prod_{r=1}^n\left[x+(r+1/2)^2\right],
 \qquad p_0(x)=1.
 \label{d6c:complete_first_moment_polynomials}
\end{align}
Dividing out the nonzero factor in $\mathscr K_d$ that
depends only on $s$, we obtain
$\int_0^\infty p_n(t^2)\,\dd\nu_h(t)=0$.
Since the $p_n$ are monic polynomials of successive degrees,
all even moments of $\nu_h$ vanish.

The exponential bound makes
$\Phi(z)=\int_0^\infty\cos(zt)\,\dd\nu_h(t)$ holomorphic
in $|\Im z|<\eta$. Its even derivatives at zero vanish
by the moment identities, and its odd derivatives vanish
because $\Phi$ is even. Thus $\Phi=0$.
Uniqueness of the Fourier transform, applied to the even
extension of this finite measure, gives $\nu_h=0$.
The strictly positive factor relating $\nu_h$ to $\omega_h$
then gives $\omega_h=0$. Its discrete and continuous parts
vanish separately:
$a(t)=0$ almost everywhere and
$\sum_{j:t_j=T}a_j\rho_j(1)=0$ for every $T$.
Only the constant component and possible cusp combinations
within repeated eigenspaces remain. Since cusp forms have
vanishing zero mode, $h_0=C_h$ holds as a distribution and,
by smoothness, for every $y>0$.

If the even cusp spectrum is simple, the nonvanishing of
$\rho_j(1)$ implies $h=C_h$. This constant vanishes if the
two candidates also share the perturbative zero mode.
Without simplicity, the complete low modes determine the
eigenspace sums $\mathcal D_f(T)=\sum_{j:t_j=T}d_j$.
Equation~\eqref{d6c:canonical} gives one assignment of
individual amplitudes consistent with these sums.
Within an even eigenspace with repeated spectral parameter
$T$, an addition preserving both low modes takes the form
\begin{align}
 \delta f_T(\tau)&=\sum_{j:t_j=T}\frac{b_j}{\rho_j(1)}u_j(\tau),
 \qquad \sum_{j:t_j=T}b_j=0, \notag\\
 (\delta f_T)_0(y)&=(\delta f_T)_1(y)=0, \notag\\
 (\delta f_T)_k(y)&=\sqrt y K_{iT}(2\pi ky)
 \sum_{j:t_j=T}b_j\lambda_j(k),
 \qquad k\ge1.
 \label{d6c:degenerate_perturbation}
\end{align}
This finite sum is smooth, even, modular and exponentially
decreasing at the cusp. Two distinct forms of the same
parity in this eigenspace must have different Hecke
eigenvalues for some $k\ge2$; otherwise their Hecke-normalized
Fourier expansions would coincide. The second Fourier mode
distinguishes them only if their $\lambda_j(2)$ differ.

If evenness is not imposed but the function remains real,
the coefficients must satisfy
$\overline{b_j}=(-1)^{\epsilon_j}b_j$, since
$\overline{u_j/\rho_j(1)}
=(-1)^{\epsilon_j}u_j/\rho_j(1)$.
Thus $b_j$ is real for even forms and purely imaginary
for odd forms. Preserving both Fourier modes $k=\pm1$
requires the sums of $b_j$ to vanish separately in the
two parity sectors. An eigenspace of dimension $r$ within
either parity therefore leaves $r-1$ real parameters.
For real functions, simplicity within each allowed parity
sector suffices. For a general complex function with only
the zero and positive first Fourier modes specified, the
corresponding assumption is simplicity of the full allowed
cusp spectrum. Diagonalizing the Hecke operators does not
imply this spectral simplicity.

\section{The spectral moment and instanton asymptotics}
\label{rev:constant}\label{d6a:finitepart}

In this appendix, we analytically continue the spectral moment
and evaluate the square-root and constant coefficients $A_k$
and $B_k$ in Section~\ref{d6a:section}. The pole at $z=1$
comes from the diagonal term of the Kuznetsov formula, and
its residue fixes $A_k$. The value $Z_k(0)$, together with
the restored Eisenstein terms, gives $B_k$.

The continuation proceeds in two stages. We first continue
the arguments of the two $L$-functions to $1/2$ and $5/2$,
keeping $\Re z$ sufficiently large for the spectral sums
to converge absolutely. We then continue the moment in $z$.

\subsection{The Kuznetsov formula and parameter continuation}
\label{d6a:kuznetsov}\label{d6a:crossing}
We use the Fourier convention~\eqref{intro:cusps} and initially
include both cusp parities. For positive index $n$ the Eisenstein
coefficient is
$\tau_t(n)=2\chi_n(t)/\Lambda(1/2+it)$, where
$\chi_n(t)=n^{it}\sigma_{-2it}(n)$. Let $\eta>0$ be sufficiently
small. For an even test function $h$,
holomorphic near $|\Im t|\leq1/2+\eta$ and satisfying
\be
 h(\pm i/2)=0,\qquad
 h(t)\ll_h(1+|t|)^{-2-\eta}
 \label{d6a:kuz_test_conditions}
\ee
uniformly there, the level-one Kuznetsov formula
reads~\cite{HLN}
\begin{align}
 &\sum_j\frac{h(t_j)}{\cosh(\pi t_j)}
 \overline{\rho_j(m)}\rho_j(n)
 +\frac1{4\pi}\int_{\mathbb R}
 \frac{h(t)\overline{\tau_t(m)}\tau_t(n)}{\cosh(\pi t)}\,\dd t
 \notag\\
 &\qquad=\delta_{mn}H_0(h)
 +\sum_{c\geq1}\frac{S(m,n;c)}c
 H^+\!\left(\frac{4\pi\sqrt{mn}}c;h\right),
 \label{d6a:kuzformula}
\end{align}
where the integral transforms are
\begin{align}
 H_0(h)&=\frac1{\pi^2}\int_{\mathbb R}
 h(t)t\tanh(\pi t)\,\dd t,
 \label{d6a:kuz_diagonal_transform}\\
 H^+(x;h)&=\frac{2i}{\pi}\int_{\mathbb R}
 J_{2it}(x)\frac{h(t)t}{\cosh(\pi t)}\,\dd t.
 \label{d6a:kuz_Bessel_transform}
\end{align}
\eqalias{d6a:main_Kuznetsov_transforms}{d6a:kuz_Bessel_transform}
Here $m,n>0$ and
\be
 S(m,n;c)=\sum_{\substack{d\bmod c\\(d,c)=1}}
 e^{2\pi i(md+n\bar d)/c},\qquad d\bar d\equiv1\pmod c,
 \qquad S(m,n;1)=1.
 \label{d6a:Kloosterman_definition}
\ee

The auxiliary Eisenstein family used as a summation weight is
$\phi_\nu=E^*_{1/2+\nu}/2$. Its coefficients are
\begin{align}
 C_{\phi_\nu}(n)&=n^\nu\sigma_{-2\nu}(n), \notag\\
 (\phi_\nu)_0&=c_\nu^+y^{1/2+\nu}+c_\nu^-y^{1/2-\nu},
 \qquad c_\nu^\pm=\tfrac12\Lambda(\tfrac12\pm\nu).
 \label{d6a:auxiliary_Eisenstein_coefficients}
\end{align}
This is the normalization of~\cite{HLN}.
The Hecke divisor identity~\eqref{d6c:period_divisor_identity} gives
\be
 \zeta(2s)\sum_{m\geq1}
 \frac{C_{\phi_\nu}(m)\rho_j(m)}{m^s}
 =\rho_j(1)L_j(s-\nu)L_j(s+\nu).
 \label{d6a:HLNmatch}
\ee
The series converges for $\Re s>3/2+|\Re\nu|$ by
\eqref{d6a:elementary_Hecke_bound}; elsewhere this is a
meromorphic identity. Similarly,
$\sum_{m\geq1}\chi_m(t)m^{-w}=\zeta(w-it)\zeta(w+it)$
for $\Re w>1$. Together with
$|\Lambda(1/2+it)|^2=|\zeta(1+2it)|^2/\cosh\pi t$, this
identifies the result of summing~\eqref{d6a:kuzformula} at $n=k$
against $\zeta(2s)C_{\phi_\nu}(m)m^{-s}$ as
\begin{align}
 \mathscr K_k(s,\nu;h)
 ={}&\sum_j\frac{h(t_j)|\rho_j(1)|^2}{\cosh\pi t_j}
 \lambda_j(k)L_j(s-\nu)L_j(s+\nu)\notag\\
 &+\frac1\pi\int_{\mathbb R}h(t)\chi_k(t)
 \frac{\prod_{\pm}\zeta(s-\nu\pm it)\zeta(s+\nu\pm it)}
 {\zeta(1+2it)\zeta(1-2it)}\,\dd t.
 \label{d6a:HLN_spectral_moment}
\end{align}
The continuous normalization follows from
$\overline{\tau_t(m)}\tau_t(k)/\cosh\pi t
=4\chi_m(t)\chi_k(t)/|\zeta(1+2it)|^2$.
On the arithmetic side the same summation gives
\begin{align}
 \mathscr K_k(s,\nu;h)
 ={}&\zeta(2s)\frac{C_{\phi_\nu}(k)}{k^s}H_0(h)\notag\\
 &+\zeta(2s)\sum_{c,m\geq1}\frac{C_{\phi_\nu}(m)S(m,k;c)}{cm^s}
 H^+\!\left(\frac{4\pi\sqrt{mk}}c;h\right).
 \label{d6a:HLN_opened_moment}
\end{align}
These identities first hold with sufficient Dirichlet and spectral
decay. HLN's first-moment formula expresses the double sum over
$c$ and $m$ containing the Kloosterman sums $S(m,k;c)$
in~\eqref{d6a:HLN_opened_moment} as additional main terms
and contour integrals involving shifted divisor sums~\cite{HLN}.

The product $L_j(1/2)L_j(5/2)$ corresponds to the endpoint
$(s,\nu)=(3/2,1)$. HLN's Eisenstein specialization starts
at $\nu\in i\mathbb R$, and their first-moment
formula~\cite{HLN} is stated for
$1/2\leq\Re s<3/2$. Reaching the required endpoint
therefore requires analytic continuation.
We begin with $\nu=i\eta$, where $0<|\eta|\ll1$, and
continue in $s$ from $1<\Re s<3/2$ to $s=\nu+1/2$.
We then vary $\nu$ along this relation until $\nu=1$,
keeping $\Re z$ sufficiently large for absolute convergence
of the spectral sums throughout. For odd cusp forms,
the functional equation gives $L_j(1/2)=0$, so only
even cusp forms contribute at the endpoint.

The two test functions differ by the factor $q_2$ that enforces
the zeros at $t=\pm i/2$ required by the Kuznetsov formula. We write
\be
 h_z^0(t)=\frac{r(t)}2(4+t^2)^{-z/2},\qquad
 h_z(t)=q_2(t)h_z^0(t).
 \label{d6a:qrepair}
\ee
The first gives $Z_k$ and the second $G_k$.
Both are analytic for $|\Im t|<2$, with the principal logarithm:
$\Re(4+t^2)>0$ there, and the apparent gamma poles of $r$ at
$\pm i/2$ cancel. Expansion of the gamma factor at zero gives
$r(i/2)=4\pi/3$, so $h_z^0(i/2)=(2\pi/3)(15/4)^{-z/2}$.
The factor $q_2$ supplies the required zeros, while uniform
Stirling estimates give
\be
 h_z(t)=O_{z,\beta}((1+|\Re t|)^{-1-\Re z}),
 \qquad |\Im t|\leq\beta<2.
 \label{d6a:corrected_test_decay}
\ee
Thus $h_z$ is admissible for $\Re z>1+\eta$; we initially
choose $\Re z>c_*$ large enough for all required exchanges of sums and integrals.
The independent polynomial bound below justifies this choice.

At the endpoint, the norm identity~\eqref{d6a:norm_identity}
and evenness give respectively
\begin{align}
 \frac{h_z^0(t_j)|\rho_j(1)|^2}{\cosh\pi t_j}
 \lambda_j(k)L_j(1/2)L_j(5/2)
 &=r(t_j)b_j(k)(4+t_j^2)^{-z/2},\notag\\
 \frac1\pi\int_{\mathbb R}h_z^0(t)P(t)\chi_k(t)\,\dd t
 &=\frac1\pi\int_0^\infty r(t)P(t)\chi_k(t)(4+t^2)^{-z/2}\,\dd t.
 \label{d6a:HLN_continuous_endpoint_match}
\end{align}
The main-text convention~\eqref{d6a:main_HLN_intro} divides
Kuznetsov by two and uses $h=2h_z$, giving the same $G_k$.
At $(s,\nu)=(3/2,1)$, the continued HLN formula with
the corrected test $h_z$ gives
\begin{align}
 G_k(z)={}&M_{{\rm diag},k,q}(z)+M_{{\rm ref},k,q}(z)
 +M_{+,k,q}(z)+M_{-,k,q}(z)\notag\\
 &+L_{{\rm short},k,q}(z)+L_{\infty,k,q}(z).
 \label{d6a:complete_corrected_identity}
\end{align}
Here $M_{\rm diag}$ is the Kuznetsov diagonal contribution,
and $M_{\rm ref}$ is the reflected main term arising from
the functional equation in HLN's derivation. The terms
$M_\pm$ come from the two terms in the zero mode of the
auxiliary Eisenstein series $\phi_\nu$.
The term $L_{\rm short}$ contains the finite shifted sum
over $1\leq m<k$, while $L_\infty$ combines the two
infinite shifted sums.

In HLN's notation,
$M=M_{\rm diag}+M_{\rm ref}$, $M^\pm=M_\pm$,
$L_{\rm short}=L_1^-$ and $L_\infty=L_1^++L_2$.
The subscript $q$ denotes evaluation with the corrected
test $h_z=q_2h_z^0$; below, $b$ denotes evaluation with
the original test function $h_z^0$.

For arithmetic contours oriented upwards, set
\be
 I_c[J]=\frac1{2\pi i}\int_{c-i\infty}^{c+i\infty}J(u)\,\dd u.
 \label{d6a:short_contour_convention}
\ee
We use $c=7/4$. The common test factor is
\be
 \frac{h_z(u/i)}{\cos\pi u}
 =\frac2\pi\frac{(1/4-u^2)^2}{(4-u^2)(49/4-u^2)}
 (4-u^2)^{-z/2}
 \Gamma(1/4+u/2)^2\Gamma(1/4-u/2)^2.
 \label{d6a:hcomplex}
\ee
Gamma recurrence and the residue at zero give the removable values
\be
 \lim_{u\to\pm1/2}\frac{h_z(u/i)}{\cos\pi u}
 =\frac8{45}(15/4)^{-z/2},\qquad
 \lim_{u\to\pm3/2}\frac{h_z(u/i)}{\cos\pi u}
 =\frac{64}{35}(7/4)^{-z/2}.
 \label{d6a:hcomplex_limits}
\ee

The continuation in $s$ and $\nu$ moves zeta poles through the
continuous spectral contour. To compute their residues, put
$a=s-\nu$, $b=s+\nu$ and let $C_{\rm st}$ denote the integral
along the real $t$ axis in~\eqref{d6a:HLN_spectral_moment}.
With $w=it$,
\begin{align}
 C_{\rm st}&=\frac1{\pi i}\int_{-i\infty}^{i\infty}
 \mathcal F_k(w)\,\dd w, \notag\\
 \mathcal F_k(w)&=h(w/i)\chi_k(w/i)
 \frac{\zeta(a+w)\zeta(a-w)\zeta(b+w)\zeta(b-w)}
 {\zeta(1+2w)\zeta(1-2w)}.
 \label{d6a:crossing_w_integrand}
\end{align}
Equivalently, hold $b=5/2$ and move $a$ across one. The poles
$w=1-a$ and $w=a-1$ exchange sides. Deform the upward contour
locally, avoiding fixed poles and denominator zeros. At
$w_0=1-a>0$, cancellation of $\zeta(2a-1)$ gives
\be
 \Res_{w=w_0}\mathcal F_k
 =h(iw_0)\chi_k(iw_0)
 \frac{\zeta(a+b-1)\zeta(b+1-a)}{\zeta(3-2a)}.
 \label{d6a:crossing_pole_residue}
\ee
Evenness makes the other residue its negative. The bulges to
the right of $w_0$ and left of $-w_0$ give $2\pi i$ times their
difference; including $1/(\pi i)$ therefore yields
\begin{align}
 C_{\rm AC}(a,b;h;k)&=C_{\rm st}(a,b;h;k)+\mathcal R_k(a,b;h),\notag\\
 \mathcal R_k(a,b;h)&=4h(i(1-a))\chi_k(i(1-a))
 \frac{\zeta(a+b-1)\zeta(b+1-a)}{\zeta(3-2a)}.
 \label{d6a:crossformula}
\end{align}
At collision points, take limits of the combined integrand.
The factor $\chi_k(i(1-a))$ follows from the residue; it is
absent from the displayed residue terms in HLN Eqs.~(6.5)--(6.6),
but is required at general $k$.

On the original parameter path, the pair of poles associated with $b$
crosses and returns to its original sides, while the pair associated
with $a$ remains exchanged.
Thus it gives the same endpoint correction. Local deformations
avoid the fixed poles at $w=\pm3/2$; at the endpoint these are
simple after a test zero cancels one order. At $(a,b)=(1/2,5/2)$,
$\chi_k(i/2)=\sigma_1(k)/\sqrt k$ and the zeta quotient is $\zeta(3)$.
Consequently the corrected residue vanishes, whereas
\be
 \mathcal R_k^0(z)=\frac{8\pi}{3}\zeta(3)
 \frac{\sigma_1(k)}{\sqrt k}(15/4)^{-z/2}.
 \label{d6a:barecross}
\ee
The continued expression for the original test function is $Z_k+\mathcal R_k^0$.
Subtracting this entire correction contributes
\be
 -\frac{\mathcal R_k^0(0)}{2\sqrt k}
 =-\frac{4\pi}{3}\zeta(3)\sigma_{-1}(k),
 \label{d6a:crossing_E1_cancel}
\ee
which cancels the restored $\widehat E_1$ constant in~\eqref{d6a:restored}.

\subsection{Analytic continuation of the shifted sums}
\label{d6a:short}\label{d6a:infinite}
The shifted terms on the arithmetic side split according to the
summation index. For the finite sum over $1\leq m<k$ we must pass
the parameter limit through the contour integral. For the infinite
sums we must also continue the sum over charges. In our conventions
the finite term is~\cite{HLN}
\begin{align}
 L_{{\rm short},k}(s,\nu;h)=-4(2\pi)^{2s-1}I_{7/4}\bigg[
 &\frac{h(u/i)u}{\cos\pi u}\sum_{m=1}^{k-1}
 C_{\phi_\nu}(m)\sigma_{1-2s}(k-m)\notag\\
 &\times m^{s-1}F_1(s,u,\nu;m/k)\bigg].
 \label{d6a:short_general}
\end{align}
To remove its apparent singularity at $\nu=1$, use the regularized
function ${}_2\mathbf F_1={}_2F_1/\Gamma(\gamma)$ and set
\begin{align}
 A(\nu;s,u,x)={}&\frac{\Gamma(1-s-\nu+u)}{\Gamma(s+\nu+u)}
 x^{1-s-\nu}\notag\\
 &\times{}_2\mathbf F_1(1-s-\nu+u,1-s-\nu-u;1-2\nu;x),\notag\\
 F_1(s,u,\nu;x)&=\frac{A(-\nu;s,u,x)-A(\nu;s,u,x)}{2\sin\pi\nu}.
 \label{d6a:short_combined_kernel}
\end{align}
This follows from~\cite{HLN} by gamma duplication
and reflection. In the regularized series at $\gamma=-1$, the
first two coefficients vanish; shifting the index by two gives
\be
 {}_2\mathbf F_1(\alpha,\beta;-1;x)
 =(\alpha)_2(\beta)_2x^2
 {}_2\mathbf F_1(\alpha+2,\beta+2;3;x).
 \label{d6a:short_series_shift}
\ee
With $\alpha=u-s$, $\beta=-u-s$, gamma recurrence then gives
$A(1;s,u,x)=A(-1;s,u,x)$. This holds for nearby $s$, so cancellation
of $\nu-1$ makes the endpoint limit independent of the path.
Equivalently, it is
$[\partial_\nu A(-1;3/2,u,x)+\partial_\nu A(1;3/2,u,x)]/(2\pi)$,
where the derivatives act on the first argument.

For $0<x<1$, put $w=1-2x$, $\ell=u-1/2$ and
$p_\nu(x)=\sqrt x[(1-x)/x]^\nu$.
Euler's transformation of the hypergeometric function,
followed by the Ferrers connection
formula~\cite{DLMF}, gives
\begin{align}
 F_1(\nu+1/2,u,\nu;x)
 &=p_\nu(x)\left[
 \sin\pi\nu\,\mathsf P_{u-1/2}^{-2\nu}(w)
 -\frac2\pi\cos\pi\nu\,\mathsf Q_{u-1/2}^{-2\nu}(w)\right],
 \label{d6a:finite_Ferrers_path}\\
 F_1(3/2,u,1;x)&=\frac2\pi\frac{1-x}{\sqrt x}
 \mathsf Q_{u-1/2}^{-2}(1-2x).
 \label{d6a:finite_Ferrers_endpoint}
\end{align}
Here $\mathsf P_\ell^\mu(w)$ and $\mathsf Q_\ell^\mu(w)$
denote the Ferrers functions of the first and second kind,
respectively: the associated Legendre functions on
$-1<w<1$, with degree $\ell$ and order $\mu$, normalized
as in~\cite{DLMF}. Their relevant integral
and reflection identities are
\begin{align}
 \mathsf P_{u-1/2}^{-2}(\cos\theta)
 &=\frac{\sqrt{2/\pi}}{\Gamma(5/2)\sin^2\theta}
 \int_0^\theta(\cos v-\cos\theta)^{3/2}\cos(uv)\,\dd v,
 \quad 0<\theta<\pi,
 \label{d6a:finite_Mehler_integral}\\
 \frac2\pi\sin\pi\ell\,\mathsf Q_\ell^{-2}(w)
 &=\cos\pi\ell\,\mathsf P_\ell^{-2}(w)-\mathsf P_\ell^{-2}(-w).
 \label{d6a:finite_Ferrers_reflection}
\end{align}
For $w=\cos\theta_x$, the integral bounds $\mathsf P$ by
$O_x(e^{\theta_x|v|})$ on $u=7/4+iv$; the reflection formula,
applied at $\theta_x$ and $\pi-\theta_x$, gives
\be
 |F_1(3/2,7/4+iv,1;x)|\ll_x e^{\theta_x|v|},\qquad \theta_x<\pi.
 \label{d6a:finite_general_charge_bound}
\ee
The estimate is uniform near $\nu=1$. The general form
of the integral representation~\eqref{d6a:finite_Mehler_integral}
is valid for order $-2\nu$ when $\Re\nu>-1/4$.
The regularized expression removes apparent singularities,
and the genuine moving poles remain to the left of
$\Re u=7/4$.
The test contributes $e^{-\pi|v|}$ times a power, dominating
this bound for each of the finitely many $x=m/k$. Thus the
endpoint limit can be taken inside the contour integral, and the finite term
is entire in $z$ for either test. At $k=1$ the finite sum is empty.

For the infinite terms, define
\begin{align}
 A_m(k)&=\sigma_{-2}(m)\sigma_{-2}(m+k), \notag\\
 B(u)&=\frac{\Gamma(u-3/2)\Gamma(u+1/2)}{\Gamma(1+2u)}, \notag\\
 G(u)&=\frac{\Gamma(1/4+u/2)}{\Gamma(3/4+u/2)}, \notag\\
 P_0(u)&=\frac{1/4-u^2}{49/4-u^2}.
 \label{d6a:infinite_factors}
\end{align}
The corresponding kernel and shifted sum are
\begin{align}
 \mathscr F_3(u;x)&=B(u)x^u\,{}_2F_1
 \!\left(u-\tfrac32,u+\tfrac12;1+2u;-x\right), \notag\\
 \mathscr S_k(u)&=\sum_{m\geq1}m^{3/2}A_m(k)
 \mathscr F_3(u;k/m).
 \label{d6a:infiniteobjects}
\end{align}
This series initially converges for $\Re u>5/2$; $x^u$ uses
the real logarithm for $x>0$. Pfaff's transformation of the
$x>1$ kernel~\cite{HLN} gives
\be
 F_1(3/2,u,1;(m+k)/k)
 =-\frac{\sin\pi u}{\pi}\left(\frac m{m+k}\right)^{3/2}
 \mathscr F_3(u;k/m).
 \label{d6a:infinite_Pfaff}
\ee
Thus HLN Eqs.~(4.35) and~(4.44) combine as
$L_1^++L_2=16\pi I_{7/4}[h_z(u/i)u(\sin\pi u-1)
\mathscr S_k(u)/\cos\pi u]$, with the contour continuation
justified below. Gamma reflection gives
\be
 (\sin\pi u-1)\Gamma(1/4-u/2)^2
 =-\frac{2\pi^2}{\Gamma(3/4+u/2)^2},
 \label{d6a:infinite_reflection}
\ee
hence the expression with the original test function is
\be
 L_{\infty,b}(z)=-64\pi^2 I_{7/4}
 [uP_0(u)G(u)^2\mathscr S_k(u)(4-u^2)^{-z/2}].
 \label{d6a:Linf}
\ee
For the corrected test, the integrand is multiplied by $q(u)=(1/4-u^2)/(4-u^2)$.
On this line $\Re(4-u^2)=15/16+v^2>0$.

To continue the sum over $m$, we count the positive integers
$m\le T$ satisfying $d\mid m$ and $e\mid m+k$.
When $(d,e)\mid k$, these integers form a single residue
class modulo $[d,e]$, so their number differs from
$T/[d,e]$ by at most one. Otherwise, there are none.
Summing with weights $d^{-2}e^{-2}$ and using
$\sum_{d,e\ge1}d^{-2}e^{-2}=\zeta(2)^2$, we obtain
\be
 \sum_{m\leq T}A_m(k)=\mathcal C_kT+R_k^{\rm ar}(T),
 \qquad |R_k^{\rm ar}(T)|\leq\zeta(2)^2,
 \label{d6a:shift_count_mean}
\ee
where the superscript ${\rm ar}$ distinguishes this arithmetic
remainder from the spectral weight $R_k(T)$, and
\begin{align}
 \mathcal C_k
 &=\sum_{\substack{d,e\geq1\\(d,e)\mid k}}
 \frac1{d^2e^2[d,e]} \notag\\
 &=\sigma_{-5}(k)\sum_{(a,b)=1}a^{-3}b^{-3}
 =\frac{\zeta(3)^2\sigma_{-5}(k)}{\zeta(6)}.
 \label{d6a:shiftmean}
\end{align}
In the second equality of~\eqref{d6a:shiftmean}, set
$d=ga$, $e=gb$, where $g\mid k$; the coprime Euler
factor is $1+2\sum_{\ell\geq1}p^{-3\ell}$.
Partial summation then continues
\be
 \mathscr D_k(w):=\sum_{m\geq1}A_m(k)m^{-w}
 =\frac{\mathcal C_kw}{w-1}
 +w\int_1^\infty R_k^{\rm ar}(T)T^{-w-1}\,\dd T
 \label{d6a:shiftDirichlet}
\ee
to $\Re w>0$, with its sole pole at one and
$\mathscr D_k(\sigma+iv)=O_{k,\sigma}(1+|v|)$ away from it.
Subtracting the leading small-$x$ term of the kernel gives
\begin{align}
 \mathscr S_k(u)={}&B(u)k^u\mathscr D_k(u-3/2)\notag\\
 &+\sum_{m\geq1}m^{3/2}A_m(k)
 [\mathscr F_3(u;k/m)-B(u)(k/m)^u].
 \label{d6a:Scontinue}
\end{align}
The remainder is locally uniformly convergent for $\Re u>3/2$,
since its summands are $O_k(m^{1/2-\Re u})$.
This continues the original series, with the sole pole
\be
 \Res_{u=5/2}\mathscr S_k(u)=\frac{k^{5/2}\mathcal C_k}{60},
 \qquad B(5/2)=\frac1{60}.
 \label{d6a:arithmeticpole}
\ee

To identify this endpoint formula with the continuation of the HLN
moment, we now carry out the same mean subtraction with $\nu$
still variable. Along $s=\nu+1/2$ set
\begin{align}
 A_{\nu,m}(k)&=\sigma_{-2\nu}(m)\sigma_{-2\nu}(m+k),\notag\\
 B_\nu(u)&=\frac{\Gamma(u+1/2-2\nu)\Gamma(u+1/2)}{\Gamma(1+2u)},\notag\\
 \mathscr F^{(\nu)}(u;x)&=B_\nu(u)x^u
 {}_2F_1(u+1/2-2\nu,u+1/2;1+2u;-x),\notag\\
 \mathscr S_{\nu,k}(u)&=\sum_{m\geq1}m^{2\nu-1/2}A_{\nu,m}(k)
 \mathscr F^{(\nu)}(u;k/m).
 \label{d6a:infinite_nu_family}
\end{align}
Applying Pfaff's transformation to the hypergeometric
kernel and combining the terms $L_1^+$ and $L_2$ gives
\begin{align}
 L_1^++L_2=I_{7/4}\bigg[
 &\frac{4(2\pi)^{2\nu}}\pi\frac{h_z(u/i)u}{\cos\pi u}\notag\\
 &\times\{\cos\pi\nu-\cos[\pi(\nu+1/2-u)]\}\mathscr S_{\nu,k}(u)\bigg].
 \label{d6a:genericcombined}
\end{align}
For $0\leq\Re\nu<5/8$ the series converges on this contour,
using $|\sigma_{-2\nu}(n)|\ll_\epsilon n^\epsilon$ at the boundary.
For $\Re\nu>1/2$, the congruence count instead gives
\begin{align}
 \sum_{m\leq T}A_{\nu,m}(k)&=\mathcal C_{\nu,k}T+R_{\nu,k}^{\rm ar}(T),\notag\\
 \mathcal C_{\nu,k}&=\frac{\zeta(2\nu+1)^2\sigma_{-4\nu-1}(k)}{\zeta(4\nu+2)},
 \qquad |R_{\nu,k}^{\rm ar}(T)|\leq\zeta(2\Re\nu)^2.
 \label{d6a:infinite_nu_mean}
\end{align}
Define $\mathscr D_{\nu,k}$ by partial summation as in
\eqref{d6a:shiftDirichlet}. The continued expression is
\begin{align}
 \mathscr S_{\nu,k}(u)={}&B_\nu(u)k^u\mathscr D_{\nu,k}(u-2\nu+1/2)\notag\\
 &+\sum_{m\geq1}m^{2\nu-1/2}A_{\nu,m}(k)
 [\mathscr F^{(\nu)}(u;k/m)-B_\nu(u)(k/m)^u].
 \label{d6a:infinite_nu_subtraction}
\end{align}
Its remainder converges locally uniformly for
$\Re u>2\Re\nu-1/2$. On $\Re u=7/4$ this gives
$1/2<\Re\nu<9/8$, overlapping the original domain and reaching one.
The moving mean pole at $u=2\nu+1/2$ cancels because
\be
 \cos\pi\nu-\cos[\pi(\nu+1/2-u)]
 =-2\sin[\pi(\nu+1/4-u/2)]\sin[\pi(u/2-1/4)]
 \label{d6a:infinite_moving_zero}
\ee
vanishes there. Its crossing of $7/4$ is away from fixed test
poles and contributes no residue. In the later rightward shift we
first set $z=0$ and restore the original test function. For generic $z$ the regulator
$(4-u^2)^{-z/2}$ has a branch point at $u=2$. At $\nu=1$ the
trigonometric factor has a double zero at $u=5/2$, which cancels the
double pole of $h_0^0(u/i)/\cos\pi u$. Their product is finite and
nonzero there, so the pole of $\mathscr S_k$ contributes to this shift.

For uniform parameter control, Euler's integral is
\be
 \mathscr F^{(\nu)}(u;x)
 =\frac{\Gamma(u+1/2-2\nu)}{\Gamma(u+1/2)}
 \int_0^1\left[\frac{xt(1-t)}{1+xt}\right]^u
 \frac{(1+xt)^{2\nu-1/2}}{\sqrt{t(1-t)}}\,\dd t.
 \label{d6a:infinite_generic_Euler}
\ee
Let $K$ be a compact subset of the strip
$\{\nu\in\mathbb C:1/2<\Re\nu<9/8\}$, and define
$\alpha_K=\min_{\nu\in K}\Re\nu$.
On the contour $u=7/4+iv$, write $V=1+|v|$.
The gamma poles lie to the left of the line. Euler's integral and
Stirling's formula bound $x^{-u}\mathscr F^{(\nu)}$ and its $x$ derivative
by $O_{K,k}(V^{-2\Re\nu})$ and $O_{K,k}(V^{1-2\Re\nu})$,
uniformly for $0\leq x\leq k$. The subtraction then leaves the
convergent sum $\sum m^{2\Re\nu-13/4}$, and gives
\begin{align}
 &\left|\frac{h_z(u/i)u}{\cos\pi u}
 \{\cos\pi\nu-\cos[\pi(\nu+1/2-u)]\}\mathscr S_{\nu,k}(u)\right|\notag\\
 &\qquad\ll_{K,k,z}V^{1-2\alpha_K-\Re z}.
 \label{d6a:infinite_nu_uniform_bound}
\end{align}
At the removable moving pole, take the limit of the combined expression.
The bound is integrable for $\Re z>2-2\alpha_K$, locally
uniformly in $z$, justifying continuation under the integral.

The preceding bound controls the continuation in $\nu$ at large
$\Re z$. At the endpoint we can use oscillation in Euler's
integral to continue the moment past $z=0$. Setting $\nu=1$ gives
\be
 \mathscr F_3(u;x)=\frac1{(u-3/2)(u-1/2)}
 \int_0^1\left[\frac{xt(1-t)}{1+xt}\right]^u
 \frac{(1+xt)^{3/2}}{\sqrt{t(1-t)}}\,\dd t.
 \label{d6a:F3Euler}
\ee
For $u=c+iv$, $c=7/4$, factor out $x^u$. The phase is
$\varphi_x(t)=\log[t(1-t)/(1+xt)]$ and the amplitude is
$a_x(t)=t^{c-1/2}(1-t)^{c-1/2}(1+xt)^{3/2-c}$.
Uniformly for $0\leq x\leq k$, the amplitude vanishes at both
ends and has bounded supremum and total variation, while
$\varphi_x''=-t^{-2}-(1-t)^{-2}+x^2/(1+xt)^2\leq-1$.
The region $|\varphi_x'|\leq|v|^{-1/2}$ has length at most
$2|v|^{-1/2}$; integration by parts on its complement gives
the same bound. Including the quadratic denominator and applying
the argument once more after an $x$ derivative yields
\begin{align}
 |\mathscr F_3(c+iv;x)|&\ll_k x^c|v|^{-5/2},
 \label{d6a:infinite_kernel_bound}\\
 |\mathscr F_3(u;x)-B(u)x^u|
 &\ll_k x^c|v|^{-5/2}\min(1,|v|x).
 \label{d6a:infinite_kernel_difference}
\end{align}
For the derivative use $H(u,x)=x^{-u}\mathscr F_3(u;x)$,
which extends to $H(u,0)=B(u)$; its derivative is
$O_k(|v|^{-3/2})$. In~\eqref{d6a:Scontinue}, the mean term is
$O_k(|v|^{-3/2})$ by~\eqref{d6a:shiftDirichlet} and
$B(c+iv)=O(|v|^{-5/2})$. Splitting the remainder at $m=|v|$ gives
\be
 |v|^{-5/2}\left[\sum_{m\leq|v|}m^{-1/4}
 +|v|\sum_{m>|v|}m^{-5/4}\right]=O(|v|^{-7/4}).
 \label{d6a:infinite_remainder_bound}
\ee
Consequently
\be
 \mathscr S_k(7/4+iv)=O_k((1+|v|)^{-3/2}).
 \label{d6a:infinite_vertical_bound}
\ee
Since $uP_0G^2$ and $uP_0G^2q$ are bounded on this line,
the regulated integrands are $O_z((1+|v|)^{-3/2-\Re z})$.
Both infinite terms are therefore holomorphic for
$\Re z>-1/2$, by locally uniform convergence.

\subsection{Pole structure and growth of the spectral moment}
\label{d6a:existence}
At the starting point $\nu=i\eta$, $1<\Re s<3/2$,
the arguments $s-\nu$ and $s+\nu$ of the two $L$-functions
both have real part greater than one, and the continuous
integral runs along the original, undeformed contour. The arithmetic contour can be shifted to $\Re u=7/4$:
the kernel poles lie to its left, and the quotient $h_z(u/i)/\cos\pi u$ is regular at $u=3/2$
by~\eqref{d6a:hcomplex_limits}.
The kernel estimates above then justify continuing this
representation along the stated parameter path.
For the continuous contribution, the contour deformations
in Section~\ref{d6a:crossing} avoid fixed poles and zeta
zeros, with colliding singularities treated together in
the complete integrand. The endpoint contour correction
vanishes for the corrected test $h_z$. Thus, for sufficiently
large $\Re z$, the continued expression equals $G_k(z)$
evaluated on the original straight spectral contour.

The remaining reflected and zero-mode terms must also be regular
on this path. Set
$\mathcal R_\nu(u)=\Gamma(u+1/2-2\nu)/\Gamma(u+1/2+2\nu)$
and $\mathcal D_{\nu,k}=k^{2\nu-1/2}\sigma_{-2\nu}(k)$.
The gamma poles $u=2\nu-1/2-n$ stay to the left of the contour
$\Re u=7/4$.
The level-one formulas~\cite{HLN}
read
\begin{align}
 \mathcal T_\nu(z)&=\frac4\pi I_{7/4}\left[
 \frac{h_z(u/i)u}{\cos\pi u}
 \cos[\pi(\nu+1/2-u)]\mathcal R_\nu(u)\right],\notag\\
 M_{{\rm ref},k,q}&=-\zeta(1-2\nu)(2\pi)^{4\nu-1}\mathcal D_{\nu,k}
 \frac{\pi\mathcal T_\nu(z)}{\cos\pi\nu},\notag\\
 M_{-,k,q}&=4\zeta(1-2\nu)(2\pi)^{4\nu-1}\mathcal D_{\nu,k}
 I_{7/4}\left[\frac{h_z(u/i)u}{\cos\pi u}\mathcal R_\nu(u)\right].
 \label{d6a:central_main_terms}
\end{align}
The auxiliary normalization is regular because
\be
 \frac{2c_\nu^\pm\pi^{1/2\pm\nu}}{\Gamma(1/2\pm\nu)}
 =\zeta(1\pm2\nu).
 \label{d6a:central_zero_mode_normalization}
\ee
At $\nu=1/2$ the denominator $\cos\pi\nu$ in the reflected
term vanishes. Here $\mathcal R_{1/2}=(u^2-1/4)^{-1}$, and
\be
 \mathcal T_{1/2}(z)
 =-\frac4\pi I_{7/4}\left[\frac{h_z(u/i)u}{u^2-1/4}\right]=0.
 \label{d6a:central_half_parameter}
\ee
The test zeros cancel the poles at $\pm1/2$, so we may shift to
the imaginary axis, where the integrand is odd. Thus the analytic
numerator $\mathcal T_\nu(z)$ vanishes at $\nu=1/2$ and the
quotient by $\cos\pi\nu$ is regular. The same contour argument gives
$M_{+,k,q}(z)=0$, since its gamma quotient is one and
$h_z(u/i)/\cos\pi u$ is holomorphic between the two lines.
At $\nu=1$,
\be
 \mathcal R_1(u)=\frac1{(u^2-1/4)(u^2-9/4)}.
 \label{d6a:existence_rational_kernel}
\ee
The integrands of the reflected and minus terms are
$O_z((1+|\Im u|)^{-4-\Re z})$, with additional exponential decay
in the minus term. Both integrals therefore extend to $\Re z>-3$.

Only the diagonal can therefore have a pole in $\Re z>-1/2$.
Its density obeys $r(t)q_2(t)t\tanh\pi t=16\pi+O(t^{-2})$
by~\eqref{d6a:stirling}. Subtracting this constant and evaluating
$\int_0^\infty(4+t^2)^{-z/2}\dd t$ by the beta integral gives
\begin{align}
 H_{0,q}(z)={}&\frac1{\pi^2}\int_0^\infty
 [r(t)q_2(t)t\tanh\pi t-16\pi](4+t^2)^{-z/2}\,\dd t\notag\\
 &+\frac{16}\pi\sqrt\pi\,2^{-z}
 \frac{\Gamma((z-1)/2)}{\Gamma(z/2)}.
 \label{d6a:H0continue}
\end{align}
The integral is holomorphic for $\Re z>-1$. The factor
$\sqrt\pi\,2^{-z}\Gamma((z-1)/2)/\Gamma(z/2)$ equals
$(z-1)^{-1}+O(1)$ at one and $-\pi z+O(z^2)$
at zero. Multiplication by the diagonal coefficient gives
\be
 \Res_{z=1}G_k(z)=\frac{16\zeta(3)\sigma_{-2}(k)}{\pi\sqrt k}=a_k.
 \label{d6a:continued_moment_residue}
\ee
Thus $G_k$ is meromorphic for $\Re z>-1/2$, with this single
simple pole and a finite value at zero.

For the Mellin shift, on $u=7/4+iv$ one has
$|\arg(4-u^2)|\leq\vartheta:=\arcsin(7/8)<\pi/2$.
This follows by maximizing
$2(7/4)|v|/[4-(7/4)^2+v^2]$. Hence
$|(4-u^2)^{-(\sigma+i\xi)/2}|
\leq|4-u^2|^{-\sigma/2}e^{\vartheta|\xi|/2}$.
The absolute bound for the subtracted diagonal integral depends
only on $\sigma$. The gamma quotient in the diagonal term has
polynomial growth.
The contour bounds above therefore give, for $0<\delta<1/2$
and $c>c_*$,
\be
 |G_k(\sigma+i\xi)|\ll_{k,\delta,c}
 (1+|\xi|)^N e^{\vartheta|\xi|/2},
 \qquad -\delta\leq\sigma\leq c,\quad |\xi|\geq1.
 \label{d6a:proved_continuation_growth}
\ee
Stirling's formula for $\Gamma(z/2)$ supplies
$e^{-\pi|\xi|/4}$, so the complete Mellin integrand decays
exponentially and the horizontal segments vanish. On
$\Re z=-\delta$, $(2X)^{z/2}$ then gives an
$O_{k,\delta}(X^{-\delta/2})$ integral. This proves the
residue expansion~\eqref{d6a:corrected_heat_asymptotic}.

\subsection{Spectral bounds and Bessel asymptotics}
\label{d6a:spectral_estimates}\label{d6a:accumulated_weight_proof}
\label{d6a:initial_spectral_convergence}\label{d6a:q2_removal_estimates}
\label{rev:bessel_estimates}\label{rev:continuous_asymptotics}
Let $|\mu_k|$ denote the measure of absolute spectral weights.
Spectral counting and the $L$-function bounds
in~\cite{BHS}, together with~\eqref{d6a:elementary_Hecke_bound},
give $|\mu_k|([0,T])\ll_k(1+T)^m$ for some finite $m$,
independently of the Gaussian asymptotic.
Splitting $t\ge1$ into intervals $[2^n,2^{n+1})$, each
interval contributes at most $O_k(2^{n(m-\Re z)})$
to the absolute spectral moment. Summing over $n$ therefore
gives absolute convergence for $\Re z>m$.
The Gaussian spectral integral also converges absolutely,
since its kernel decays faster than any power.
Choosing $c_*>\max\{1,m\}$ and using the exponential decay
of $\Gamma(z/2)$ along vertical lines then justifies
interchanging the Mellin contour integral with the cusp
sum and the continuous spectral integral
in~\eqref{d6a:heat_mellin}
and~\eqref{d6a:corrected_heat_mellin}.

To remove the factor $q_2$ and estimate the Bessel tails, we need
the sharper bound
\be
 |\mu_k|([0,T])
 =\sum_{\substack{j\ {\rm even}\\t_j\leq T}}|d_j\lambda_j(k)|
 +\frac1\pi\int_0^T|W(t)\chi_k(t)|\,\dd t
 \ll_k1+T.
 \label{d6a:M}
\ee
\eqalias{d6a:absolute_spectral_weight}{d6a:M}
For even cusp forms, $L_j(1/2)\ge0$~\cite{KatokSarnak};
see also~\cite{BHS}.
Together with $L_j(5/2)>0$, the positive norm denominator,
$r(t)>0$ and $P(t)\geq0$, this gives
\be
 b_j(1)\geq0,\qquad \mu_1\geq0.
 \label{d6a:positive_joint_measure}
\ee
The bounds $|\lambda_j(k)|\leq\sigma_1(k)/\sqrt k$ and
$|\chi_k(t)|\leq d(k)$ imply
\be
 |\mu_k|\leq C_k^{\rm cmp}\mu_1,\qquad
 C_k^{\rm cmp}=\max\{\sigma_1(k)/\sqrt k,d(k)\}.
 \label{d6a:charge_domination}
\ee
Here the superscript ``cmp'' labels the comparison constant. For $T\ge1$ and $0\le t\le T$, we have
$q_2(t)\ge1/16$ and $e^{-t^2/(2T^2)}\ge e^{-1/2}$.
Positivity of $\mu_1$ and the asymptotic
formula~\eqref{d6a:corrected_heat_asymptotic}
for the Gaussian spectral integral therefore give
$\mu_1([0,T])\le16e^{1/2}\mathscr H_{1,q}(T^2)\ll T$.
The comparison~\eqref{d6a:charge_domination} extends this
first-mode bound to every fixed positive $k$,
proving~\eqref{d6a:M}. The initial Mellin inversion is
justified independently by the polynomial bound established
at the start of this subsection.

The identity $q_2(t)=1-15/[4(t^2+4)]$ gives
\eqref{d6a:addback} in the initial domain of absolute
convergence. By~\eqref{d6a:M}, the absolute contribution
from each interval $[T,2T]$, with $T\ge1$, is
$O_k(T^{1-\Re z})$. Summing over $T=1,2,4,\ldots$
shows that $Z_k(z)$ converges absolutely for $\Re z>1$.
Hence $Z_k(z+2)$ is holomorphic for $\Re z>-1$.
The relation
$Z_k(z)=G_k(z)+\tfrac{15}{4}Z_k(z+2)$
then continues $Z_k$ to $\Re z>-1/2$, with the same
residue at $z=1$ as $G_k$, and gives its value at zero
in~\eqref{d6a:existence_value_at_zero}.
In particular, $Z_k(2)$ is absolutely convergent.

To remove $q_2$ from the Gaussian spectral integral,
we estimate the integral of
$(1-e^{-t^2/(2X)})/(4+t^2)$ against $|\mu_k|$.
For $t\le\sqrt X$, the integrand is at most $1/(2X)$,
so~\eqref{d6a:M} bounds this contribution by
$O_k(X^{-1/2})$.
For $t>\sqrt X$, we bound the numerator by one and
split the range into intervals
$[2^n\sqrt X,2^{n+1}\sqrt X]$, $n\ge0$.
Each contributes $O_k(2^{-n}X^{-1/2})$, giving the same
bound after summation.
This proves~\eqref{d6a:heat_addback} and, together with
\eqref{d6a:corrected_heat_asymptotic}, gives the Gaussian
asymptotic used in~\eqref{d6a:jointheat}.

To compare the Bessel and Gaussian kernels, write the rescaled integrand in
\eqref{d6a:rescaled_Bessel_integral} as
$e^{-v^2/2}e^{-R_X(v)}$, where
$R_X(v)=X[\cosh(v/\sqrt X)-1-v^2/(2X)]\geq0$.
Since the oscillatory factor has modulus one and
$1-e^{-R_X}\leq R_X$, Gaussian averaging gives
\begin{align}
 |H_X(t)-e^{-t^2/(2X)}|
 &\leq X[e^{1/(2X)}-1-1/(2X)]\notag\\
 &\leq\frac{e^{1/2}}{8X},\qquad X\geq1,\quad t\in\mathbb R.
 \label{d6a:gaussian}
\end{align}
We apply this bound only up to $R=\sqrt X\log X$, since the total absolute
mass need not be finite. Equation~\eqref{d6a:M} makes the
integrated remainder $O_k((1+R)/X)=O_k(\log X/\sqrt X)$.

Beyond this cutoff, the Bessel and Gaussian tails are smaller than every inverse power of $X$.
For the Bessel tail,~\eqref{d6c:Mellin_Bessel_decay} gives
$|K_{it}(X)|\leq e^{-t\theta}K_0(X\cos\theta)$, while the integral
representation gives $K_0(q)\leq\sqrt{\pi/(2q)}e^{-q}$.
Choose $\theta=t/(2X)$ for $t\leq X$ and $\theta=1/2$ otherwise;
using $\cos\theta\geq1-\theta^2/2$ yields
\be
 |K_{it}(X)|\leq c_0X^{-1/2}e^{-X}
 \begin{cases}
 e^{-3t^2/(8X)},&0\leq t\leq X,\\
 e^{-c_1t},&t\geq X,
 \end{cases}
 \qquad X\geq1.
 \label{d6a:uniformtail}
\ee
Here $c_0=\sqrt{\pi/[2\cos(1/2)]}$ and
$c_1=\cos(1/2)-1/2>0$. The normalization in $H_X$ removes $X^{-1/2}e^{-X}$.
Integrating the remaining decay beyond $R$ against the linear
mass bound controls the infinite tail, as does the Gaussian
decay. This gives the $O_k(\log X/\sqrt X)$ remainder in the first line
of~\eqref{d6a:jointheat} for the replacement of the Bessel kernel.

Finally, $|\chi_k|\leq d(k)$,~\eqref{d6a:Wbound} and
\eqref{d6a:uniformtail} give, with $X=2\pi ky$,
\be
 e^X|C_k(y)|\ll_{k,\epsilon}
 \int_0^X(1+t)^{-1/2+\epsilon}e^{-3t^2/(8X)}\,\dd t
 +\int_X^\infty(1+t)^{-1/2+\epsilon}e^{-c_1t}\,\dd t.
\ee
The second integral is exponentially small. Setting $t=\sqrt X\,v$
in the first gives
\be
 C_k(y)=O_{k,\epsilon}\!\left(e^{-2\pi ky}y^{1/4+\epsilon}\right),
 \qquad 0<\epsilon<\tfrac14.
 \label{d6a:continuous_fixed_charge_bound}
\ee
Together with the asymptotics of the restored term $S_k(y)$, this implies
\be
 \lim_{y\to\infty}
 \frac{e^{2\pi ky}[S_k(y)+C_k(y)]}{\sqrt y}=0.
 \label{d6a:nosqrtcontinuous}
\ee
The leading $\sqrt y$ coefficient therefore comes from the cusp
spectrum; the constant requires the joint moment and restored terms.

\subsection{Removing the auxiliary factor from the trace formula}
\label{d6a:barecorrection}
The identity~\eqref{d6a:addback} removes $q_2$ on the spectral
side. On the arithmetic side, removing its zero at $u=1/2$
introduces a residue when we shift the Bessel contour. To compute
that residue we first keep $\Re z$ large and approximate the
original test function $h=h_z^0$ by tests satisfying $h_\varepsilon(\pm i/2)=0$.
Take
\be
 h_\varepsilon(t)=\frac{t^2+1/4}{t^2+1/4+\varepsilon}h(t),
 \qquad\varepsilon>0.
 \label{d6a:bare_admissible_approximation}
\ee
Each test is admissible in a strip wider than $1/2$, with nearest
possible poles at $\pm i\sqrt{1/4+\varepsilon}$.
Fix $1/4<\beta<1/2$. On $t=a+ib$, $|b|\leq\beta$,
$\Re(t^2+1/4+\varepsilon)\geq a^2+1/4-\beta^2>0$, and
$h_\varepsilon-h=-\varepsilon h/(t^2+1/4+\varepsilon)$.
Thus convergence and polynomial decay are uniform on this fixed strip.
Changing variables in the Bessel transform gives
\be
 H^+(x;h)=4I_0[J_{2u}(x)h(u/i)u/\cos\pi u].
 \label{d6a:bare_initial_Barnes_transform}
\ee
For $h_\varepsilon$ shift to $\Re u=\beta$ without crossing a pole.
The Bessel series and Stirling estimates give
$J_{2\beta+2iv}(x)\ll_\beta
x^{2\beta}(1+|v|)^{-2\beta-1/2}e^{\pi|v|}$ for $0<x\leq1$.
Hence $H^+(x;h_\varepsilon)\ll_{z,\beta}x^{2\beta}$ uniformly in
$\varepsilon$. Weil's bound
$|S(m,n;c)|\leq d(c)\sqrt{c\gcd(m,n,c)}$
\cite{HLN} makes the modulus summand
$O_{m,n,z,\beta,\eta}(c^{-1/2-2\beta+\eta})$.
Choose $0<\eta<2\beta-1/2$ to pass the limit through that sum.
On the real axis the multiplier lies in $[0,1]$, so absolute
convergence also passes the limit through the spectral side
and diagonal. This establishes the Kuznetsov formula for the original test function, with its contour
left of $u=1/2$.

Shifting this contour to the right crosses
\be
 \Res_{u=1/2}[J_{2u}(x)h(u/i)u/\cos\pi u]
 =-\frac{h(i/2)}{2\pi}J_1(x).
 \label{d6a:bare_Bessel_residue}
\ee
Since both vertical lines run upwards, $I_c-I_0$ is the residue
sum. Therefore
\be
 H^+_b(x;h)
 =4I_c[J_{2u}(x)h(u/i)u/\cos\pi u]
 +\frac{2h(i/2)}\pi J_1(x).
 \label{d6a:J1correction}
\ee
For our original test function this reaches $c=7/4$ without a second test
residue:~\eqref{d6a:hcomplex_limits} gives
\be
 \frac{h_z^0(u/i)}{\cos\pi u}
 =\frac{4-u^2}{1/4-u^2}\frac{h_z(u/i)}{\cos\pi u},
 \qquad
 \lim_{u\to3/2}\frac{h_z^0(u/i)}{\cos\pi u}
 =-\frac85(7/4)^{-z/2}.
 \label{d6a:bare_test_threehalves}
\ee
Arithmetic kernel poles are treated separately.

The Petersson trace formula relates Fourier coefficients
of holomorphic cusp forms to a diagonal term and a
Kloosterman--Bessel sum. Since there are no holomorphic
cusp forms of weight two at level one, its weight-two
version~\cite{DiScipio}
reduces to
\be
 \sum_{c\geq1}\frac{S(m,n;c)}c
 J_1\!\left(\frac{4\pi\sqrt{mn}}c\right)
 =\frac{\delta_{mn}}{2\pi}.
 \label{d6a:weighttwo}
\ee
The residual and continuous terms vanish in this holomorphic
trace formula. Its sum is absolutely convergent by Weil's bound
and $J_1(x)=O(x)$ near zero. To insert it into the first moment,
use $J_1(x)\ll\min(x,x^{-1/2})$ and split at $c=\sqrt m$:
\be
 \sum_{c\geq1}\frac{|S(m,k;c)|}{c}
 \left|J_1\!\left(\frac{4\pi\sqrt{mk}}c\right)\right|
 \ll_{k,\eta}m^{1/4+\eta}.
 \label{d6a:bare_rankin_absolute_bound}
\ee
For imaginary $\nu$, $|C_{\phi_\nu}(m)|\leq d(m)\ll_\eta m^\eta$,
so $\Re s>5/4+2\eta$ justifies exchanging the two sums and
overlaps the HLN region. The residue then contributes
$\mathcal J_k(s,\nu;h)=h(i/2)\zeta(2s)C_{\phi_\nu}(k)/(\pi^2k^s)$,
which continues along the same parameter path. At the endpoint,
\be
 \mathcal J_k(z)=\frac{2\zeta(3)\sigma_{-2}(k)}{3\pi\sqrt k}
 (15/4)^{-z/2}.
 \label{d6a:rankinJ}
\ee

The zero-mode formula~\cite{HLN}, using
\eqref{d6a:central_zero_mode_normalization}, is
\be
 M_{+,k}(\nu+1/2,\nu;h)
 =\frac{\zeta(1+2\nu)\sigma_{-2\nu}(k)}{2\pi\sqrt k}
 4I_{7/4}[h(u/i)u/\cos\pi u].
 \label{d6a:bare_plus_central_formula}
\ee
The imaginary-axis integral is odd. For $h_z^0$, its shift
therefore equals the residue $-h_z^0(i/2)/(2\pi)$, with no
residue at $3/2$. This proves
\be
 M_{+,k,b}(z)=-\mathcal J_k(z).
 \label{d6a:bare_plus_exact_cancellation}
\ee
Both vanish separately for the corrected test.

The shifted integrals also lose their auxiliary factor. On the
fixed contour the identity
\begin{align}
 &q(u)+\frac{15}{4(4-u^2)}=1, \notag\\
 &L_{\infty,q}(0)+\frac{15}{4}L_{\infty,b}(2)=L_{\infty,b}(0)
 \label{d6a:infinite_addback_identity}
\end{align}
holds between convergent continued integrals. It removes the
artificial pole at $u=2$: its two residues are $15/16$ and
$(15/4)(-1/4)$. The remaining factor is regular there by
\eqref{d6a:Scontinue}, although its original charge series does
not converge there. At the genuine poles $5/2$ and $7/2$, the
multiplier is restored to one.

Combining the Bessel correction with the subtraction of the
continuous correction~\eqref{d6a:barecross} yields
\begin{align}
 Z_k(z)={}&M_{{\rm diag},k,b}(z)+M_{{\rm ref},k,b}(z)
 +M_{+,k,b}(z)+M_{-,k,b}(z)\notag\\
 &+L_{{\rm short},k,b}(z)+L_{\infty,k,b}(z)
 +\mathcal J_k(z)-\mathcal R_k^0(z).
 \label{d6a:bare_complete_identity}
\end{align}
Each term uses the same continued contour with $h_z^0$ replacing
$h_z$. For the original test function, the shift in~\eqref{d6a:central_half_parameter}
crosses $u=1/2$ and the value of $\mathcal T_{1/2}(z)$ is
$-2h_z^0(i/2)/\pi$.
Consequently $M_{{\rm ref},k,b}$ has residue
$2h_z^0(i/2)\chi_k(i/2)$ at $\nu=1/2$ along $s=\nu+1/2$.
The factor $\zeta(2\nu)$ in~\eqref{d6a:crossformula} gives the
same residue in the continuous correction. Their difference is
regular there. The identity holds initially at large $\Re z$ and
continues to zero.

\subsection{Contour contributions to the constant term}
\label{d6a:mainterms}
To evaluate $Z_k(0)$ in~\eqref{d6a:bare_complete_identity},
we begin with the diagonal, the two remaining main terms and the
infinite shifted sum. Write
$D_k=k^{3/2}\sigma_{-2}(k)$ and
$E_k=\zeta(3)\sigma_{-2}(k)/\sqrt k$ for scalar coefficients.
For the diagonal term with the original test function, we subtract the limiting density as in
\eqref{d6a:H0continue}. The resulting gamma quotient vanishes at
$z=0$, leaving
\be
 \mathcal H=H_{0,b}(0)=\frac1{\pi^2}\int_0^\infty
 [r(t)t\tanh\pi t-16\pi]\,\dd t,\qquad
 M_{{\rm diag},k}=E_k\mathcal H.
 \label{d6a:Hfinite}
\ee
Set $F(u)=uP_0(u)G(u)^2$. Gamma reflection~\eqref{d6a:infinite_reflection}
gives $h_0^0(u/i)u(\tan\pi u-1/\cos\pi u)=-4\pi F(u)$.
The secant term is odd and integrable on the imaginary axis,
so the change of variables in~\eqref{d6a:Hfinite} gives
\be
 \mathcal H=-\frac2\pi I_0[h_0^0(u/i)u\tan\pi u+8\pi]
 =8I_0[F(u)-2].
 \label{d6a:diagonal_subtracted_contour}
\ee
The subtracted terms must remain inside the contour integral. The gamma-ratio
expansion~\cite{DLMF} gives
$G(u)^2=(2/u)(1+O(u^{-2}))$ in the closed right half-plane;
the relative $u^{-1}$ coefficient vanishes since $1/4+3/4=1$.
Hence $F-2=O(u^{-2})$ and the right semicircle contributes zero.
The sole enclosed pole is at $7/2$, with
$G(7/2)=4/(3\sqrt\pi)$ and $\Res F=32/(3\pi)$.
Closing an upward line to the right is clockwise, so
\be
 \mathcal H=-\frac{256}{3\pi},\qquad
 M_{{\rm diag},k}=-\frac{256\zeta(3)\sigma_{-2}(k)}{3\pi\sqrt k}.
 \label{d6a:Hexact}
\ee

Specializing~\eqref{d6a:central_main_terms} gives
\begin{align}
 M_{{\rm ref},k}(h)&=\frac{8\pi^3}{3}D_kI_{7/4}
 [h(u/i)u\tan\pi u\,\mathcal R_1(u)],\notag\\
 M_{-,k}(h)&=-\frac{8\pi^3}{3}D_kI_{7/4}
 [h(u/i)u\mathcal R_1(u)/\cos\pi u].
 \label{d6a:main_endpoint_contours}
\end{align}
At $z=0$ their combined kernel with the original test function is
$\mathcal K(u)=F(u)\mathcal R_1(u)
=uG(u)^2/[(u^2-49/4)(u^2-9/4)]$.
It is $O(u^{-4})$ in the right half-plane. The pole $3/2$
lies left of the contour; the enclosed pole $7/2$ has residue
$G(7/2)^2/20=4/(45\pi)$. Consequently
\be
 M_{{\rm ref},k,b}(0)+M_{-,k,b}(0)
 =-\frac{32\pi^4}{3}D_kI_{7/4}[\mathcal K]
 =\frac{128\pi^3}{135}D_k.
 \label{d6a:Msum}
\ee

The integrand in~\eqref{d6a:Linf} has only two poles
to the right of $7/4$, at $5/2$ and $7/2$. To justify closing the contour,
Euler's integral involves the following maximum:
\be
 \beta(x)=\max_{0\leq t\leq1}\frac{xt(1-t)}{1+xt}
 =\frac{x}{(\sqrt{1+x}+1)^2}<1,\qquad \beta(x)\leq x/4.
 \label{d6a:infinite_beta}
\ee
For $u=R+iv$, $R>5$, use $A_m(k)\leq\zeta(2)^2$ and
three powers of $\beta(k/m)\leq k/(4m)$ to obtain
\be
 |\mathscr S_k(R+iv)|\ll_k
 \frac{\sum_{m\geq1}m^{3/2}\beta(k/m)^R}{R^2+v^2}
 \ll_k\frac{\beta(k)^{R-3}}{R^2+v^2}.
 \label{d6a:infinite_right_bound}
\ee
Since $uP_0G^2$ is bounded, the integral over the right vertical line is
$O_k(R^{-1}\beta(k)^{R-3})$.
For each fixed $R$, the phase and amplitude-variation bounds
used in~\eqref{d6a:infinite_kernel_bound}--\eqref{d6a:infinite_remainder_bound}
are uniform for $u=c+iv$, $7/4\leq c\leq R$.
Together with~\eqref{d6a:shiftDirichlet} this gives
$\mathscr S_k(c+iv)=O_{k,R}((1+|v|)^{-3/2})$, so the horizontal
integrals vanish. Send their heights to infinity, then $R$ to infinity.
The upward-contour identity $I_{7/4}=I_R-\sum\Res$ makes the
coefficient of each residue in $L_{\infty,b}(0)$ equal to $64\pi^2$.

At $u=5/2$, $F(5/2)=-5\pi/8$, so~\eqref{d6a:arithmeticpole}
contributes $-2\pi^3k^{5/2}\mathcal C_k/3$.
After division by $2\sqrt k$, this is
\be
 -\frac{\pi^3}{3}k^2\mathcal C_k
 =-\frac{315\zeta(3)^2}{\pi^3}k^2\sigma_{-5}(k),
 \label{d6a:infinite_E3_normalization}
\ee
canceling the restored $E_3$ constant. The original test function is essential:
the corrected residue alone would retain $q(5/2)$.
At $u=7/2$, the residue of $F$ is $32/(3\pi)$ as above. Therefore
\be
 L_{\infty,b}(0)=-\frac{2\pi^3}{3}k^{5/2}\mathcal C_k
 +\frac{2048\pi}{3}\mathscr S_k(7/2).
 \label{d6a:Linfresult}
\ee
The remaining value has the absolutely convergent positive series
\be
 \mathscr S_k(7/2)=\frac{k^{7/2}}{840}\sum_{m\geq1}
 \frac{A_m(k)}{m^2}{}_2F_1(2,4;8;-k/m).
 \label{d6a:Spositive}
\ee
Euler's integral bounds the hypergeometric factor between zero
and one, giving an $O_k(M^{-1})$ tail after $M$ terms.

The Bessel-contour residue cancels $M_{+,k,b}$. The continuous-contour
correction $-\mathcal R_k^0/(2\sqrt k)$ cancels the restored
$\widehat E_1$ constant, and the pole at $u=5/2$ cancels the
restored $E_3$ constant. Thus~\eqref{d6a:Bzeta} reduces to
\be
 B_k=\frac1{2\sqrt k}\left[
 M_{{\rm diag},k}+M_{{\rm ref},k}+M_{-,k}
 +L_{{\rm short},k}+\frac{2048\pi}{3}\mathscr S_k(7/2)\right].
 \label{d6a:reducedB}
\ee
Here the arithmetic terms are evaluated at zero with the original test function.

\subsection{Evaluation of the constant term}
\label{d6a:exactB}\label{d6a:positive_pair_evaluation}\label{d6a:opposite_pair_cancellation}
It remains to combine the finite sum over pairs of positive indices
with the value $\mathscr S_k(7/2)$ from the infinite sum.
For $a+b=k$, put $x=a/k$ and $w=1-2x$.
Equations~\eqref{d6a:short_general} and~\eqref{d6a:finite_Ferrers_endpoint}
give the finite sum at the endpoint with the original test function
\begin{align}
 L_{{\rm short},k}
 ={}&-32\pi k^{3/2}
 \sum_{\substack{a,b\geq1\\a+b=k}}x(1-x)\sigma_{-2}(a)\sigma_{-2}(b)
 \notag\\
 &\qquad\times I_{7/4}\left[\frac{h_0^0(u/i)u}{\cos\pi u}
 \mathsf Q_{u-1/2}^{-2}(w)\right].
 \label{d6a:short_all_charges_endpoint}
\end{align}
The coefficient is symmetric under $a\leftrightarrow b$,
which sends $w$ to $-w$. We therefore average the
contributions at $w$ and $-w$ and define
\begin{align}
 \mathcal P_u(w)&=\frac{\mathsf P_{u-1/2}^{-2}(w)+\mathsf P_{u-1/2}^{-2}(-w)}2,\notag\\
 \mathcal T(w)&=I_{7/4}\left[\frac{uP_0(u)G(u)^2}{\cos\pi u}\mathcal P_u(w)\right].
 \label{d6a:short_symmetric_transform}
\end{align}
Reflection~\eqref{d6a:finite_Ferrers_reflection} and
\eqref{d6a:infinite_reflection} give
\be
 \frac{h_0^0(u/i)}{\cos\pi u}
 \frac{\mathsf Q_{u-1/2}^{-2}(w)+\mathsf Q_{u-1/2}^{-2}(-w)}2
 =\frac{2\pi^2P_0(u)G(u)^2}{\cos\pi u}\mathcal P_u(w).
 \label{d6a:short_symmetrization_normalization}
\ee
Thus
\be
 L_{{\rm short},k}=-64\pi^3k^{3/2}
 \sum_{\substack{a,b\geq1\\a+b=k}}
 x(1-x)\sigma_{-2}(a)\sigma_{-2}(b)\mathcal T(1-2a/k).
 \label{d6a:short_symmetric_sum}
\ee

To close the contour defining $\mathcal T$ to the right,
integrate~\eqref{d6a:finite_Mehler_integral} twice by parts.
The boundary terms vanish and the second derivative of
$(\cos v-\cos\theta)^{3/2}$ is integrable, giving
$\mathsf P_{u-1/2}^{-2}(\cos\theta)
=O_\theta(|u|^{-2}e^{\theta|\Im u|})$.
Apply this at $\theta$ and $\pi-\theta$. On integer right edges,
the secant decays as $e^{-\pi|\Im u|}$ and $uP_0G^2=O(1)$;
the right-edge integral is $O_w(R^{-2})$ and horizontal edges
vanish. The closure is clockwise.

At the even-degree poles $u_n=2n+1/2$, $n\geq1$, put
$\beta_n=((1/2)_n/n!)^2$. Then $G(u_n)^2=\pi\beta_n$ and
$\Res_{u_n}\sec\pi u=-1/\pi$.
Odd-degree secant poles cancel against $\mathcal P_u$, except
at $7/2$, where $P_0$ also has a pole. Differentiating
\eqref{d6a:finite_Ferrers_reflection} gives
$\partial_u\mathcal P_u|_{7/2}=\mathsf Q_3^{-2}(w)$.
Using $\Res_{7/2}(uP_0G^2)=32/(3\pi)$ and
$\Res_{7/2}\sec\pi u=1/\pi$, we obtain
\be
 \mathcal T(w)=\sum_{n\geq1}
 \frac{(2n+1/2)\beta_n(2n)(2n+1)}{(2n)(2n+1)-12}
 \mathsf P_{2n}^{-2}(w)-\frac{32}{3\pi^2}\mathsf Q_3^{-2}(w).
 \label{d6a:short_general_residues}
\ee
The bound from the integral representation also gives absolute convergence on compact
subintervals of $(-1,1)$.

The denominator of this residue series comes from the rational factor
$P_0(u)$ in~\eqref{d6a:infinite_factors}. To sum the series, introduce
a Legendre operator acting on the auxiliary variable $w=(b-a)/k$.
The Ferrers functions $\mathsf P_{2n}^{-2}(w)$ have
eigenvalues $(2n)(2n+1)$ under this operator.
Subtracting $12$ therefore reproduces the denominator
in the residue series.
\begin{align}
 \mathcal L_2&=-\frac{\dd}{\dd w}\left[(1-w^2)\frac{\dd}{\dd w}\right]
 +\frac4{1-w^2},\notag\\
 \Phi(w)&=-\frac{\sqrt{1-w^2}(1-2w^2)}{3\pi}, \notag\\
 (\mathcal L_2-12)\Phi&=\frac1{\pi\sqrt{1-w^2}}.
 \label{d6a:short_elementary_sum}
\end{align}
The last identity follows by differentiation. For $\ell=2n$ with $n\ge1$,
Ferrers orthogonality~\cite{DLMF} gives
\be
 \|\mathsf P_\ell^{-2}\|_2^2
 =\frac1{(\ell+1/2)(\ell-1)\ell(\ell+1)(\ell+2)}.
 \label{d6a:Ferrers_norm}
\ee
The Legendre generating function further yields
\begin{align}
 \int_{-1}^1\frac{P_{2n}(w)}{\sqrt{1-w^2}}\,\dd w
 &=\pi\beta_n, \notag\\
 \frac1\pi\int_{-1}^1
 \frac{\mathsf P_{2n}^{-2}(w)}{\sqrt{1-w^2}}\,\dd w
 &=\frac{\beta_n}{(2n-1)(2n+2)}.
 \label{d6a:short_source_projection}
\end{align}
For the first integral in~\eqref{d6a:short_source_projection},
set $w=\cos\theta$ in the generating function and extract the
constant Fourier coefficient. The second follows by integration
by parts from
$\mathsf P_\ell^{-2}=(1-w^2)P_\ell''/
[(\ell-1)\ell(\ell+1)(\ell+2)]$ and the Legendre equation.
Pair~\eqref{d6a:short_elementary_sum} with $\mathsf P_{2n}^{-2}$;
boundary terms vanish. Dividing the last projection by
$[\ell(\ell+1)-12]\|\mathsf P_\ell^{-2}\|_2^2$ reproduces the
coefficient in~\eqref{d6a:short_general_residues}.
The eigenvalue $12$ occurs only at odd degree three, absent
from this even expansion. Completeness therefore identifies
its sum with $\Phi$, giving
\be
 \mathcal T(w)=-\frac{\sqrt{1-w^2}(1-2w^2)}{3\pi}
 -\frac{32}{3\pi^2}\mathsf Q_3^{-2}(w).
 \label{d6a:short_general_transform_value}
\ee

For comparison with the divisor identity, define
\be
 Q_1^{(2,2)}(w)=\frac1{4(w-1)^2(w+1)^2}
 \int_{-1}^1\frac{(1-t^2)^3}{(w-t)^2}\,\dd t,
 \qquad |w|>1.
 \label{d6a:exact_Jacobi_definition}
\ee
For $-1<w<1$, define $Q_1^{(2,2)}(w)$ by averaging
its analytic boundary values from above and below the
real axis, following~\cite{FKR}. This defines an even function and is essential
for positive pairs. Differentiating
$\mathsf Q_3(w)=(5w^3-3w)\log[(1+w)/(1-w)]/4-5w^2/2+2/3$
and the defining integral gives
\begin{align}
 Q_1^{(2,2)}&=\tfrac15\mathsf Q_3'', \notag\\
 \mathsf Q_3^{-2}(w)&=\frac{1-w^2}{120}\mathsf Q_3''(w)
 =\frac{1-w^2}{24}Q_1^{(2,2)}(w),\qquad -1<w<1.
 \label{d6a:short_Jacobi_conversion}
\end{align}
The derivative identity for the Ferrers function
$\mathsf Q_3^{-2}$ follows from~\cite{DLMF}. Set
\be
 P_k^{\rm Jac}=\sum_{\substack{a,b\geq1\\a+b=k}}
 Q_1^{(2,2)}((b-a)/k)\sigma_2(a)\sigma_2(b).
 \label{d6a:exact_positive_pairs}
\ee
Substitute~\eqref{d6a:short_general_transform_value} into
\eqref{d6a:short_symmetric_sum}, using
$1-w^2=4ab/k^2$, $1-2w^2=(6ab-a^2-b^2)/k^2$ and
$\sigma_2(a)=a^2\sigma_{-2}(a)$. The result is
\begin{align}
 L_{{\rm short},k}={}&\frac{128\pi^2}{3k^{7/2}}
 \sum_{\substack{a,b\geq1\\a+b=k}}
 \frac{\sigma_2(a)\sigma_2(b)}{\sqrt{ab}}(6ab-a^2-b^2)\notag\\
 &+\frac{1024\pi}{9k^{5/2}}P_k^{\rm Jac}.
 \label{d6a:shortexact_general}
\end{align}

The divisor-convolution theorem of Fedosova, Klinger-Logan and
Radchenko~\cite{FKR}, at $r_1=r_2=2$, $d=1$, sums
over nonzero signed $a,b$ with $a+b=k$; for negative arguments
use $\sigma_2(a)=\sigma_2(|a|)$. Its elementary coefficient is
$Z_1^{(2,2)}(k)=\zeta(2)k^2/40+3\zeta'(-2)/2$.
The holomorphic cusp term vanishes because its weight is eight
and the level-one cusp space is zero. The two elementary terms
therefore give
\be
 \sum_{\substack{a+b=k\\ab\ne0}}
 Q_1^{(2,2)}((b-a)/k)\sigma_2(a)\sigma_2(b)
 =-\left[\frac{\zeta(2)k^2}{20}+3\zeta'(-2)\right]\sigma_2(k).
 \label{d6a:FKR}
\ee
For $a=-m$, $b=m+k$, substitute $t=1-2v$ into
\eqref{d6a:exact_Jacobi_definition} and use Euler's integral:
\begin{align}
 &Q_1^{(2,2)}((2m+k)/k)\sigma_2(m)\sigma_2(m+k)\notag\\
 &\qquad=\frac{k^6A_m(k)}{2m^2}
 \int_0^1\frac{v^3(1-v)^3}{(1+(k/m)v)^2}\,\dd v\notag\\
 &\qquad=\frac{k^6}{280m^2}A_m(k){}_2F_1(2,4;8;-k/m).
 \label{d6a:kernelmatch}
\end{align}
Comparison with~\eqref{d6a:Spositive} gives
$3k^{5/2}\mathscr S_k(7/2)$ for this branch, and the same for
the other by evenness. Absolute convergence justifies separating
these branches from the finite positive pairs:
\be
 6k^{5/2}\mathscr S_k(7/2)+P_k^{\rm Jac}
 =-\left[\frac{\zeta(2)k^2}{20}+3\zeta'(-2)\right]\sigma_2(k).
 \label{d6a:exact_FKR_split}
\ee
Using $\zeta(2)=\pi^2/6$ and $\zeta'(-2)=-\zeta(3)/(4\pi^2)$ gives
\be
 \frac{2048\pi}{3}\mathscr S_k(7/2)
 =-\frac{128\pi^3}{135}k^{3/2}\sigma_{-2}(k)
 +\frac{256\zeta(3)}{3\pi\sqrt k}\sigma_{-2}(k)
 -\frac{1024\pi}{9k^{5/2}}P_k^{\rm Jac}.
 \label{d6a:opposite_general_value}
\ee
The three terms cancel, respectively,~\eqref{d6a:Msum},
\eqref{d6a:Hexact} and the last term of~\eqref{d6a:shortexact_general}.
Equation~\eqref{d6a:reducedB} becomes
\be
 B_k=\frac{64\pi^2}{3k^4}
 \sum_{\substack{a,b\geq1\\a+b=k}}
 \frac{\sigma_2(a)\sigma_2(b)}{\sqrt{ab}}(6ab-a^2-b^2),
 \label{d6a:B_general_derivation}
\ee
proving~\eqref{d6a:constant_general} without higher-charge input.
Equivalently,
\be
 \frac{Z_k(0)}{2\sqrt k}
 =B_k-\frac{315\zeta(3)^2}{\pi^3}k^2\sigma_{-5}(k)
 -\frac{4\pi\zeta(3)}3\sigma_{-1}(k).
 \label{d6a:finite_value_general}
\ee

\section{Modular reconstruction with partial Fourier data}
\label{d6c:neutral_appendix}

This appendix develops the estimates and constructions used in
Section~\ref{rev:incomplete}. We first examine how neutral zero-mode
profiles contribute to the continuous spectrum and whether finite
neutral truncations are compatible with modularity. We then derive
spectral bounds sufficient for omitted contributions to leave a
leading instanton coefficient unchanged. Finally, two explicit
modular constructions show how higher-mode coefficients can vary
while different parts of the low-mode data remain fixed.

\subsection{Neutral profiles and the continuous spectrum}
\label{d6c:neutral_response_details}
\label{d6c:exponential_profile_details}
Let $h=\widetilde f-f$ be the difference of two smooth,
real, even modular functions with the same perturbative
subtraction. Assume that $h$ decreases rapidly at the cusp
and satisfies the unfolding conditions of
Section~\ref{rev:framework}. Writing $h_0=\delta f_0$
for the zero-mode difference, we obtain
\begin{align}
 \delta R(s)&=\int_{\mathcal F}hE_s\,\dd\mu
 =\int_0^\infty\delta f_0(y)y^{s-2}\,\dd y,
 \qquad\Re s>1,\label{d6c:neutral_response_unfolding_proof}\\
 \Lambda(s)\delta R(s)&=\Lambda(1-s)\delta R(1-s).
 \label{d6c:neutral_response_scattering}
\end{align}
The second identity follows from the completed Eisenstein functional
equation. Continuation to the spectral line gives
$\langle h,E_s\rangle=\delta R(1-s)$ and hence the continuous
contribution~\eqref{d6c:neutral_response_continuous}. At positive charge the
full Fourier-mode difference also includes the cusp term, with
$\delta d_j=\langle h,u_j\rangle\rho_j(1)$ in
\eqref{d6c:missing_exact_spectrum}; the constant spectral projection
contributes only to charge zero. Knowing only a profile's
large-$y$ expansion does not determine it for all $y$ or establish
the compatibility condition~\eqref{d6c:neutral_response_scattering}.

Consider the exact neutral profile $e^{-4\pi My}$, $M\ge1$.
Its Mellin transform is
$\delta R(s)=(4\pi M)^{1-s}\Gamma(s-1)$ for $\Re s>1$.
We first evaluate the corresponding continuous contribution,
then check whether this profile can be the zero mode of a
modular function.
At fixed $k,M$, put $X=2\pi ky$ and shift the integration contour to
$1/2<c<1$:
\be
 \delta C_k(y)=\frac{\sqrt y}{\sqrt k}\frac1{2\pi i}
 \int_{(c)}\frac{\Gamma(-s)}{\Gamma(s)}
 (4\pi^2Mk)^s\frac{\sigma_{1-2s}(k)}{\zeta(2s)}
 K_{s-1/2}(X)\,\dd s.
 \label{d6c:exponential_shifted_contour}
\ee
No pole is crossed; $1/\zeta(2s)$ vanishes at $s=1/2$, and the
Bessel factor controls the horizontal segments. On $s=c+it$,
$|K_{c-1/2+it}(X)|\le K_{c-1/2}(X)$, while the gamma ratio is
$O_c((1+|t|)^{-2c})$. The arithmetic factors are bounded there.
Thus dominated convergence gives
\be
 b_{k,M}:=\lim_{y\to\infty}e^{2\pi ky}\delta C_k(y)
 =\frac1{2k}\frac1{2\pi i}\int_{(c)}
 \frac{\Gamma(-s)}{\Gamma(s)}(4\pi^2Mk)^s
 \frac{\sigma_{1-2s}(k)}{\zeta(2s)}\,\dd s.
 \label{d6c:exponential_constant_contour}
\ee
Expand $\sigma_{1-2s}(k)=\sum_{d\mid k}d^{1-2s}$ and
$1/\zeta(2s)=\sum_{n\ge1}\mu(n)n^{-2s}$, both absolutely convergent
on this contour. The Mellin--Barnes identity
\cite{DLMF}
\be
 \frac1{2\pi i}\int_{(c)}\frac{\Gamma(-s)}{\Gamma(s)}z^s\,\dd s
 =-\sqrt z\,J_1(2\sqrt z),\qquad z>0,
 \label{d6c:exponential_Barnes_J}
\ee
follows from the Mellin--Barnes representation of $J_\nu$
by setting $\nu=1$, shifting the integration variable by one
and using
$\Gamma(1-s)/\Gamma(1+s)=-\Gamma(-s)/\Gamma(s)$.
It gives
\begin{align}
 b_{k,M}
 &=-\pi\sqrt{\frac Mk}\sum_{d\mid k}\sum_{n\ge1}
   \frac{\mu(n)}n J_1\!\left(\frac{4\pi\sqrt{Mk}}{dn}\right)\notag\\
 &=\frac1{2k}\sum_{\ell\ge1}
   \frac{(-1)^\ell(4\pi^2Mk)^\ell}{\ell!(\ell-1)!}
   \frac{\sigma_{1-2\ell}(k)}{\zeta(2\ell)}.
 \label{d6c:exponential_constant_power_series}
\end{align}
Both series converge absolutely. In the first, the expansion
$J_1(x)=x/2+O(x^3)$ gives an $O_{k,M}(n^{-2})$ summand.
The factorial denominators in the power series for $J_1$
allow the sums over $n$ and $\ell$ to be taken in either
order, yielding the second expression.

The bound $|J_1(x)|\ll\min(x,x^{-1/2})$
\cite{DLMF} gives, by splitting at $n=A$,
$\sum_{n\ge1}|J_1(A/n)|/n\ll1$ for $A\ge1$, and $O(A)$ for
$A<1$. Consequently
\be
 |b_{k,M}|\ll\sqrt{\frac Mk}\,d(k),\qquad
 b_{k,M}=O_{M,\epsilon}(k^{-1/2+\epsilon}),
 \qquad0<\epsilon<\tfrac12.
 \label{d6c:exponential_charge_bound}
\ee
To control the remainder, move~\eqref{d6c:exponential_shifted_contour}
to $3/2<c'<2$. The crossed pole at $s=1$ contributes exactly
$-12M\sigma_{-1}(k)e^{-X}$. For $\eta=c'-1/2$, the integral
representation of $K$ gives
\be
 \left|\sqrt{\frac{2X}{\pi}}e^XK_{\eta+it}(X)-1\right|
 \le\frac{C_\eta(1+t^2)}X,\qquad X\ge1,
 \label{d6c:exponential_uniform_kernel}
\ee
where $C_\eta$ depends only on $\eta$. Indeed, use
$|\cosh((\eta+it)u)-1|\le C_\eta(1+t^2)u^2e^{|\eta|u}$ and
$\cosh u-1\ge u^2/2$; the normalized $K_0$ term differs from one
by $O(X^{-1})$. The remaining contour weight is $O((1+|t|)^{-2c'})$.
Since $c'>3/2$, multiplying it by $1+t^2$ still gives
an absolutely convergent integral over $t\in\mathbb R$.
This proves the $O_{k,M}(y^{-1})$ remainder in
\eqref{d6c:exponential_continuous_asymptotic}.

The coefficient $b_{k,M}$ gives the constant term in the
instanton prefactor of the corresponding continuous contribution.
For the chosen profile to be the zero mode of a modular function,
its transform must satisfy~\eqref{d6c:neutral_response_scattering}.
However, $\delta R$ has a pole at $s=-1$ and is regular at $s=2$,
while $\Lambda(-1)$ and $\Lambda(2)$ are finite and nonzero.
The functional equation therefore requires additional zero-mode
terms. Adding cusp forms cannot supply them, since their zero
modes vanish.

\subsection{Truncating the GMV neutral expansion}
\label{d6c:neutral_truncation_details}
A  finite sum of the exact GMV neutral profiles
cannot be the zero mode of a modular function, because its
Mellin transform fails the reflection relation required by
modularity. Combining the contributions from the charge
pairs $(m,-m)$ and $(-m,m)$ gives
\be
 N_m(y)=\frac{64\pi}{315}\frac{\sigma_2(m)^2}{m^3}
 \sum_{i,j=0}^1r^{ij}(\pi my)K_i(2\pi my)K_j(2\pi my),
 \qquad m\ge1.
 \label{d6c:neutral_exact_profile_app}
\ee
Its Mellin transform is
\be
 R_m(s):=\int_0^\infty N_m(y)y^{s-2}\,\dd y
 =\frac{8\pi^{2-s}}{315}\mathcal G(s)
   \frac{\sigma_2(m)^2}{m^{s+2}},\qquad\Re s>4.
 \label{d6c:neutral_profile_transform_app}
\ee
Here $\mathcal G$ is the gamma sum of Appendix~\ref{d6c:gamma};
the transform follows by $X=2\pi my$ and
\eqref{d6c:zero_Bessel_Mellin}. The factor of two is already included.
The exponential $e^{-4\pi my}$ comes from the opposite charges in the
GMV input~\cite{GMV}, not from modular invariance alone.

Let $R^{(L)}=\sum_{m=1}^L R_m$, $L\ge1$. Equations
\eqref{d6c:Q_definition}--\eqref{d6c:gammarecurrence} give
$\Res_{s=4}\mathcal G=-30$ and $\mathcal G(-3)=-1024\pi^2$.
Thus
\be
 \Res_{s=4}R^{(L)}(s)
 =-\frac{16}{21\pi^2}\sum_{m=1}^L\frac{\sigma_2(m)^2}{m^6}\ne0,
 \label{d6c:neutral_truncation_scattering_test}
\ee
whereas $R^{(L)}$ is regular at $-3$. Since $\Lambda(4)$ and
$\Lambda(-3)$ are finite and nonzero, every nonempty finite truncation
fails~\eqref{d6c:neutral_response_scattering}. The corresponding Eisenstein
integral is modular but does not have the prescribed truncated zero mode.
The complete transform, summed before continuation, does satisfy the
reflection identity~\eqref{d6c:RS_reflection}.

The continuous contribution associated with the finite truncation is
\begin{align}
 W^{(L)}(t)&:=\Re\frac{R^{(L)}(1/2-it)}{\Lambda(1/2+it)},\notag\\
 C_k^{(L)}(y)&:=\frac{\sqrt y}{\pi}\int_0^\infty
 W^{(L)}(t)\chi_k(t)K_{it}(2\pi ky)\,\dd t.
 \label{d6c:neutral_formal_weight}
\end{align}
The real part is required: conjugation sends the overlap quotient to
its value at $-t$, whereas $\chi_k$ and $K_{it}$ are real and even.
Stirling's formula gives $|\mathcal G(1/2+it)|\ll t^{-3}e^{-\pi t/2}$.
Together with
$|\Lambda(1/2+it)|=|\zeta(1+2it)|/\sqrt{\cosh\pi t}$, this yields
\be
 |W^{(L)}(t)|\ll
 \left(\sum_{m=1}^L\frac{\sigma_2(m)^2}{m^{5/2}}\right)
 t^{-3}\log(2+t),\qquad t\ge2.
 \label{d6c:neutral_finite_weight_bound}
\ee
There is no singularity at $t=0$, where $1/\Lambda$ vanishes.
Thus $W^{(L)}\in L^1$, and $|H_X|\le1$, $H_X(t)\to1$ give
\begin{align}
 C_k^{(L)}(y)&=e^{-2\pi ky}[b_k^{(L)}+o(1)],\notag\\
 b_k^{(L)}&=\frac1{2\pi\sqrt k}\int_0^\infty
 W^{(L)}(t)\chi_k(t)\,\dd t.
 \label{d6c:neutral_formal_constant}
\end{align}

The continued charge series cannot be summed ordinarily on the
spectral line: there
$|\sigma_2(m)^2/m^{s+2}|=m^{3/2}\sigma_{-2}(m)^2$, with
$1\le\sigma_{-2}(m)\le\zeta(2)$, so its terms do not tend to zero.
This does not establish divergence of the integrated $C_k^{(L)}(y)$
at fixed $y$, or exclude a regularized sum. The change of variable in
\eqref{d6c:neutral_profile_transform_app} probes $y=O(1/m)$; large-$y$
exponential suppression therefore gives no uniform Mellin-tail bound.

Define the omitted transform as $\delta R^{(L)}=R_f-R^{(L)}$, first
in $\Re s>4$ and then by continuation, and set
$\delta W^{(L)}=W-W^{(L)}$, $\delta C_k^{(L)}=C_k-C_k^{(L)}$.
The bound on the finite truncation, together with regularity at the
lower endpoint, gives
\be
 \int_0^\infty|W^{(L)}(t)|\,\dd t
 \ll\sum_{m=1}^L\frac{\sigma_2(m)^2}{m^{5/2}}
 \ll L^{5/2},
 \label{d6c:neutral_finite_uniform_mass}
\ee
uniformly in $L$. Hence $e^{2\pi ky}|C_k^{(L)}(y)|\ll_kL^{5/2}$
for $y>0$. Combining with~\eqref{d6a:continuous_fixed_charge_bound} and
\eqref{d6a:continuous_mass_bound} gives
\begin{align}
 e^{2\pi ky}|\delta C_k^{(L)}(y)|
 &\ll_{k,\epsilon}y^{1/4+\epsilon}+L^{5/2},
 \qquad y,L\ge1,\label{d6c:neutral_tail_leading_test}\\
 \frac1\pi\int_0^T|\delta W^{(L)}(t)\chi_k(t)|\,\dd t
 &\ll_{k,\epsilon}(1+T)^{1/2+\epsilon}+L^{5/2}.
 \label{d6c:neutral_omitted_mass}
\end{align}
At fixed $L$, the omitted continuous contribution is
$o(\sqrt y e^{-2\pi ky})$, taking $0<\epsilon<1/4$; its cumulative
absolute weight is $o(T)$, taking $\epsilon<1/2$. The same normalized
large-$y$ limit holds for $L(y)=o(y^{1/5})$. These bounds do not improve
with $L$, supply no approximation theorem at fixed $y$, and do not
control the cusp contribution.

\subsection{Stability of instanton coefficients}
\label{d6c:stability_proof}
\label{d6c:residual_stability}
For a fixed Fourier mode $k\ge1$, let $R_k$ and $V_k$ denote
the cumulative signed and absolute weights of the spectral
difference, defined in~\eqref{d6c:missing_cumulative_weights}.
Taking absolute values before summing the cusp contributions
ensures that $V_k(T)$ bounds the total variation of the spectral
measure on $[0,T]$, even when several forms share the same
spectral parameter. If $V_k(\infty)<\infty$, dominated
convergence gives
\begin{align}
 |e^{2\pi ky}\delta f_k(y)|&\le\frac{V_k(\infty)}{2\sqrt k}, \notag\\
 e^{2\pi ky}\delta f_k(y)&=\frac{R_k(\infty)}{2\sqrt k}+o(1).
 \label{d8e:finite_mass_bound}
\end{align}
The constant term in the instanton prefactor, when defined,
is preserved precisely when $R_k(\infty)=0$. This holds
also for infinite cusp sums, provided $V_k(\infty)<\infty$.
The remainder is $o(1)$, with no specified rate of decay.

Now assume that $V_k(T)=O(1+T^2)$. We integrate the identity
\be
 e^{-u^2/(2X)}=\frac1X\int_u^\infty t e^{-t^2/(2X)}\,\dd t
\ee
against the spectral weights. Fubini's theorem allows the order of integration to be reversed
when the double integral converges absolutely. Here this follows
from
$X^{-1}\int_0^\infty t e^{-t^2/(2X)}V_k(t)\dd t<\infty$.
Performing the spectral sum and integral first gives the
Gaussian integral in~\eqref{d6c:missing_exact_bound}.
To compare it with the exact Bessel kernel $H_X$, we apply
\eqref{d6a:gaussian} on $0\le t\le T=\sqrt X\log X$:
\be
 \int_{[0,T]}|H_X(t)-e^{-t^2/(2X)}|\,\dd V_k(t)
 \le\frac{e^{1/2}}{8X}V_k(T)=O((\log X)^2).
 \label{d6c:stability_kernel_difference}
\ee
Above $T$, the Bessel bound~\eqref{d6a:uniformtail} and Gaussian decay
dominate the quadratic growth of $V_k$. Summing over intervals with
successively doubled endpoints bounds both tails by every fixed inverse
power of $X$. This proves
\eqref{d6c:missing_exact_bound}. If $|R_k(T)|\le C(1+T^\beta)$,
$\beta>0$, then
\be
 \left|\frac1X\int_0^\infty t e^{-t^2/(2X)}R_k(t)\,\dd t\right|
 \le C[1+(2X)^{\beta/2}\Gamma(1+\beta/2)],
 \label{d8e:rembound}
\ee
proving~\eqref{d6c:missing_power_bound}. More generally, suppose
$R_k(T)=o(T^{2\alpha})$ with $\alpha>0$. For each $\varepsilon>0$,
choose $T_\varepsilon$ such that
$|R_k(t)|\le\varepsilon t^{2\alpha}$ for all $t\ge T_\varepsilon$.
The Gaussian integral is then bounded in absolute value by
$C_{k,\varepsilon}X^{-1}
+\varepsilon(2X)^\alpha\Gamma(1+\alpha)$.
Dividing by $X^\alpha$ and taking first $X\to\infty$ and then
$\varepsilon\to0$ proves~\eqref{d6c:missing_leading_coefficient},
since the remainder in the comparison with the Bessel kernel
is also $o(X^\alpha)$. This remainder need not tend to zero,
so the argument does not establish a constant limit.

We now apply these estimates to the $D^6R^4$ reconstruction
with a fixed spectral cutoff. Let $\mu_k$ denote the full
spectral measure, including both cusp and continuous
contributions, and set
$a_k=\frac{16\zeta(3)\sigma_{-2}(k)}{\pi\sqrt k}$.
The linear bound on the absolute weights~\eqref{d6a:M}
and the relation~\eqref{d6a:addback} between $Z_k$ and $G_k$
imply that
\be
 Z_k(z)=\int_0^\infty(4+t^2)^{-z/2}\,\dd\mu_k(t),
 \qquad\Re z>1,
 \label{d6c:d6_tauberian_input}
\ee
has a meromorphic continuation to $\Re z>-1/2$, with
$Z_k(z)-a_k/(z-1)$ holomorphic there. This follows from
the continuation of $G_k$ in~\eqref{d6a:moment_properties}
and the holomorphy of $Z_k(z+2)$ for $\Re z>-1$.

We use the Wiener--Ikehara theorem~\cite{Ikehara1931}
to relate this residue to the large-$T$ behavior of the
cumulative spectral weight. To apply the theorem to
nonnegative measures, use
$\mu_1\ge0$ and $|\mu_k|\le C_k^{\rm cmp}\mu_1$ from
\eqref{d6a:positive_joint_measure}--\eqref{d6a:charge_domination}.
For $b>C_k^{\rm cmp}$, the measures
$\nu_\pm=b\mu_1\pm\mu_k$ are therefore nonnegative.
Define their cumulative weights by
$A_\pm(x)=\nu_\pm([0,\sqrt{x^2-4}])$ for $x\ge2$ and
$A_\pm(x)=0$ for $1\le x<2$.
Their Mellin--Stieltjes transforms integrate against
$\dd A_\pm(x)$, including both the continuous weights
and the discrete jumps from cusp forms:
\be
 \int_1^\infty x^{-z}\,\dd A_\pm(x)=bZ_1(z)\pm Z_k(z),
 \qquad\Re z>1.
 \label{d6c:d6_positive_combinations}
\ee
The transforms converge absolutely for $\Re z>1$ and, after
subtracting $(ba_1\pm a_k)/(z-1)$, extend holomorphically
across the line $\Re z=1$. The theorem therefore gives
$A_\pm(x)=(ba_1\pm a_k)x+o(x)$.
Subtracting these two asymptotics and setting
$x=\sqrt{T^2+4}$ yields $\mu_k([0,T])=a_kT+o(T)$.

For a fixed spectral cutoff $T_0$, define $\delta f$ as
the full reconstruction minus the truncated one.
The omitted spectral weights satisfy
\be
 R_k(T)=\mu_k((T_0,T])=a_kT+o(T),
 \qquad V_k(T)\le C_k(1+T).
 \label{d6c:d6_fixed_cutoff_weights}
\ee
The continuous absolute weight is $o(T)$ by
\eqref{d6a:continuous_mass_bound}, so the cusp weights alone have this
linear asymptotic. Keeping $S_k$ exact and truncating both spectra at
$T_0$, insert $R_k(t)=a_kt+o(t)$ into
\eqref{d6c:missing_exact_bound}. Dominated convergence under
$t=\sqrt X\,v$ gives
\begin{align}
 e^{2\pi ky}\delta f_k(y)
 &=\frac{a_k}{2\sqrt k}\sqrt{\frac{\pi X}{2}}+o(\sqrt X)
 \notag\\
 &=\frac{8\zeta(3)\sigma_{-2}(k)}{\sqrt k}\sqrt y+o(\sqrt y),
 \notag\\
 e^{2\pi ky}\delta C_k(y)&=o(\sqrt y),\qquad X=2\pi ky.
 \label{d6c:d6_tail_Gaussian_details}
\end{align}
Here the initial finite interval contributes $O(X^{-1})$, and
$|R_k(t)|\ll1+t$ controls the rescaled integral. The retained spectral
prefactor is bounded by $|\mu_k|([0,T_0])/(2\sqrt k)$, while $S_k$ is
bounded at this order. Hence every fixed cusp truncation misses the
entire leading square-root term, even with the continuum exact. This
comparison does not preserve the complete low-mode data.

We can also control the cusp contribution through the complete
first Fourier mode after subtracting the restored Eisenstein
and continuous terms. Assume that the two functions share the
restored Eisenstein term and that the difference of their cusp
projections lies in $L^2$. Their first-mode cusp difference is
then $\delta h_1=\delta f_1-\delta C_1$.
The damped Mellin argument of
Appendix~\ref{d6c:Mellin_justification} gives
\begin{align}
 \delta\mathcal M_{h,d}(s)
 &=\int_0^\infty y^{s-3/2}e^{-2\pi y}\delta h_1(y)\,\dd y,
 \qquad\Re s>\tfrac12,\notag\\
 \delta d_j&=\frac{2\pi^{it_j}}{\Gamma(it_j)}
 \Res_{s=it_j}\delta\mathcal M_{h,d}(s),
 \label{d6c:missing_residual_residue}
\end{align}
where the second line uses the standing even-sector simplicity
assumption. Choose $0<\eta_j<1/4$ so that the closed disk about $it_j$
contains no other possible kernel pole; no uniform lower bound on
$\eta_j$ is required. If, for $0<q<2$,
\be
 \eta_j\sup_{|s-it_j|=\eta_j}|\delta\mathcal M_{h,d}(s)|
 \ll e^{-\pi t_j/2}(1+t_j)^{-q-1/2},
 \label{d6c:missing_residual_contour_bound}
\ee
uniformly in $j$, the residue theorem and
$|\Gamma(it)|^2=\pi/[t\sinh(\pi t)]$ give
$|\delta d_j|\ll(1+t_j)^{-q}$.
The Weyl bound states that the number of cusp forms with
$t_j\le T$, counted with multiplicity, is
$O((1+T)^2)$~\cite{IwaniecSpectral}.
Combining this with the Hecke bound
\eqref{d6a:elementary_Hecke_bound} and summing over intervals
$[2^n,2^{n+1})$, with the bounded range $t_j<1$ treated
separately, gives
\begin{align}
 \sum_{t_j\le T}|\delta d_j\lambda_j(k)|&\ll_k(1+T)^{2-q},\notag\\
 e^{2\pi ky}|\delta U_k(y)|&\ll_k y^{1-q/2},\qquad y\ge1.
 \label{d6c:missing_residual_cusp_control}
\end{align}
Thus $q>1$ protects a square-root coefficient and $q>0$ a linear one
against the cusp difference. A continuous absolute weight
$O(1+T^\gamma)$, $0<\gamma\le2$, combines with this estimate through
$\beta=\max(2-q,\gamma)$ in~\eqref{d6c:missing_power_bound}.
Condition~\eqref{d6c:missing_residual_contour_bound} concerns analytic
continuation, not finite asymptotic data: converting the residue to a
cusp amplitude introduces a factor of order $t_j^{1/2}e^{\pi t_j/2}$.
Indeed, the form with unit first Fourier amplitude
$u_j/\rho_j(1)$ has
$\sup_{1\le y\le2}|\sqrt yK_{it_j}(2\pi y)|
\ll t_j^{-1/2}e^{-\pi t_j/2}\to0$ by
\eqref{d6c:complete_first_order_bound}, precluding a uniform bound on cusp
amplitudes in terms of this norm.

\subsection{Modular ambiguity from an infinite cusp sum}
\label{d6c:infinite_cusp_proof}
We establish the convergence and Fourier asymptotics of the
cusp sum $\Psi_m$ defined in~\eqref{d6c:infinite_cusp_construction}.
Its zero mode vanishes, and the coefficient of $y e^{-2\pi ky}$
in its $k$th Fourier mode is $\delta_{mk}$ for $k\ge1$.
For fixed $m,k\ge1$, define the positive weights
$\omega_j=q_2(t_j)|\rho_j(1)|^2/\cosh(\pi t_j)$ and the
Gaussian spectral sum
$\mathcal S_{m,k}(X)=\sum_{j\ {\rm even}}
\omega_j\lambda_j(m)\lambda_j(k)e^{-t_j^2/(2X)}$.
Averaging the Kuznetsov formulas for $(m,k)$ and $(m,-k)$
selects the even cusp forms and halves the diagonal term.
With $a=4\pi\sqrt{mk}$, our normalization gives
\begin{align}
 &\sum_{j\ {\rm even}}
 \frac{\overline{\rho_j(m)}\rho_j(k)}{\cosh(\pi t_j)}h(t_j)
 +\frac1\pi\int_{\mathbb R}
 \frac{h(t)\chi_m(t)\chi_k(t)}{|\zeta(1+2it)|^2}\,\dd t\notag\\
 &\quad=\frac{\delta_{mk}}2H_0(h)
 +\frac12\sum_{c\ge1}\frac{
 S(m,k;c)H^+(a/c;h)+S(m,-k;c)H^-(a/c;h)}c.
 \label{d6c:infinite_cusp_even_Kuznetsov}
\end{align}
This identity follows from~\cite{BlomerCorbett}
after multiplying by two to match our normalization
\eqref{d6a:cusp_norm_for_weight}; see also
\cite{HumphriesKhanRandomWave}.
The transforms $H_0,H^+$ and the Kloosterman sum $S$ are
defined in~\eqref{d6a:kuz_diagonal_transform}--
\eqref{d6a:Kloosterman_definition}, while the opposite-sign
transform is
\begin{align}
 H^-(x;h)&=\frac4{\pi^2}\int_{\mathbb R}
 K_{2it}(x)t\sinh(\pi t)h(t)\,\dd t\notag\\
 &=\frac{2i}{\pi}\int_{\mathbb R}
 I_{2it}(x)\frac{th(t)}{\cosh(\pi t)}\,\dd t.
 \label{d6c:infinite_cusp_minus_transform}
\end{align}
The second expression follows by expressing $K_{2it}$ in terms
of $I_{\pm2it}$~\cite{DLMF}, then changing
$t\to-t$ in the $I_{-2it}$ term and using the evenness of $h$.

Take $h_X=q_2e^{-t^2/(2X)}$, $X\ge1$, which is even, holomorphic in
$|\Im t|\le2/3$ and rapidly decreasing there. Since
$q_2=1-15/[4(t^2+4)]$,
\begin{align}
 &\tfrac12H_0(h_X)=X/\pi^2+O(\log(2+X)), \notag\\
 &\left|\frac1\pi\int_{\mathbb R}
 \frac{h_X(t)\chi_m(t)\chi_k(t)}{|\zeta(1+2it)|^2}\,\dd t\right|
 \ll_{m,k}\sqrt X\log^2(2+X).
 \label{d6c:infinite_cusp_continuous_bound}
\end{align}
The first estimate uses $\int_{\mathbb R}|t|e^{-t^2/(2X)}\dd t=2X$;
replacing $t\tanh\pi t$ by $|t|$ introduces an $O(1)$ remainder, while
$q_2-1$ contributes $O(\log X)$. The second uses $|\chi_n|\le d(n)$ and
$|\zeta(1+2it)|^{-1}\ll\log(2+|t|)$~\cite{Yang}.

To estimate $H^+$ and $H^-$, we shift the contours in their
$J_{2it}$ and $I_{2it}$ representations to $t=u-i$.
The zero of $q_2$ at $t=-i/2$ cancels the pole of
$1/\cosh(\pi t)$, while the poles of $q_2$ at $\pm2i$
lie outside the strip crossed by the contour.
For $0<x\le a$, the defining power series of
$J_\nu$ and $I_\nu$~\cite{DLMF} give
$|J_{2+2iu}(x)|+|I_{2+2iu}(x)|
\ll_a x^2/|\Gamma(3+2iu)|$.
Indeed, after extracting the leading factor, each remaining
series is bounded in absolute value by
$\sum_{r\ge0}(a^2/4)^r/[r!(3)_r]$.
Stirling's formula then bounds the shifted integrands by
$x^2(1+|u|)^{-3/2}e^{(1-u^2)/(2X)}$,
whose integral over $u\in\mathbb R$ is bounded uniformly
for $X\ge1$. The Gaussian ensures that the integrals along
the vertical sides vanish as the contour extends to infinity.
Consequently,
$|H^+(x;h_X)|+|H^-(x;h_X)|\ll_a x^2$.
Together with $|S(m,\pm k;c)|\le c$, this bounds the
$c$th term in the arithmetic sum by $O_{m,k}(c^{-2})$,
so the sum is $O_{m,k}(1)$. It follows that
\be
 \mathcal S_{m,k}(X)=\frac{\delta_{mk}X}{\pi^2}
 +O_{m,k}(\sqrt X\log^2(2+X)).
 \label{d6c:infinite_cusp_Gaussian_asymptotic}
\ee

Positivity at $m=k=n$, with $X=\max(1,T^2)$, gives
$M_n(T):=\sum_{t_j\le T}\omega_j\lambda_j(n)^2\ll_n1+T^2$.
Cauchy--Schwarz therefore implies
\be
 V_{m,k}(T):=\sum_{t_j\le T}\omega_j|\lambda_j(m)\lambda_j(k)|
 \le\sqrt{M_m(T)M_k(T)}\ll_{m,k}1+T^2.
 \label{d6c:infinite_cusp_absolute_weight}
\ee
For the coefficients of~\eqref{d6c:infinite_cusp_construction}, omitting
its overall $\pi/\sqrt m$, put
$\alpha_j=q_2(t_j)\overline{\rho_j(m)}/\cosh(\pi t_j)$. Then
\be
 |\alpha_j|^2=\frac{q_2(t_j)}{\cosh(\pi t_j)}
 \omega_j\lambda_j(m)^2,
 \qquad\sum_j(1+t_j^2)^{2r}|\alpha_j|^2<\infty\quad(r\ge0).
 \label{d6c:infinite_cusp_Ltwo}
\ee
Thus the series converges in the cuspidal $L^2$ subspace.
To control derivatives, we use the extension of the contour
bound~\eqref{d6c:Mellin_Bessel_decay},
$|K_{b+it}(z)|\le e^{-\theta t}K_{|b|}(z\cos\theta)$,
valid for integer $b$, $z>0$, $t\ge0$ and $0<\theta<\pi/2$.
Combining this with
$K_b(z)\le\sqrt{\pi/(2z)}e^{-z+b^2/(2z)}$
and the Hecke bound gives the following bound on the sum
of the absolute values of the differentiated terms, for
each fixed derivative order and uniformly for $y\ge y_0>0$:
\be
 \left(\sum_j\omega_j|\lambda_j(m)|e^{-\theta t_j}\right)
 \left(\sum_{n\ge1}n^A d(n)e^{-2\pi ny\cos\theta}\right)<\infty,
 \label{d6c:infinite_cusp_derivative_majorant}
\ee
where $A$ depends on the derivative orders. Local uniform convergence
passes modularity and evenness from partial sums to the limit and
allows termwise Fourier integration. The zero mode vanishes; reality
follows from the real products $\omega_j\lambda_j(m)\lambda_j(k)$.
The same bound gives uniform decay of every fixed derivative as
$O(e^{-cy})$, $0<c<2\pi$, independently of the phases of $u_j$.

Finally, with $X=2\pi ky$,
\begin{align}
 e^X(\Psi_m)_k(y)
 &=\frac{\pi}{2\sqrt{mk}}\sum_{j\ {\rm even}}
 \omega_j\lambda_j(m)\lambda_j(k)H_X(t_j)\notag\\
 &=\delta_{mk}y+O_{m,k}(\sqrt y\log^2(2+y)).
 \label{d6c:infinite_cusp_exact_asymptotic}
\end{align}
The quadratic bound~\eqref{d6c:infinite_cusp_absolute_weight} permits
replacing the Bessel kernel by the Gaussian with an $O(\log^2X)$ remainder.
The diagonal contribution is
$\pi X/(2\pi^2\sqrt{mk})=y$ for $m=k$, proving
\eqref{d6c:infinite_cusp_limits} without a simplicity assumption.
This addition preserves the complete zero mode and the leading
first-mode coefficient for $m\ge2$, not the complete first mode.

\subsection{Preserving the full one-instanton expansion}
\label{d8e:full_sector_ambiguity}
We determine how $H_{\rm amb}$ in
\eqref{d8e:ambiguity_full_sector_construction} changes the
zero, first and second Fourier modes. Expanding both its
numerator and its reciprocal denominator identifies the
Fourier index and exponential suppression of each contribution.
With $q=e^{2\pi i\tau}$, define
\begin{align}
 \Delta_{\rm hol}(\tau)&=q\prod_{n\ge1}(1-q^n)^{24},\notag\\
 E_{24}^{\rm hol}(\tau)&=\frac1{2\zeta(24)}
 \sum_{(c,d)\ne(0,0)}(c\tau+d)^{-24}
 =1+\gamma_{24}\sum_{n\ge1}\sigma_{23}(n)q^n,\notag\\
 \gamma_{24}&=-48/B_{24}=131040/236364091.
 \label{d8e:ambiguity_holomorphic_factors}
\end{align}
Here $B_{24}$ is the 24th Bernoulli number. The discriminant has weight $12$~\cite{DLMF}, while reindexing
the lattice sum shows that $E_{24}^{\rm hol}$ has weight $24$.
Each product in the numerator of
\eqref{d8e:ambiguity_full_sector_construction} therefore has
weights $(24,24)$, canceled by the factor $y^{24}$.
The denominator $E_{23}$ is modular invariant and strictly
positive by its convergent defining series.
Hence $H_{\rm amb}$ is modular and real analytic throughout
the upper half-plane. The conjugate terms make it real,
and the real Fourier coefficients of its factors make it even.
The cusp decay established below also implies
$H_{\rm amb}\in L^2$.

Separate the zero Fourier mode of $E_{23}$ from its nonzero
modes and expand the reciprocal for sufficiently large $y$:
\begin{align}
 E_{23}(x+iy)&=y^{23}A(y)+B(x,y),
 \qquad A(y)=1+\varphi(23)y^{-45},\notag\\
 \frac{y^{24}}{E_{23}(x+iy)}
 &=\frac{y}{A(y)}\sum_{\ell\ge0}
 \left[-\frac{B(x,y)}{y^{23}A(y)}\right]^\ell.
 \label{d8e:ambiguity_denominator_expansion}
\end{align}
Here $B$ contains only nonzero Fourier modes.
Equation~\eqref{intro:Ezero0} and the finite expression for
$K_{45/2}$ give
$B_n(y)=O((1+|n|)^{22}e^{-2\pi|n|y})$
for $n\ne0$ and $y\ge1$.
Thus $B(x,y)=O(e^{-2\pi y})$ uniformly in $x$, ensuring
absolute and uniform convergence of the reciprocal expansion
for sufficiently large $y$.
A product of denominator modes with total Fourier index $n$
has an exponential factor no larger than $e^{-2\pi|n|y}$,
since the sum of their absolute indices is at least $|n|$.

A numerator monomial $q^r\bar q^s$ in the first branch has $r\ge2$,
$s\ge0$ and charge $r-s$. Its contribution at charge $k$ therefore
has exponential index at least $r+s+|k-r+s|$. Including the conjugate
branch, the minimum indices at $k=0,1,2$ are $4,3,2$. Only $q^2$ with
the constant denominator attains index two at charge two; the next
index is four. Absolute convergence gives
\begin{align}
 (H_{\rm amb})_0(y)&=O(ye^{-8\pi y}),\notag\\
 (H_{\rm amb})_1(y)&=O(ye^{-6\pi y}),\notag\\
 (H_{\rm amb})_2(y)&=
 \frac{ye^{-4\pi y}}{1+\varphi(23)y^{-45}}+O(ye^{-8\pi y}).
 \label{d8e:ambiguity_full_sector_modes}
\end{align}
The conjugate $\bar q^2$ belongs to charge $-2$, so the leading
charge-two coefficient is one. Moreover,
\begin{align}
 \lim_{y\to\infty}\frac{e^{6\pi y}}y(H_{\rm amb})_1(y)&=\gamma_{24}, \notag\\
 \lim_{y\to\infty}\frac{e^{8\pi y}}y(H_{\rm amb})_0(y)
 &=2\gamma_{24}(1+2^{23}).
 \label{d8e:ambiguity_changed_low_modes}
\end{align}
These coefficients come from the numerator terms $q^2\bar q$
and $q^2\bar q^2$. Contributions involving nonzero denominator
modes at the same exponential orders carry an additional
factor $y^{-23}$.
Thus $cH_{\rm amb}$ preserves the specified asymptotic sectors
of the zero and first modes while shifting the coefficient of
$y e^{-4\pi y}$ in the second mode by $c$.
Both low modes change at more suppressed exponentials.
This construction does not address whether the same shift is
possible with both the complete zero mode and the full
one-instanton expansion fixed, nor does it establish the
physical admissibility of the addition.

\section{Normalization of the \texorpdfstring{$D^8R^4$}{D8R4} amplitude}
\label{d8e:amplitude_details}

We derive the perturbative terms and leading one-instanton
coefficient quoted in Section~\ref{d8e:input} from the corresponding
string amplitudes, in the normalization of~\eqref{d8e:ampdef}.
We also determine how the local coefficient depends on the
threshold subtraction scale. We first expand the amplitudes
in momenta at fixed $y$, then study the weak-coupling behavior
of the resulting coefficients.

Let $s_{\rm phys},t_{\rm phys},u_{\rm phys}$ denote the dimensionful
Mandelstam invariants and define
\begin{align}
 a&=\frac{\alpha's_{\rm phys}}4,\qquad
 b=\frac{\alpha't_{\rm phys}}4,\qquad
 c=\frac{\alpha'u_{\rm phys}}4,
 \qquad a+b+c=0,
 \label{d8e:stringinvariants}\\
 a&=\frac{\sqrt y}{4}S,\qquad
 b=\frac{\sqrt y}{4}T,\qquad
 c=\frac{\sqrt y}{4}U.
 \label{d8e:main_input_conversion}
\end{align}
The second line follows from $\ell_P^8=g_s^2\alpha'^4$ and
$S=\ell_P^2s_{\rm phys}$, with analogous definitions for
$T$ and $U$~\cite{Agmon}.
After factoring out the gravitational coupling and the
four-graviton kinematic tensor, the sphere amplitude
takes the form~\cite{GreenVanhove9910056}
\be
 \mathcal A_0(a,b,c)=\frac1{abc}
 \prod_{z=a,b,c}\frac{\Gamma(1-z)}{\Gamma(1+z)}.
 \label{d8e:VS}
\ee
Define $\sigma_p=a^p+b^p+c^p$. Since $a+b+c=0$, we have
$\sigma_3=3abc$, $\sigma_4=\sigma_2^2/2$ and
$\sigma_7=7\sigma_2^2\sigma_3/12$. Expanding the gamma
quotient then gives
$\mathcal A_0\big|_{D^8R^4}=\zeta(7)\sigma_4$.

At genus $h$, normalization by the tree-level supergravity
amplitude introduces a factor $abc$, while the relative
coupling factor is $g_s^{2h}=y^{-2h}$.
A polynomial term $A_{8,h}\sigma_4$ in the scalar amplitude
therefore contributes
\be
 \mathcal F_{8;h}=4^{-7}y^{7/2-2h}A_{8,h}.
 \label{d8e:polynomial_conversion}
\ee

\paragraph{Perturbative thresholds and scale dependence.} The genus-one polynomial and threshold terms combine to
give~\cite{DHokerGreen,CD2412}
\be
 \mathcal A_1\big|_{D^8R^4}
 =K_1\sum_{z=a,b,c}z^4[C-\log(-z-i0)],
 \qquad K_1=\frac{8\pi^2\zeta(3)}{45},
 \label{d8e:g1CD}
\ee
where $C$ is given in~\eqref{d8e:g1DG} and $K_1$ includes
the overall factor $2\pi^2/3$.
At a fixed dimensionless subtraction scale $\mu>0$,
$\log(-a-i0)=\log[(-S-i0)/\mu]+\log(\mu\sqrt y/4)$.
The first term is the threshold logarithm, leaving the local
contribution
\be
 \mathcal F_{8,0;h=1}^{(\mu)}(y)
 =\kappa_1y^{3/2}
   \left[C+\log\frac4\mu-\frac12\log y\right],
 \qquad \kappa_1=\frac{K_1}{4^7}
 =\frac{\pi^2\zeta(3)}{92160}.
 \label{d8e:g1local}
\ee
Here $\mu$ is held fixed in Einstein units. A $y$-dependent
scale would change the modular transformation law of the
separated local coefficient.

The genus-two logarithm follows from the nonseparating
degeneration of the regulated genus-two integral
$I_2(L)=(\pi/4)\int_{\mathcal F_2(L)}\dd\mu_2\,
\mathcal B_{(2,0)}^{(2)}$.
The modular invariant $\mathcal B_{(2,0)}^{(2)}$ is the
coefficient of $\sigma_2^2/2!=\sigma_4$ in the low-energy
expansion of the genus-two four-graviton integrand
\cite{Pioline1810}.
Here $L$ cuts off the long-tube length
$t=\Im\sigma-(\Im v)^2/\Im\rho$ in the period matrix, and
$\rho$ is the remaining torus modulus.
For $\mathcal B_{(2,0)}^{(2)}=(Z_1-2Z_2+Z_3)/2$, with the
$Z_i$ defined in~\cite{DGP1806}, the
boundary average is $8\pi^2t^2/45+O_\rho(1)$
\cite{DGP1806,Pioline1810}.
The genus-two integral has a logarithmic dependence on the
cutoff $L$ as $L\to\infty$. The degeneration analysis
of~\cite{Pioline1810} gives the coefficient of $\log L$:
\be
 K_2:=\lim_{L\to\infty}L\partial_LI_2(L)
 =\frac\pi2\frac{8\pi^2}{45}\frac{2\pi}{3}
 =\frac{8\pi^4}{135}.
 \label{d8e:K2}
\ee
The cutoff logarithm cancels against the contribution from
the excluded long-tube region, fixing the overall threshold
coefficient. The individual channel
logarithms follow from the massless cut with a supergravity
amplitude on one side and an $R^4$ amplitude on the
other~\cite{HPS2407}.
Together these results give
\begin{align}
 \mathcal A_2\big|_{D^8R^4}
 &=\sum_{z=a,b,c}z^4[b_2-K_2\log(-z-i0)], \notag\\
 b_2&=\lim_{L\to\infty}[I_2(L)-K_2\log L]+b_{\mathrm{tube}}.
 \label{d8e:b2}
\end{align}
The finite contributions from the interior and tube regions
in $b_2$ are not fixed by the supplied input.
Using~\eqref{d8e:polynomial_conversion}, we obtain
\be
 \mathcal F_{8,0;h=2}^{(\mu)}(y)
 =\frac{y^{-1/2}}{4^7}
   \left[b_2+K_2\log\frac4\mu-\frac{K_2}{2}\log y\right].
 \label{d8e:g2local}
\ee
At $\mu=1$, this reproduces~\eqref{d8e:zeroinput}, with
$a_2=(b_2+K_2\log4)/4^7$.

The logarithmic threshold arises from intermediate massless
supergravity states going on shell. At this derivative order,
its discontinuity is determined by a two-particle unitarity
cut, with a supergravity amplitude on one side and an amplitude
containing one $R^4$ vertex on the other~\cite{HPS2407}.
The threshold is therefore linear in the $R^4$ coefficient
$\mathcal F_0$. Its normalization is fixed by the perturbative
terms
\be
 \left.(\mathcal F_0)_0(y)\right|_{h=0,1}
 =\frac{\zeta(3)}{32}y^{3/2}
   +\frac{\pi^2}{96}y^{-1/2}
 \label{d8e:R4normalization}
\ee
and the two logarithmic coefficients derived
above~\cite{GreenVanhove9910056,DHokerGreen,Pioline1810}.
The full threshold coefficient involves the exact modular
$R^4$ coupling $\mathcal F_0$, including its D-instanton
contributions~\cite{Pioline1810}.
The threshold therefore contributes $\mathcal F_0H_8^{(\mu)}$
to $M/STU$. Its momentum dependence and the compensating
change of the local coefficient are
\begin{align}
 H_8^{(\mu)}(S,T,U)
 &=-\frac{\pi^2}{2880}\sum_{P=S,T,U}P^4
       \log\frac{-P-i0}{\mu},
 \label{d8e:thresholdkernel}\\
 \mathcal F_8^{(e^\delta\mu)}-\mathcal F_8^{(\mu)}
 &=-\frac{\pi^2}{2880}\delta\mathcal F_0.
 \label{d8e:scaleshift}
\end{align}
This determines the scale dependence, while the local
coefficient at a reference scale remains an independent input.

\paragraph{Single-instanton contribution.} We extract the leading single-instanton contribution to
$D^8R^4$ from the result of Agmon et al.~\cite{Agmon}.
With the four-graviton kinematic tensor factored out, the
scalar amplitude contains
\be
 \frac{F^{(1,0)}}{\mathcal P}
 \supset \frac{e^{2\pi i\tau}}y
   \left[C_{\rm disk}-\frac18\sum_{z=a,b,c}z\psi(1-z)\right],
 \qquad
 \mathcal P=2^{-9/2}\pi^{-21/4}\kappa^{7/2}.
 \label{d8e:Agmon}
\ee
Here $\kappa$ is the ten-dimensional gravitational coupling,
$\psi=\Gamma'/\Gamma$ is the digamma function, and $(1,0)$
labels the sector containing a single D$(-1)$-instanton
and no anti-instanton. The momentum-independent constant
$C_{\rm disk}$ contributes only to the $R^4$ coupling.

To convert this result to the supergravity-normalized
amplitude $M$, we use the gravitational and instanton
normalizations of~\cite{Agmon,SenSingle}:
\begin{align}
 F_{\rm SUGRA}
 &=\frac{\kappa^2}{s_{\rm phys}t_{\rm phys}u_{\rm phys}},
 \qquad \kappa^2=64\pi^7\alpha'^4y^{-2},
 \label{d8e:gravitationalnormalization}\\
 \mathcal P&=64\pi^7\alpha'^7y^{-7/2},
 \qquad F_{\rm SUGRA}=\frac{\mathcal P}{STU},
 \qquad M=STU\frac{F}{\mathcal P}.
 \label{d8e:Pcancel}
\end{align}
For $|a|,|b|,|c|<1$, the Taylor series of the digamma
function~\cite{DLMF} gives the quartic term
$\zeta(4)\sigma_4/8$ in the bracket
of~\eqref{d8e:Agmon}. Converting to Planck units with
$a=\sqrt y\,S/4$ and the analogous relations for $b$ and
$c$ gives
\be
 \frac{M}{STU}\bigg|_{D^8R^4,\,(1,0)}
 \supset e^{2\pi ix}e^{-2\pi y}\frac{\zeta(4)y}{2048}
      (S^4+T^4+U^4),
 \label{d8e:main_disk_extraction}
\ee
recovering the leading first-mode coefficient
in~\eqref{d8e:firstasymp}.

To evaluate the scale shift in~\eqref{d8e:scaleshift},
we also need the leading first Fourier mode of the $R^4$
coupling. In the same normalization, this is~\cite{Agmon}
\be
 (\mathcal F_0)_1(y)=\frac\pi{16}e^{-2\pi y}
       +O(y^{-1}e^{-2\pi y}).
 \label{d8e:R4firstmode}
\ee
Substituting into~\eqref{d8e:scaleshift} gives
\be
 \mathcal F_{8,1}^{(e^\delta\mu)}(y)
 -\mathcal F_{8,1}^{(\mu)}(y)
 =-\frac{\pi^3}{46080}\delta e^{-2\pi y}
     +O_\delta(y^{-1}e^{-2\pi y}).
 \label{d8e:firstscaleshift}
\ee
For a $y$-independent change of scale, this shift is
subleading to $y e^{-2\pi y}$. The leading coefficient
in~\eqref{d8e:firstasymp} is therefore unchanged.

\section{Comparison with the GMV Fourier modes}
\label{rev:selected_numerics}
\label{check:general_constant}
\label{check:GMV_analytic}
\label{check:GMV_numerical}
\label{check:finite_y_reconstruction}

We compare the reconstruction with GMV, first through the square-root
and constant terms at weak coupling for every positive charge, then numerically at charges
$k=5,10$ and finite $y$.

\paragraph{Analytic comparison.}
In the GMV expression, the terms proportional to
$\sqrt yK_{7/2}(2\pi ky)$ cancel exactly, before any large-$y$
expansion:
\be
 2\alpha_{0,k}
 +\sum_{\substack{a+b=k\\ab\ne0}}\alpha_{a,b}=0,
 \qquad k\ge1.
 \label{check:GMV_general_homogeneous_cancellation}
\ee
This cancellation follows from the divisor identity
of~\cite{FKR}, applied
to the coefficients in~\cite{GMV}. For $k=1$, it was derived explicitly
in Appendix~\ref{d6c:input_coefficients}.

The remaining terms with one zero index give
$8\zeta(3)\sigma_2(k)/k^{5/2}$ as the coefficient of
$\sqrt y e^{-2\pi ky}$, with no constant term in the
prefactor multiplying $e^{-2\pi ky}$
\cite{GMV}.
The Bessel-product terms with opposite-sign indices have total
contribution $e^{-2\pi ky}O_k(e^{-4\pi y})$.
Indeed, the Bessel bounds of
Appendix~\ref{d6c:input_coefficients} apply with $p=m+k$,
while the coefficients grow at most polynomially in $m$
at fixed $k$, using
$\sigma_{-2}(m)\sigma_{-2}(m+k)\le\zeta(2)^2$.

For integers $a,b\ge1$ satisfying $a+b=k$, expansion
of~\cite[Eqs.~(2.35)--(2.39)]{GMV} gives
\be
 e^{2\pi ky}f^P_{a,b}(y)
 =\frac{64\pi^2\sigma_2(a)\sigma_2(b)}{3k^4\sqrt{ab}}
   (6ab-a^2-b^2)+O_{a,b}(y^{-1}).
 \label{check:GMV_positive_pair_constant}
\ee
Summing over these ordered pairs reproduces
$B_k$ in~\eqref{d6a:constant_general}, while
$\sigma_2(k)=k^2\sigma_{-2}(k)$ identifies the square-root
coefficient. Both coefficients in~\eqref{d6a:opening_result}
therefore agree analytically with GMV for every positive charge.

\paragraph{Numerical comparison.}
We compare the Fourier modes at $k=5,10$ and finite $y$ using
the exact spectral representation~\eqref{d6a:split} and the
full GMV expression. No large-$y$ expansion is used in either
calculation. Retaining the first $N$ even Maass cusp forms
for $SL(2,\mathbb Z)$, ordered by spectral parameter, we evaluate
\be
 f_k^{[N]}(y)
 =S_k(y)+C_k(y)
 +\sqrt y\sum_{j=1}^{N}
 d_j\lambda_j(k)K_{it_j}(2\pi ky).
 \label{check:fifteen_form_reconstruction}
\ee
The restored Eisenstein terms, continuous contribution and
Bessel functions retain their full $y$ dependence.

For the first thirty forms, we use Fourier coefficients
downloaded from the LMFDB at their full available precision
\cite{LMFDB,LMFDBreliability,LMFDBcomplete}.
For subsequent forms, we use the coefficients rounded to
six decimal places in the published LMFDB
dataset~\cite{LMFDBMaassData}. The spectral parameters are
taken from the full-precision LMFDB metadata throughout.
The calculations are performed in Python using
\texttt{mpmath} and \texttt{python-flint}. We compute the
completed $L$-values by modular inversion and the norms by
integration over the fundamental domain. We use 50--60 decimal digits of working precision for the
Maass integrals and 65 digits for the continuous contribution.

After the cancellation
in~\eqref{check:GMV_general_homogeneous_cancellation},
the GMV expression is
\be
 f_k(y)=2f^P_{0,k}(y)
 +\sum_{a=1}^{k-1}f^P_{a,k-a}(y)
 +2\sum_{m\ge1}f^P_{m+k,-m}(y).
 \label{check:GMV_finite_y_sum}
\ee
The finite sum includes both orderings of unequal positive
indices; an equal-index pair occurs once. The factors of two
account for reversed orderings in the other terms.
We evaluate this expression with 100 decimal digits of
working precision, truncating the opposite-sign sum at
$m=10$. The cutoffs $m=6$ and $m=12$ give the same
displayed GMV values.

\begin{table}[H]
\centering
\small
\setlength{\tabcolsep}{4pt}
\begin{tabular}{cccccc}
 $k$ & $y$ & $N$ & $t_N$ & GMV value $e^{2\pi ky}f_k(y)$
 & $10^9|f_k^{[N]}-f_k|/|f_k|$\\\hline
 $5$ & $1$ & $10$ & $28.530748$
 & $409.78371724022864001$ & $0.948$\\
 $5$ & $2$ & $30$ & $41.555578$
 & $392.30896724269344176$ & $2.34$\\
 $5$ & $3$ & $50$ & $50.089705$
 & $387.24917229085089503$ & $1.05$\\\hline
 $10$ & $1$ & $25$ & $39.168085$
 & $1597.3750311147924238$ & $1.79$\\
 $10$ & $2$ & $60$ & $53.707586$
 & $1560.9911374361263289$ & $1.86$\\
 $10$ & $3$ & $100$ & $65.776665$
 & $1549.5631128568051178$ & $3.69$
\end{tabular}
\caption{Finite-$y$ comparison with the GMV Fourier modes at
$k=5,10$. The GMV column displays $e^{2\pi ky}f_k(y)$.
Here $N$ is the number of retained even cusp forms
and $t_N$ is the largest retained spectral parameter, rounded
to six decimal places. The continuous integral is truncated
at $t=160$ and the GMV opposite-sign sum at $m=10$.
The last column gives the observed relative differences
multiplied by $10^9$.}
\label{check:finite_y_table}
\end{table}

Table~\ref{check:finite_y_table} compares the reconstruction
with GMV for $k=5,10$ at $y=1,2,3$, with observed relative
differences below $4\times10^{-9}$ in all six cases.
For $k=5$, retaining the same ten cusp forms at all three
values of $y$ gives relative differences of
$9.48\times10^{-10}$, $4.75\times10^{-6}$ and
$6.13\times10^{-5}$, respectively.
Increasing the number of retained forms as $y$ grows
restores comparable agreement.
This behavior is consistent with the analysis in
Section~\ref{d6a:section}: the square-root term comes from
the infinite cusp sum, through the growing range
$t_j=O(\sqrt y)$ at fixed charge, whereas every finite
cusp sum has a bounded instanton prefactor.

\bibliographystyle{JHEP}
\bibliography{references}
\end{document}